\pdfoutput=1
\documentclass[aps, prd, floats, floatfix, superscriptaddress, twocolumn, PRD, nofootinbib, amsmath, amssymb]{revtex4-2}

\usepackage[T1]{fontenc}
\usepackage[utf8]{inputenc}
\usepackage{lmodern}
\usepackage{microtype}
\usepackage[english]{babel}

\usepackage{MnSymbol}
\usepackage{eufrak}
\usepackage{bm}
\usepackage{slashed}
\usepackage{phoenician}

\usepackage{graphicx}
\usepackage{subcaption}
\usepackage{multirow}
\usepackage{ragged2e}
\usepackage{enumerate}
\usepackage{appendix}
\usepackage{import}
\usepackage{makeidx}
\usepackage{xspace}

\usepackage{lipsum}
\usepackage{blindtext}

\usepackage[dvipsnames]{xcolor}
\usepackage[normalem]{ulem}   

\usepackage{listings}
\definecolor{codegreen}{rgb}{0,0.6,0}
\definecolor{codegray}{rgb}{0.5,0.5,0.5}
\definecolor{codepurple}{rgb}{0.58,0,0.82}
\definecolor{backcolour}{rgb}{0.95,0.95,0.92}
\lstdefinestyle{mystyle}{
    backgroundcolor=\color{backcolour},
    commentstyle=\color{codegreen},
    keywordstyle=\color{magenta},
    numberstyle=\tiny\color{codegray},
    stringstyle=\color{codepurple},
    basicstyle=\ttfamily,
    breakatwhitespace=false,
    breaklines=true,
    captionpos=b,
    keepspaces=true,
    numbers=left,
    numbersep=5pt,
    showspaces=false,
    showstringspaces=false,
    showtabs=false,
    tabsize=2
}
\usepackage{tikz}
\usetikzlibrary{decorations.pathreplacing}

\definecolor{linkcolor}{rgb}{0.0,0.3,0.5}
\usepackage[hypertexnames=false, unicode, colorlinks=true, linkcolor=linkcolor,
            citecolor=linkcolor, filecolor=linkcolor, urlcolor=linkcolor,
            pdfusetitle]{hyperref}
\usepackage[all]{hypcap}

\renewcommand{\vec}[1]{\bm{#1}}
\newcommand{\vhat}[1]{\vec{\hat{#1}}}
\def\VEV#1{\left\langle #1 \right\rangle}
\newcommand{\Legendre}[1]{\mathcal{L}_{#1}}

\newcommand{\cP}{\mathcal P}
\newcommand{\cS}{\mathcal S}
\newcommand{\cG}{\mathcal G} 
\newcommand{\bk}{{\vec{k}}} 
\newcommand{\bq}{{\vec{q}}}
\newcommand{\bx}{{\vec{x}}} 

\begin{document}

\makeatletter
\renewcommand{\@makecaption}[2]{%
  \vskip\abovecaptionskip
  \noindent
  \justifying
  \textbf{#1.} #2\par
  \vskip\belowcaptionskip
}
\makeatother

\title{Extracting non-Gaussian information of line-intensity maps with weighted skew spectra}

\newcommand\elehome{
\affiliation{Department of Physics, Ben-Gurion University of the Negev, Be'er Sheva 84105, Israel}
\affiliation{School of Physics and Astronomy, Tel-Aviv University, Tel-Aviv 69978, Israel}
}

\newcommand\azadehhome{
\affiliation{Laboratoire d'Annecy de Physique Theorique (LAPTh), CNRS/USMB, 99 Chemin de Bellevue BP110, Annecy F-74941, France}
}

\newcommand\elyhome{
\affiliation{Department of Physics, Ben-Gurion University of the Negev, Be'er Sheva 84105, Israel}
}

\author{Eleonora Vanzan}
\email{evanzan@tauex.tau.ac.il}
\elehome

\author{Azadeh Moradinezhad Dizgah}
\email{azadeh.moradinezhad@lapth.cnrs.fr}
\azadehhome

\author{Ely D. Kovetz}
\email{kovetz@bgu.ac.il}
\elyhome

\begin{abstract}
    The bispectrum of line-intensity maps carries cosmological and astrophysical information beyond the power spectrum, but its analysis of survey data is hampered by the large size of the data vector. Skew spectra address this by compressing the bispectrum into a handful of power-spectrum-like observables, cross-spectra of the observed field with filtered squares of it. We present the first application of optimally weighted skew spectra to line-intensity mapping of galaxy-hosted emission, where all clustering and shot-noise contributions are luminosity weighted and the mean intensity is unknown. We perform idealized real-space Fisher forecasts, without instrumental smoothing or survey-window effects, for a SPHEREx-like survey of H$\alpha$ emission at~$0.2 \lesssim z \lesssim 4.6$, in its deep fields and in an all-sky configuration with higher instrumental noise. First, the compression carries over to intensity maps: on perturbative scales and for a fixed redshift evolution of the emitters, four skew spectra, one of them matched to the shot noise, recover the bispectrum constraints on the bias parameters, the shot noise and the cosmological parameters to within a few percent. Second, the joint analysis of the power spectrum and the skew spectra measures the ratios of the second-order biases to the linear bias in each redshift bin. Within a bin the individual biases are degenerate with the unknown mean intensity, but their ratios are not, so their redshift evolution is measured without a model for the emitters. 
    For emission that depends on halo mass alone, these ratios are determined by luminosity-weighted averages of the halo biases. Their redshift evolution can therefore test the halo--emission connection, including dependence on assembly history or environment. The bias ratios are not accessible to the tree-level real-space power spectrum alone.
    We find that applying skew spectra to line intensity maps at~$z \lesssim 3$, an all-sky SPHEREx-like configuration can detect departures of order~$0.1$ in the bias ratios at~$3\sigma$, while  deeper fields perform better in the highest-redshift bins.

\end{abstract}

\maketitle

\section{Introduction}
\label{sec:intro}

Line intensity mapping (LIM) is a fast-growing observational technique to efficiently map the cosmic large-scale structure (LSS) over large comoving volumes by measuring spatial fluctuations in the aggregate emission of atomic and molecular spectral lines, without resolving individual sources~\cite{Chang:2026ake,Bernal:2022jap,Kovetz:2017agg}. Spatial fluctuations in the line intensity trace the underlying dark matter distribution, and the observed frequency of a line, compared with its known rest-frame frequency, provides a spectroscopic determination of the redshift of the emission. Since the observed intensity is the cumulative emission from all sources, including faint populations that are inaccessible to traditional galaxy surveys, LIM is also a powerful probe of astrophysics, sensitive to the cosmic star-formation history, galaxy evolution, the epoch of reionization, and the properties of the interstellar and intergalactic media. These prospects have motivated a broad observational program targeting multiple lines across a wide range of wavelengths and redshifts: 21-cm experiments in the radio~\cite{HERA:2022wmy,Cunnington:2022uzo,CHIME:2025cee,Cunnington:2026opd,EorCDScienceWorkingGroup:2026trp}, CO and [CII] surveys at (sub-)millimeter wavelengths~\cite{Keating:2016pka,Ade:2019ril,Sun:2020mco,Keating:2020wlx,CCAT-Prime:2021lly,Bethermin:2022lmd,COMAP:2024pkb}, and Ly$\alpha$, H$\alpha$, [OII] and [OIII] mapping in the optical and near-infrared~\cite{SPHEREx:2014bgr,Gong:2017,LujanNiemeyer:2025yow}.
In this work, we focus on lines emitted by galaxies, such as CO, [CII], H$\alpha$, [OII], [OIII], and post-reionization 21-cm emission from neutral hydrogen in halos, whose fluctuations on large scales trace the matter field as luminosity-weighted biased tracers with shot noise from discrete emitters. This excludes the 21-cm signal during cosmic dawn and reionization, where the non-Gaussianity is driven by the morphology of ionization and heating in the diffuse intergalactic medium.

While the power spectrum has been the main summary statistic in forecasts and analyses of LIM surveys, the nonlinear nature of structure formation, as well as the astrophysical processes that determine the line emission, imply that it is a lossy compression of intensity maps, and, as for galaxy clustering, the cosmological and astrophysical information is spread across the hierarchy of $N$-point correlation functions. In the context of LIM, the bispectrum, the lowest-order non-Gaussian statistic, has been studied mainly for 21-cm intensity mapping, both during cosmic dawn and reionization~\cite{Shimabukuro:2016viy,Majumdar:2017tdm,Watkinson:2018efd,Kamran:2020pkr,Watkinson:2021ctc} and after reionization~\cite{Sarkar:2019ojl,Durrer:2020orn,Cunnington:2021czb} (see Ref.~\cite{Majumdar:2026dgt} for a recent overview), as well as in cross-correlation with [CII] maps~\cite{Beane:2018pmx}. For other lines, the voxel intensity distribution has been shown to complement the power spectrum in constraining the line luminosity function~\cite{Ihle:2018dky} and physics beyond $\Lambda$CDM~\cite{Libanore:2022ntl,Sabla:2024dxz}.

The power spectrum of the line intensity is commonly modeled with the halo model~\cite{Cooray:2002dia}, in which the two-halo term is given on large scales by the linear matter power spectrum multiplied by the luminosity-weighted linear bias of the host halos, $P(k)=\langle I\rangle^2 b_1^2 P_m(k)$, and the shot noise is assumed to be Poissonian and set by the second moment of the line luminosity,\footnote{In general, the shot noise deviates from the Poisson prediction~\cite{MoradinezhadDizgah:2021dei}, in which case its amplitude is no longer given by the second moment of the line luminosity alone.} so that, within a parametric emission model, the power spectrum constrains the parameters of the luminosity function. Unlike the mean number density of a galaxy sample, however, the mean intensity is in general not directly observable, and since it multiplies the whole fluctuation field, the linear power spectrum measures only $\langle I\rangle b_1$, which is in turn degenerate with the amplitude of matter fluctuations. Including nonlinear corrections to the clustering~\cite{MoradinezhadDizgah:2021dei,Modi:2019ewx} reduces the degeneracy with the amplitude of matter fluctuations, since the loop terms scale with the square of the linear matter power spectrum, but not the $\langle I\rangle$--$b_1$ degeneracy, since the higher-order biases also appear multiplied by the mean intensity; these biases, which enter the power spectrum only through the loop terms, remain weakly constrained. In redshift space, the velocity contribution to the observed fluctuations is not weighted by the bias, so that the quadrupole of the power spectrum can break the $\langle I\rangle$--$b_1$ degeneracy for a known growth rate and amplitude of matter fluctuations~\cite{Bernal:2019jdo}. In practice, however, current measurements constrain only a single clustering amplitude, due to their limited signal-to-noise~\cite{Cunnington:2022uzo,CHIME:2025cee}. Alternatively, the combination of the power spectrum with the voxel intensity distribution has been shown to break this degeneracy within a single redshift bin, without redshift-space information~\cite{Breysse:2022alx}.

The bispectrum carries information complementary to the power spectrum, which can help lift these degeneracies and probe the population of emitters. First, since it scales with a different power of the amplitude of matter fluctuations, its combination with the power spectrum separates this amplitude from the clustering amplitude $\langle I\rangle b_1$, as established for galaxy clustering~\cite{Sefusatti:2006pa,Sefusatti:2007ih,Eggemeier:2021cam,Euclid:2026glh}. Second, the bispectrum of line intensity maps contains shot-noise contributions involving higher moments of the line luminosity~\cite{Sato-Polito:2022fkd}, which depend on the mean intensity differently from the clustering terms. Lastly, the second-order biases of the perturbative expansion~\cite{Desjacques:2016bnm}, $b_2$ and $b_{\cG_2}$, which multiply the square of the matter density and the tidal operator $\cG_2$, respectively, enter the bispectrum already at tree level. Since the mean intensity multiplies all the bias parameters, the bispectrum, in combination with the power spectrum, determines the ratios $b_2/b_1$ and $b_{\cG_2}/b_1$ independently of the mean intensity. Measured in each redshift bin, without a model for the redshift evolution of the emitters, these ratios can be compared with the relations predicted for dark matter halos~\cite{Lazeyras:2015lgp,Sheth:2012fc}; since the bias parameters of the line are luminosity-weighted averages over the host halos, departures from these relations carry information on the distribution of the line emission over halo mass and on any dependence of the emission on halo properties beyond mass~\cite{Barreira:2021ukk,Lazeyras:2021dar}.

In galaxy clustering, the benefits of the bispectrum are by now well established: joint analyses of the power spectrum and bispectrum of SDSS~\cite{Gil-Marin:2016wya,Philcox:2021kcw,DAmico:2022osl,Ivanov:2023qzb,Cagliari:2025rqe} and, more recently, DESI~\cite{NovellMasot:2025fju,Chudaykin:2025aux,Novell-Masot:2026sgg, Guandalin:2026yom} data have shown that it improves the constraints on cosmological parameters and tightens those on the nonlinear bias parameters of the observed galaxy samples. Despite decades of theoretical work~\cite{Fry:1992vr,Matarrese:1997sk,Scoccimarro:1999ed}, however, the use of the bispectrum in survey analyses long lagged behind that of the power spectrum, owing to the limited signal-to-noise of earlier measurements, the challenges in modeling the signal and the observational systematics, and its high dimensionality, which complicates its measurement, the estimation of its covariance, and the parameter inference. For LIM, these challenges are further compounded by observational effects: foreground removal and the instrumental beam substantially reduce the signal-to-noise of the bispectrum~\cite{Cunnington:2021czb}, and residual systematics are expected to hinder its detection~\cite{Noble:2026oqm}. It is therefore desirable to compress the bispectrum into a small set of statistics that retain its information on the parameters of interest, at a computational cost comparable to that of the power spectrum, and that can, in turn, simplify the treatment of observational effects in realistic analyses.

Weighted skew spectra provide one such compression~\cite{Schmittfull:2014tca,MoradinezhadDizgah:2019xun,Schmittfull:2020hoi}.\footnote{Harmonic-space skew spectra were first introduced in the context of the CMB as skew-$C_\ell$~\cite{Komatsu:2003iq,Munshi:2009ik} and more recently extended to LSS observables~\cite{Munshi:2020ofi,Chakraborty:2022aok,Harscouet:2024hnq,Roskill:2025eot}.}
They are defined as cross-spectra of the observed overdensity field with quadratic fields, each obtained as the product of two filtered copies of the overdensity itself.\footnote{Skew spectra belong to the broader class of composite-field spectra~\cite{Gao:2025yqd}, defined as two-point functions constructed from cross-correlations of appropriately filtered products of the observed fields evaluated at the same location. This class also includes kurto spectra, which probe the trispectrum~\cite{Lazeyras:2017hxw,Abidi:2018eyd,Jamieson:2024mau,Gao:2025yqd,Gao:2026zaz}, which have also been extended to harmonic space~\cite{Munshi:2021uwn}.}
The filters, or mode-coupling kernels, are chosen to mirror the structure of the tree-level bispectrum of biased tracers: in real space, they correspond to the squared density, shift, and tidal terms, while in redshift space additional kernels capture the contributions from the mapping to redshift space~\cite{Schmittfull:2020hoi}.
Since the tree-level bispectrum is a sum of separable components, each skew spectrum, with an appropriate inverse-variance weighting, approximates the maximum-likelihood estimator of the amplitude of the corresponding component, so that together the skew spectra capture the full information of the tree-level bispectrum on amplitude-like parameters, such as the bias parameters~\cite{Schmittfull:2014tca,MoradinezhadDizgah:2019xun}.
Being functions of a single wavenumber, skew spectra can be estimated with fast Fourier transforms at a cost comparable to that of the power spectrum, their covariance requires far fewer simulations to estimate than that of the bispectrum, and the treatment of the survey window is expected to be simpler.
Following their introduction in Ref.~\cite{Schmittfull:2014tca}, skew spectra were used in forecasts for galaxy clustering in real space~\cite{Schmittfull:2014tca,MoradinezhadDizgah:2019xun}, extended to redshift space~\cite{Schmittfull:2020hoi}, employed in simulation-based Fisher forecasts of cosmological parameters on halo and galaxy catalogs~\cite{Hou:2022rcd}, and applied to BOSS data, both with simulation-based inference~\cite{Hou:2024blc} and with a perturbative model to constrain primordial non-Gaussianity~\cite{Chen:2024bdg}.
They have recently also been extended to the Ly$\alpha$ forest, where they offer a route to breaking the degeneracy between the mean transmitted flux and the power-spectrum amplitude~\cite{deBelsunce:2025edy}, an analogue of the degeneracy between mean intensity and bias in LIM.
In 21-cm intensity mapping of cosmic dawn and reionization, the unweighted skew spectrum, i.e.,~the cross-spectrum of the squared brightness-temperature field with the field itself, has been measured on semi-numerical simulations and forecast to be detectable with SKA-Low~\cite{Ma:2023eml,Cook:2024bps}.
Similar compressions of a three-point statistic into a two-point-like one have been used to recover information lost to foreground removal, since long–short mode coupling imprints the removed low-$k_{\parallel}$ modes onto the small-scale modes that survive: e.g.,~the squeezed-limit HI--HI--CMB-lensing bispectrum of~\cite{Moodley:2023lmu} and the squared-21cm$\times$NIRB cross-spectrum of~\cite{Sun:2024vhy}.
Skew spectra generalize these, by summing over all configurations with optimal weights.

In this paper, we present the first application of weighted skew spectra to LIM, for lines emitted by galaxies, whose fluctuations on large scales are a biased tracer of the matter field, so that the kernels can be matched to the templates of the tree-level bispectrum.
We extend the formalism of~\cite{Schmittfull:2014tca,MoradinezhadDizgah:2019xun} to LIM, in which the amplitudes are luminosity-weighted, the shot noise is set by moments of the luminosity function rather than by number counts, and the instrumental noise enters the optimal weights. We show that the compression carries over to intensity maps, provided that a dedicated kernel is introduced to match the shot noise terms. While the formalism is applicable to any emission line, as a concrete example we consider H$\alpha$ intensity maps from a SPHEREx-like survey~\cite{SPHEREx:2014bgr,Bock:2025ijl} over $0.2\lesssim z\lesssim 4.6$, both in its deep-field configuration and in an all-sky configuration with a correspondingly higher instrumental noise, the latter going beyond the baseline SPHEREx intensity-mapping program.

As a first exploration, we perform Fisher forecasts in real space, deferring the inclusion of redshift-space distortions (RSD) and observational systematics to future work.
First, we compare the information content of the bispectrum and the skew spectra on the bias, shot-noise, and cosmological parameters, assuming known luminosity moments and a fixed redshift evolution of the biases.
Second, we let the luminosity moments vary, assuming only their redshift evolution. 
Third, we combine the skew spectra with the power spectrum, whose joint analysis can break the~$\langle I\rangle$--$b_1$ degeneracy bin by bin through the shot-noise terms of the bispectrum, once the linear bias weighted by the squared luminosity is tied to the luminosity-weighted one through the emission model, and we quantify what this delivers for a SPHEREx-like survey.
Lastly, we drop any assumption on the redshift evolution of both the biases and the luminosity moments, and analyze each redshift bin separately.
The generalization to the cross-correlations of multiple lines observed in the same volume~\cite{Sarkar:2026vmx} is developed in a companion paper~\cite{Vanzan:2026pp1}.

The rest of the paper is organized as follows.
In Sec.~\ref{sec:method}, we review the skew-spectrum formalism and present its extension to LIM, including the dedicated shot-noise kernel.
In Sec.~\ref{sec:fisher}, we describe the survey specifications, the line-emission model, and our Fisher setup.
In Sec.~\ref{sec:results}, we present our forecasts, following the four steps outlined above, and we conclude in Sec.~\ref{sec:conclusions}.
Technical details, including the constraints on cosmological parameters, are presented in the appendices.

\section{Skew spectra of line intensity fluctuations}
\label{sec:method}

We work at tree level in perturbation theory, in real space, and with a single line; the instrumental noise is taken to be Gaussian. Skew spectra and their optimal weighting were developed for galaxy clustering~\cite{Schmittfull:2014tca,MoradinezhadDizgah:2019xun}, and we only recall the definitions needed below (Sec.~\ref{sec:formalism}), referring to Appendix~\ref{app:formalism} for the full expressions of the signal and covariance. Three ingredients are specific to line-intensity mapping: the amplitudes are the mean intensity times luminosity-weighted biases (Sec.~\ref{sec:lim}); the shot noise is set by the higher moments of the line luminosity and by a bias weighted by the luminosity squared (Sec.~\ref{sec:shot}), which calls for an additional filtering kernel (Sec.~\ref{sec:shot_kernels}); and the resulting parameter content fixes which degeneracies each statistic can and cannot break (Sec.~\ref{sec:degeneracies}).

\subsection{Weighted skew spectra}
\label{sec:formalism}

Skew spectra are defined as cross-spectra of an observed field~$\delta_{\rm obs}$, e.g.,~the galaxy overdensity or the line intensity fluctuations, with quadratic composite fields constructed from it. In configuration space, these are products of two filtered copies of~$\delta_{\rm obs}$; in Fourier space, they correspond to convolutions of two copies of the field weighted by a mode-coupling kernel~$D_2^{(n)}$,
\begin{equation}
\label{eq:quadratic_field}
    \cS_n(\bk) = \int \frac{d^3 \bq}{(2\pi)^3}\, D_2^{(n)}(\bq,\bk-\bq)\, \delta_{\rm obs}(\bq)\, \delta_{\rm obs}(\bk-\bq) \, .
\end{equation}
Their isotropic cross-spectrum with the observed field is a weighted integral of its bispectrum,
\begin{align}
\label{eq:skew_definition}
    &\VEV{ \cS_n(\bk)\, \delta_{\rm obs}(\bk') } = (2\pi)^3 \delta_D^{(3)}(\bk+\bk')\, \cP_{\cS_n,\delta_{\rm obs}}(k) \\
    & = \int \frac{d^3 \bq}{(2\pi)^3}\, D_2^{(n)}(\bq,\bk-\bq) \VEV{ \delta_{\rm obs}(\bq)\, \delta_{\rm obs}(\bk-\bq)\, \delta_{\rm obs}(\bk') } \, . \notag
\end{align}
The kernels~$D_2^{(n)}$ can be tailored to target specific contributions to the bispectrum. In particular, if the theoretical bispectrum is a sum of separable templates, the kernels can be chosen such that, in the weakly non-Gaussian regime, the weighted sum over~$k$ of each skew spectrum reproduces the maximum-likelihood estimator of the corresponding template amplitude~\cite{Schmittfull:2014tca,Gao:2025yqd}, while the skew spectra themselves, as functions of~$k$, retain additional information. This is the construction we adopt in this paper, and the starting point for defining the kernels is therefore the tree-level bispectrum of the biased tracer.

In the perturbative bias expansion~\cite{Desjacques:2016bnm}, the clustering part of the overdensity of a given dark matter tracer, such as galaxies or line intensity fluctuations, is described up to second order by 
\begin{equation}
\label{eq:bias_expansion}
    \delta(\bx) = b_1 \delta_m(\bx) + \frac{b_2}{2} \delta_m^2(\bx) + b_{\cG_2} \cG_2(\bx) \, ,
\end{equation}
where~$\cG_2$ is the Galileon operator built from the tidal field,
\begin{equation}
\label{eq:G2}
    \cG_2(\bx) = \left[ \partial_i \partial_j \Phi(\bx) \right]^2 - \left[ \nabla^2 \Phi(\bx) \right]^2 \, ,
\end{equation}
with~$\Phi$ the gravitational potential normalized such that~$\nabla^2\Phi = \delta_m$. 
The observed field~$\delta_{\rm obs}$ additionally contains stochastic contributions, which we discuss in Secs.~\ref{sec:lim} and~\ref{sec:shot}.
At tree level, the bispectrum of the clustering part is
\begin{align}
\label{eq:B_tree}
    & B(\bk_1,\bk_2,\bk_3) = 2 b_1^2 P_m(k_1) P_m(k_2) \notag \\
    & \qquad \times \left[ b_1 F_2(\bk_1,\bk_2) + \frac{b_2}{2} + b_{\cG_2}\left( \mu_{12}^2 - 1 \right) \right] + 2~{\rm perm.} \, ,
\end{align}
where~$\mu_{12} = \hat{\bk}_1 \cdot \hat{\bk}_2$ and~$F_2$ is the second-order kernel of standard perturbation theory, which decomposes into Legendre polynomials of~$\mu_{12}$ as
\begin{equation}
\label{eq:F2_legendre}
    F_2(\bk_1,\bk_2) = \frac{17}{21}\Legendre{0}(\mu_{12}) + F_2^1(k_1,k_2)\, \Legendre{1}(\mu_{12}) + \frac{4}{21}\Legendre{2}(\mu_{12}) \, ,
\end{equation}
with
\begin{equation}
    F_2^1(k_1,k_2) = \frac{1}{2} \left( \frac{k_1}{k_2} + \frac{k_2}{k_1} \right) \, .
\end{equation}
Since~$\mu_{12}^2 - 1 = \frac{2}{3}\left[\Legendre{2}(\mu_{12}) - \Legendre{0}(\mu_{12})\right]$, the bracket in Eq.~\eqref{eq:B_tree} is a combination of three fixed templates, $\Legendre{0}$, $F_2^1 \Legendre{1}$, and~$\Legendre{2}$, with amplitudes set by the bias parameters.

We therefore use the three kernels
\begin{equation}
\label{eq:kernels}
    \mathbf{D}_2(\bk_1,\bk_2) = \left\{ \Legendre{0}(\mu_{12}), \; F_2^1(k_1,k_2)\, \Legendre{1}(\mu_{12}), \; \Legendre{2}(\mu_{12}) \right\} \, ,
\end{equation}
with components~$D_2^{(n)}$, $n \in \{{\rm dens},{\rm disp},{\rm tide}\}$, which, to realize the maximum-likelihood construction, are weighted by the inverse power spectrum of the observed field~\cite{Schmittfull:2014tca},
\begin{equation}
\label{eq:weighted_kernel}
    \tilde{D}_2^{(n)}(\bq,\bk-\bq) = \frac{P_m(q)}{P_{\rm obs}(q)} \frac{P_m(|\bk-\bq|)}{P_{\rm obs}(|\bk-\bq|)}\, D_2^{(n)}(\bq,\bk-\bq) \, ,
\end{equation}
where~$P_m$ is the linear matter power spectrum and~$P_{\rm obs}$ the power spectrum of the observed field, including all noise contributions. In what follows, $\cS_n$ denotes the quadratic field built with the weighted kernel~$\tilde{D}_2^{(n)}$, and its skew spectrum is the sum of a clustering and a shot-noise contribution,
\begin{equation}
\label{eq:skew_decomposition}
    \cP_{\cS_n,\delta_{\rm obs}}(k) = \cP^{\rm clust}_{\cS_n, \delta_{\rm obs}}(k) + \cP^{\rm shot}_{\cS_n, \delta_{\rm obs}}(k) \, .
\end{equation}
Since the instrumental noise is assumed to be Gaussian, it does not contribute to the skew spectra. The clustering contribution is
\begin{align}
\label{eq:skew_clust}
    \cP^{\rm clust}_{\cS_n, \delta_{\rm obs}}(k) & =  \; 2 b_1^3 \, \mathcal{I}^{(n)}_{F_2}(k) \notag \\
    &+  \; 2 b_1^2 \left( \frac{b_2}{2} - \frac{2 b_{\cG_2}}{3} \right) \mathcal{I}^{(n)}_{\Legendre{0}}(k) \notag \\
    & +  \; \frac{4}{3} b_1^2 b_{\cG_2} \, \mathcal{I}^{(n)}_{\Legendre{2}}(k) \, ,
\end{align}
where~$\mathcal{I}^{(n)}_Y = I^{(n)}_Y + 2 I^{(n),{\rm bare}}_Y$ are mode-coupling integrals of the weighted kernel against the linear power spectrum (Eqs.~\eqref{eq:I_DE}--\eqref{eq:I_DE_bare}). For line intensity maps, each term is multiplied by~$\VEV{I}^3$ (Sec.~\ref{sec:lim}). 

We derive the shot-noise contribution~$\cP^{\rm shot}_{\cS_n,\delta_{\rm obs}}$ for line intensity maps in Sec.~\ref{sec:shot}, together with its galaxy-clustering limit. The Gaussian covariance of the data vector~$\{\cP_{\cS_n,\delta_{\rm obs}}(k)\}$, including the cross-covariance between kernels, is given in Eq.~\eqref{eq:skew_cov}.

\subsection{Line intensity mapping model}
\label{sec:lim}

We write the intensity of the line as~$I(\bx) = \VEV{I}\left[1 + \delta_I(\bx)\right]$, and adopt a simple model that relates its mean and fluctuations to the halos hosting the emitting galaxies~\cite{Visbal:2010rz,Gong:2011ts,Lidz:2011dx,Silva:2014ira}. Its basic ingredients are the moments of the line luminosity~$L(M,z)$ over the halo mass function~$dn/dM$, converted to intensity units, 
\begin{equation}
\label{eq:moments}
\begin{split}
    \VEV{I^n}(z) &= p_{n,\sigma} \int_{M_{\rm min}}^{M_{\rm max}} dM \, \frac{dn}{dM} \left[ \frac{L(M,z)}{4\pi D_L^2(z)} \left( \frac{dl}{d\theta} \right)^2 \frac{dl}{d\nu} \right]^n \, , \\
    p_{n,\sigma} &= \int dx \, \frac{10^{nx}}{\sqrt{2\pi}\sigma} \, e^{-x^2/2\sigma^2} \, ,
\end{split}
\end{equation}
where~$D_L$ is the luminosity distance, $dl/d\theta = \chi(z)$ and~$dl/d\nu = c(1+z)/[\nu_{\rm obs} H(z)]$ convert comoving lengths into angles and frequencies, and~$p_{n,\sigma}$ accounts for a log-normal scatter~$\sigma$ (in dex) of the luminosity at fixed halo mass.
The first moment is the mean intensity, while the second and third moments, with units of~$({\rm Jy/sr})^2\,{\rm Mpc}^3$ and~$({\rm Jy/sr})^3\,{\rm Mpc}^6$, set the Poisson shot noise of the power spectrum and bispectrum (Eq.~\eqref{eq:Pshot} and Sec.~\ref{sec:shot}); since they are not moments of the intensity field, we denote them by~$\VEV{I^n}$ with a slight abuse of notation.

In this model, the bias parameters of the line intensity signal are the biases of the halos hosting the emitting galaxies, weighted by the luminosity of the emitters,
\begin{equation}
\label{eq:L-weighted_biases}
    b_{\alpha}(z) = \frac{\int_{M_{\rm min}}^{M_{\rm max}} dM \frac{dn}{dM} b_{\alpha,h}(M,z) L(M,z)}{\int_{M_{\rm min}}^{M_{\rm max}} dM \frac{dn}{dM} L(M,z)} \, ,
\end{equation}
$\alpha \in \{ 1, 2, \cG_2 \}$.
In what follows, we compute the halo mass function and the halo biases~$b_{1,h}$ and~$b_{2,h}$ with the Sheth--Tormen model~\cite{Sheth:1999mn,Scoccimarro:2000gm}, and the tidal bias~$b_{\cG_2,h}$ with the coevolution relation~\cite{Baldauf:2012hs,Chan:2012jj} (Appendix~\ref{app:biases}).
All quantities are evaluated at the central redshift of each bin, and to avoid cluttering the equations we drop the explicit redshift dependence of the moments, biases, and power spectra in what follows.

The clustering part of~$\delta_I$ then follows the bias expansion of Eq.~\eqref{eq:bias_expansion} with the luminosity-weighted biases of Eq.~\eqref{eq:L-weighted_biases}.
Since the mean intensity is not measured, the observed field is the intensity fluctuation in units of Jy/sr, $\delta_{\rm obs}(\bx) = I(\bx) - \VEV{I} = \VEV{I}\,\delta_I(\bx)$, plus shot and instrumental noise.
Its clustering $N$-point functions are therefore those of~$\delta_I$ multiplied by~$\VEV{I}^n$, and each bias parameter appears only in the combination~$\VEV{I} b_\alpha$, which is the origin of the degeneracies discussed in Sec.~\ref{sec:degeneracies}.

The power spectrum of the observed field is the sum of three contributions, clustering, shot noise, and instrumental noise,
\begin{equation}
\label{eq:Ptot}
    P_{\rm obs}(k,z) = P_{\rm clust}(k,z) + P_{\rm shot}(z) + P_{\rm N}(z) \, ,
\end{equation}
in units of~$({\rm Jy/sr})^2 {\rm Mpc}^3$.
At tree level, the clustering contribution is
\begin{equation}
\label{eq:Pclust}
    P_{\rm clust}(k,z) = \VEV{I}^2 b_1^2 P_m(k,z) \, .
\end{equation}
The Poisson shot noise is the second moment of luminosity,
\begin{equation}
\label{eq:Pshot}
    P_{\rm shot}(z) = (1+\alpha_{\rm P}) \VEV{I^2} \, ,
\end{equation}
where~$\VEV{I^2}$ is the Poisson prediction and~$\alpha_{\rm P}$ parametrizes a fractional deviation from it~\cite{MoradinezhadDizgah:2021dei}.
The instrumental noise is set by the per-voxel noise~$\sigma_{\rm N}$ and by the comoving volume of a voxel of solid angle~$\Omega_{\rm vox}$ and bandwidth~$\delta\nu$,
\begin{equation}
\label{eq:PN}
    P_{\rm N} = \sigma_{\rm N}^2 \, \Omega_{\rm vox} \, \delta\nu \left( \frac{dl}{d\theta} \right)^2 \frac{dl}{d\nu} \, .
\end{equation}
Since we take the instrumental noise to be Gaussian, it enters the skew spectra only through the weights in Eq.~\eqref{eq:weighted_kernel} and the covariance, and does not contribute to the three-point signal.

\subsection{Shot noise of the line intensity bispectrum and skew spectra}
\label{sec:shot}

The derivation of the shot-noise contribution to the line intensity bispectrum, sketched in Appendix~\ref{app:shot}, follows the galaxy-clustering case~\cite{Chan:2016ehg,Sato-Polito:2022fkd} with one difference: the contribution from two galaxies that are at the same point enters with the square of their luminosity, so a correlator containing pairs located at the same position carries a bias weighted by~$L^2$ rather than~$L$,
\begin{equation}
\label{eq:b1L2}
    b_1^{[L^2]}(z) = \frac{\int dM \, \frac{dn}{dM} \, b_{1,h}(M,z) \, L^2(M,z)}{\int dM \, \frac{dn}{dM} \, L^2(M,z)} \, .
\end{equation}
The observed line intensity bispectrum is given by
\begin{align}
\label{eq:Bobs_tot}
    & B_{\rm obs}(\bk_1,\bk_2,\bk_3) = \VEV{I}^3 B_I(\bk_1,\bk_2,\bk_3) \notag \\
    & \hspace{0.1in} + (1+\alpha_1) \VEV{I} \VEV{I^2} \, b_1 b_1^{[L^2]} \left[ P_m(k_1) + P_m(k_2) + P_m(k_3) \right] \notag \\
    & \hspace{0.1in} + (1+\alpha_2) \VEV{I^3} \, ,
\end{align}
where~$B_I$ is the tree-level bispectrum of the biased field and~$\alpha_{1,2}$ are common fractional deviations from the Poisson shot-noise predictions.
Since the line luminosity grows with halo mass and more massive halos are more biased, the~$L^2$ weighting gives~$b_1^{[L^2]} > b_1$. For the H$\alpha$ model of Sec.~\ref{sec:emission} and the halo mass function and biases of Appendix~\ref{app:biases}, the ratio~$r = b_1^{[L^2]}/b_1$ grows from~$1.04$ to~$1.22$ across the survey, so that approximating~$b_1^{[L^2]} \simeq b_1$, adopted in earlier works \cite{Sato-Polito:2022fkd}, would misweight the mixed shot-clustering term by up to~$20\%$.
In what follows, we treat~$b_1^{[L^2]}$ in two ways: \emph{tied}, with~$r$ fixed to its fiducial value in each bin so that~$b_1^{[L^2]} = r_{\rm fid} \, b_1$ follows the variation of~$b_1$; or \emph{free}, as an independent parameter in our forecasts.

Filtering Eq.~\eqref{eq:Bobs_tot} with a weighted kernel, the skew spectra inherit the shot-noise contributions
\begin{equation}
\label{eq:skew_shot}
\begin{split}
    \cP^{\rm shot}_{\cS_n,\delta_{\rm obs}}(k) &= (1+\alpha_1) \VEV{I} \VEV{I^2} \, b_1 b_1^{[L^2]} \left[ P_m(k) \, J^{(n)}(k) \right. \\ 
    &+ \left. 2 \Tilde{J}^{(n)}(k) \right] + (1+\alpha_2) \VEV{I^3} J^{(n)}(k) \, ,
\end{split}
\end{equation}
with the kernel integrals
\begin{equation}
\label{eq:J_integrals}
\begin{split}
    J^{(n)}(k) &= \int \frac{d^3 q}{(2\pi)^3}  \tilde{D}_2^{(n)}(\bq,\bk-\bq) \, , \\
    \tilde{J}^{(n)}(k) &= \int \frac{d^3 q}{(2\pi)^3}  \tilde{D}_2^{(n)}(\bq,\bk-\bq) \, P_m(q) \, .
\end{split}
\end{equation}
In our baseline analysis, we make the simplifying assumption~$\alpha_{\rm P}=\alpha_1=\alpha_2\equiv\alpha$, as all three originate from the discreteness of the emitters, and we discuss the impact of treating~$\alpha_1$ and~$\alpha_2$ as independent in Sec.~\ref{sec:results_perbin}.
The galaxy-clustering expressions are recovered for~$\VEV{I} \to 1$, $\VEV{I^2} \to 1/\bar n$, $\VEV{I^3} \to 1/\bar n^2$, and~$b_1^{[L^2]} \to b_1$, since all galaxies carry the same weight.

\subsection{Dedicated shot-noise kernels}
\label{sec:shot_kernels}

The kernels of Eq.~\eqref{eq:kernels} are matched to the clustering templates of the bispectrum, and in forecasts for galaxy surveys they were found to recover the bispectrum constraints on the biases and the shot-noise amplitude~\cite{Schmittfull:2014tca,MoradinezhadDizgah:2019xun}.\footnote{The residual difference in the shot-noise constraint between skew spectra and bispectrum reported in Ref.~\cite{MoradinezhadDizgah:2019xun} has since been traced to numerical noise.}
However, the bispectrum of Eq.~\eqref{eq:Bobs_tot} contains two further contributions, a constant and a term proportional to linear matter power spectrum, whose amplitudes are free parameters through~$\alpha$ or the moments.
With three kernels, the four parameters (three biases and one shot-noise amplitude) can then be separated only through the $k$-dependence of the skew spectra.
The large volumes of galaxy surveys provide enough $k$-leverage for this, while the smaller volume per redshift bin of the LIM survey considered here does not.
To recover the bispectrum constraints, the data vector must therefore also contain a skew spectrum matched to the shot-noise templates.

By the same logic that produces the standard set, the optimal weighted kernel for a template~$T$ is~$\tilde{D}_2 \propto T/[P_{\rm obs}(q) P_{\rm obs}(|\bk-\bq|)]$, which for the constant template is pure inverse-variance weighting.
In terms of the bare kernel to which the weighting of Eq.~\eqref{eq:weighted_kernel} is applied,
\begin{equation}
\label{eq:shot_kernels_def}
\begin{split}
    D_2^{({\rm shot})}(\bq,\bk-\bq) &= \frac{1}{P_m(q) \, P_m(|\bk-\bq|)} \, , \\
    D_2^{({\rm mixed})}(\bq,\bk-\bq) &= \frac{P_m(q)+P_m(|\bk-\bq|)}{P_m(q)\,P_m(|\bk-\bq|)} \, ,
\end{split}
\end{equation}
up to a normalization that does not affect the Fisher analysis.\footnote{The kernel matched to the mixed template, $D_2^{({\rm mixed})}$, yields constraints consistent with~$D_2^{({\rm shot})}$ at the~$2$--$3\%$ level, and a second shot-sector kernel would be a linear combination of directions already present in the data vector at the level of accuracy of our covariance (Sec.~\ref{sec:results}); we therefore use a single one. $D_2^{({\rm mixed})}$ is numerically disfavored by its~$0.94$ correlation with the density kernel, against~$0.8$ for~$D_2^{({\rm shot})}$.}

Since~$D_2^{({\rm shot})}$ overlaps with the density kernel, we define the fourth kernel to be its component orthogonal to the clustering set in the covariance inner product,
\begin{equation}
\label{eq:orth_kernel}
\begin{gathered}
    D_2^{({\rm orth})} = D_2^{({\rm shot})} - \sum_{j \in \{{\rm dens},{\rm disp},{\rm tide}\}} \gamma_j D_2^{(j)} \, , \\ 
    \sum_j G^{(i,j)} \gamma^{(j)} = G^{(i,{\rm shot})} \, ,
\end{gathered}
\end{equation}
with $G^{(i,j)} = I^{(i,j)}(k_\star)$ the Gram matrix of the weighted kernels, where~$I^{(i,j)}$ is defined in Eq.~\eqref{eq:I_nm}, evaluated at a representative scale~$k_\star$, chosen to be the midpoint of the~$[k_{\rm min}(z), k_{\rm max}]$ range in each redshift bin.
This redefinition is merely a choice of basis: its advantage is the fact that the orthogonalized kernel carries only the shot-sector information that is not already spanned by the clustering kernels.

\begin{figure*}[t]
    \centering
    \includegraphics[height=4.8cm]{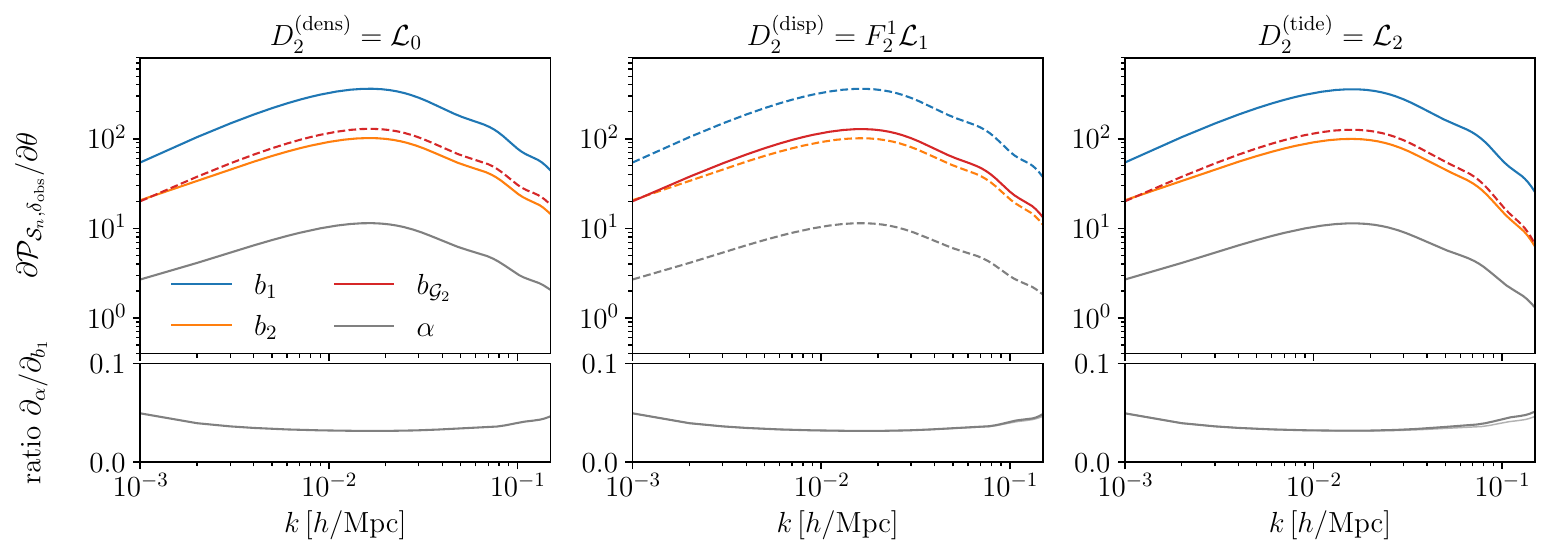} \\
    \includegraphics[height=4.8cm]{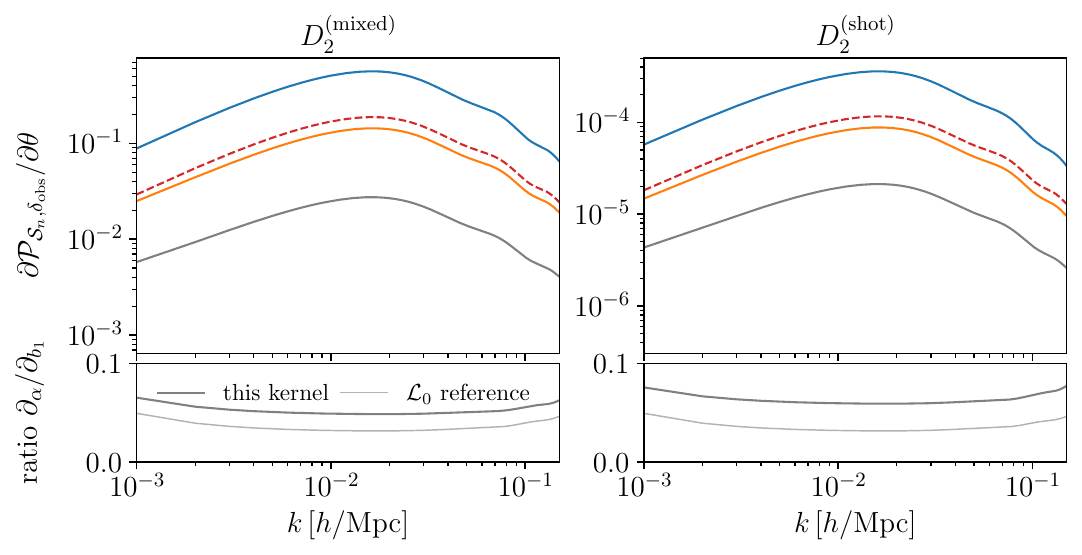}
    \caption{Responses of the weighted skew spectra to the biases and the shot-noise parameter, $\partial \cP_{\cS_n,\delta_{\rm obs}}/\partial\theta$.
    Solid (dashed) curves denote positive (negative) response.
    Each kernel responds to all four parameters. The lower panels show the ratio of the~$\alpha$ response to the~$b_1$ response, $(\partial \cP_{\cS_n,\delta_{\rm obs}}/\partial\alpha)/(\partial \cP_{\cS_n,\delta_{\rm obs}}/\partial b_1)$, as a function of~$k$, with the~$\Legendre{0}$ ratio repeated in each panel as a reference (light curve). For the three clustering kernels, the ratio takes a common value~$\simeq 0.038$ across the analysis range, while for the shot-sector kernels it is larger by a factor~$\simeq 1.7$.}
    \label{fig:responses}
\end{figure*}

The response of each skew-spectrum to variations of the four parameters~$\{b_1, b_2, b_{\cG_2}, \alpha\}$ is shown in Fig.~\ref{fig:responses}.
First, within each kernel the four responses share nearly the same shape in~$k$: a single skew-spectrum constrains one linear combination of the parameters well and the orthogonal combinations only through small shape differences.
The information on individual parameters therefore comes from \emph{combining} kernels, each of which contributes a distinct pattern of response amplitudes.
Second, every kernel responds to~$\alpha$ at a non-negligible level, but with a ratio to the~$b_1$ response that is nearly universal across the three clustering kernels: therefore, among themselves, the clustering kernels provide almost no leverage to separate~$\alpha$ from~$b_1$.
The shot-matched kernel does not share this feature: its~$\alpha$-to-$b_1$ response ratio is larger by a factor of~$1.7$ than the clustering kernels' common value, and this excess is the response direction that the clustering set lacks.
The case for a fourth, shot-sector kernel can be read off this figure, with its benefit quantified in Sec.~\ref{sec:results}.

\subsection{Parameter content and degeneracies}
\label{sec:degeneracies}

At a given redshift, Eqs.~\eqref{eq:skew_clust},~\eqref{eq:Ptot},  and~\eqref{eq:skew_shot} fix the combinations of moments and biases that each statistic measures
\begin{equation}
\label{eq:amplitudes}
\begin{aligned}
    P: \quad & \VEV{I}^2 b_1^2 \, , \quad (1+\alpha_{\rm P})\VEV{I^2} \, ; \\
    B, \mathbf{S}: \quad & \VEV{I}^3 b_1^3 \, , \quad \VEV{I}^3 b_1^2 b_2 \, , \quad \VEV{I}^3 b_1^2 b_{\cG_2} \, , \\
    & (1+\alpha_1) \VEV{I}\VEV{I^2} \, b_1 b_1^{[L^2]} \, , \quad (1+\alpha_2) \VEV{I^3} \, .
\end{aligned}
\end{equation}
First, assuming the three shot corrections to be the same~$\alpha = \alpha_{\rm P} = \alpha_1 = \alpha_2$, the shot correction~$\alpha$ is exactly degenerate with a common rescaling of the second and third moments, $(1+\alpha) \, \partial_\alpha = \partial_{\ln\VEV{I^2}} + \partial_{\ln\VEV{I^3}}$: we therefore either vary~$\alpha$ with the moments fixed to the emission model, or vary the moments and drop~$\alpha$.
Second, for each statistic taken alone the mean intensity and the linear bias are degenerate: the power spectrum measures only~$\VEV{I} b_1$, and the three-point statistics are exactly invariant under the rescaling\footnote{In the case where~$b_1^{[L^2]}$ is fixed, the transformation acts on it automatically through~$b_1$.}
\begin{equation}
\label{eq:invariant_rescaling}
\begin{gathered}
    \VEV{I} \to \lambda \VEV{I} \, , \qquad \VEV{I^2} \to \lambda \VEV{I^2} \, , \\
    \left\{ b_1, b_2, b_{\cG_2}, b_1^{[L^2]} \right\} \to \left\{ b_1, b_2, b_{\cG_2}, b_1^{[L^2]} \right\}/\lambda \, .
\end{gathered}
\end{equation}
For each statistic taken alone, this degeneracy can be lifted by combining redshift bins if the redshift evolution of the parameters is known, or by RSD if the growth rate is known~\cite{Kovetz:2017agg,Bernal:2019jdo}.
Third, if the ratio~$r$ of the two luminosity-weighted biases is fixed by the emission model, the joint analysis of power spectrum and skew spectra breaks the degeneracy within a single bin: the mixed shot-clustering amplitude divided by the product of the two power-spectrum amplitudes,
\begin{equation}
\label{eq:joint_identity}
    \frac{(1+\alpha)\VEV{I}\VEV{I^2} \, b_1 b_1^{[L^2]}}{\VEV{I}^2 b_1^2 \cdot (1+\alpha)\VEV{I^2}} = \frac{b_1^{[L^2]}}{b_1} \, \frac{1}{\VEV{I}} = \frac{r}{\VEV{I}} \, ,
\end{equation}
is a direct measurement of the mean intensity, bin by bin and without redshift-space information.
If instead~$b_1^{[L^2]}$ is free, the joint analysis measures~$b_1^{[L^2]}$ itself, but~$\VEV{I}$ and~$b_1$ stay degenerate within the bin.
Equation~\eqref{eq:joint_identity} relies on the common correction to the Poisson prediction assumed in Sec.~\ref{sec:shot}; if the corrections in the power spectrum and in the mixed term of the bispectrum are treated as independent, the mean intensity is determined only up to the factor~$(1+\alpha_1)/(1+\alpha_{\rm P})$.
This mechanism is quantified in Sec.~\ref{sec:results_Ibarb1}.

\section{Forecast setup}
\label{sec:fisher}

We assess the constraining power of the LIM skew spectra with a Fisher analysis, in two stages.
First, we compare the skew spectra with the bispectrum for a fixed emission model, to establish that the skew spectra recover the bispectrum constraints.
Second, we compare the power spectrum, the skew spectra, and their combination under increasingly general parametrizations of the emission model. The survey and the emission model are specified in Secs.~\ref{sec:survey} and~\ref{sec:emission}, the data vectors and Fisher matrices in Sec.~\ref{sec:fisher_matrices}, and the parameter sets of the two stages in Sec.~\ref{sec:parameters}.

We extend the code developed in~\cite{MoradinezhadDizgah:2019xun} for galaxy clustering to line-intensity maps, including the luminosity-weighted signal and covariance, the shot-noise kernels of Sec.~\ref{sec:shot}, FFTLog evaluation of the mode-coupling integrals (Appendix~\ref{app:fftlog}), combination with the power spectrum, and the covariance regularization of Appendix~\ref{app:covreg}.
The code will be released with the companion multi-line paper~\cite{Vanzan:2026pp1}.
The linear matter power spectrum is computed with the Boltzmann solver CLASS~\cite{Blas:2011rf} for a fiducial flat $\Lambda$CDM cosmology with $h=0.6774$, $\omega_c=0.1188$, $\omega_b=0.0223$, $n_s=0.9667$, and $\ln(10^{10}A_s)=3.1127$.
 
\subsection{Survey specifications}
\label{sec:survey}

As a concrete proof-of-concept target we consider a SPHEREx-like survey~\cite{SPHEREx:2014bgr} observing the H$\alpha$ line.
We use a representative constant spectral resolution~$R=41$ across the H$\alpha$ redshift range, rather than the wavelength-dependent SPHEREx band specifications, and divide the volume into~11 bins uniformly spaced over~$z=0.2$--$4.6$.
Each bin is treated as an independent volume with its own fundamental mode, mean intensity, and noise level.
The survey covers the two SPHEREx deep fields, with a total area of~$2 \times 100 \, {\rm deg}^2$ corresponding to a sky fraction~$f_{\rm sky} = 4.8 \times 10^{-3}$.
The per-voxel instrumental noise entering Eq.~\eqref{eq:PN} is obtained from the deep-field $5\sigma$ point-source depth~$m_{\rm AB} = 22$~\cite{Cheng:2018hox}, converting the corresponding flux uncertainty to a single pixel, over the~$n_{\rm pix}=3$ pixels effectively covered by the point-spread function~\cite{Schaan:2021hhy}, and dividing by the pixel solid angle~$\Omega_{\rm pix} = (6.2'')^2$,
\begin{equation}
\label{eq:sigmaN}
    \sigma_{\rm N} = \frac{F_\nu^{5\sigma}}{5\,\sqrt{n_{\rm pix}}\,\Omega_{\rm pix}} \, , \qquad F_\nu^{5\sigma} = 3631\,{\rm Jy}\times 10^{-0.4\,m_{\rm AB}} \, ,
\end{equation}
which gives~$\sigma_{\rm N} = 735\,{\rm Jy/sr}$.
In our forecasts, we adopt a single representative value throughout the redshift range.
We use the spectral resolution to set the voxel volume and noise power, but do not include the spectral response, beam smoothing, or angular survey window in the signal or covariance.
The forecasts therefore describe an idealized isotropic volume with SPHEREx-like noise and volume, rather than the sensitivity of the full instrumental response.
While~$\sigma_{\rm N}$ is redshift-independent, the noise power~$P_{\rm N}$ increases with redshift through the comoving voxel volume.

We adopt the deep-field configuration as the natural setting for a proof of concept.
However, the parameters we target scale differently with area and depth: e.g.,~biases and shot-noise parameters are limited by the number of modes in a small volume.
Therefore, we also consider an all-sky configuration of the same instrument, covering~$30\,000 \, {\rm deg}^2$ ($f_{\rm sky} = 0.73$~\cite{Feder:2023rqg,SPHEREx:2014bgr}) with the same spectral resolution and redshift binning, and a per-voxel noise~$\sigma_{\rm N} = 5500 \, {\rm Jy/sr}$ that accounts for the shorter integration time per pixel: with the conversion of Eq.~\eqref{eq:sigmaN} it corresponds to a~$5\sigma$ point-source depth~$m_{\rm AB} \simeq 19.8$.\footnote{The values are representative of the SPHEREx public sensitivity estimates in the first redshift bins, \url{https://github.com/SPHEREx/Public-products}.}

\subsection{Line emission model and fiducial signal}
\label{sec:emission}

The luminosity in H$\alpha$ (in units of erg\,s$^{-1}$) is obtained from the SFR (in units of~$M_{\odot} {\rm yr}^{-1}$) as~\cite{Kennicutt:1998zb}
\begin{equation}
    L_{{\rm H}\alpha}(M,z) = \frac{{\rm SFR}(M,z)}{7.9 \times 10^{-42}}.
\end{equation}
We use the empirical fit of Behroozi et al.~\cite{Behroozi:2012sp,Behroozi:2013SFR} to relate the SFR and the halo mass and redshift.
The integrals in Eq.~\eqref{eq:moments} span from~$M_{\rm min}=10^9\,M_{\odot}$ to~$M_{\rm max}=10^{15}\,M_{\odot}$, and the parameters~$p_{n,\sigma}$, which capture the scatter in the SFR-to-luminosity relation, are computed using~$\sigma = 0.37$, which was calibrated on CO scatter~\cite{Li:2015gqa}.
Recent work has pointed out that scatter might be significantly larger due to bursty star formation~\cite{Kovetz:2026xfs}.
Within the convention that the mean luminosity is calibrated on observations, the burstiness only affects the higher-order moments of luminosity,~$\VEV{I^2}$ and~$\VEV{I^3}$: any forecast's dependence on the scatter is largely absorbed into the fiducial point once the intensity moments are varied (Sec.~\ref{sec:results_moments}).
A mean-preserving scatter does not affect the bias parameters that are linear in the luminosity; however, the pair-weighted bias~$b_1^{[L^2]}$ which depends on~$L^2(M)$ would change in the presence of a mass-dependent scatter, which re-weights the halo mass kernel contributing to the integrand.

The line-emission model, together with the halo-bias relations of Appendix~\ref{app:biases}, fixes the fiducial values of the mean intensity, the luminosity-weighted biases, and the shot-noise moments in each bin; they are collected in Table~\ref{tab:specs} in Appendix~\ref{app:survey_specs}.

\begin{figure}[t]
    \centering
    \includegraphics[width=\columnwidth]{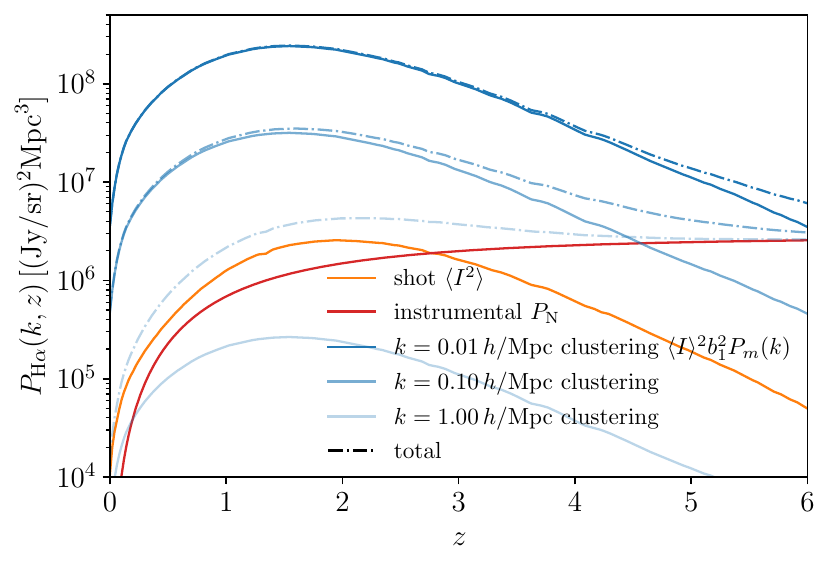}
    \caption{Contributions to the total observed H$\alpha$ power spectrum for the SPHEREx-like survey, as a function of redshift: clustering, shot-noise, and instrumental-noise terms.
    On the scales entering our analysis, $k \leq k_{\rm max}=0.15\,h/{\rm Mpc}$, the intensity maps are clustering-dominated at all but the highest redshifts, while shot noise and instrumental noise control the inverse-variance weights of Eq.~\eqref{eq:weighted_kernel}.}
    \label{fig:contributions_pk}
\end{figure}
 
Figure~\ref{fig:contributions_pk} summarizes the resulting noise budget.
On the scales entering the analysis the maps are clustering-dominated over most of the redshift range, so the survey is a meaningful target for a higher-order-statistics forecast.
The redshift evolution of the clustering contribution at a fixed value of~$k$ closely tracks the star-formation rate, while its scale-dependence follows the shape of the linear matter power spectrum.

Contrary to the power spectrum, the amplitude of the skew spectra signal as a function of redshift does not closely track the star-formation rate, because of the inverse-variance weighting: in the clustering-dominated regime, the amplitude scales as~$[\VEV{I}b_1]^{-1}$ (see Appendix~\ref{app:formalism}).
Figure~\ref{fig:contributions_skew} shows the contributions to the skew spectra as a function of scale: the clustering terms of Eq.~\eqref{eq:skew_clust} and the shot noise terms of Eq.~\eqref{eq:skew_shot}.

\begin{figure*}
    \centering
    \includegraphics[width=\linewidth]{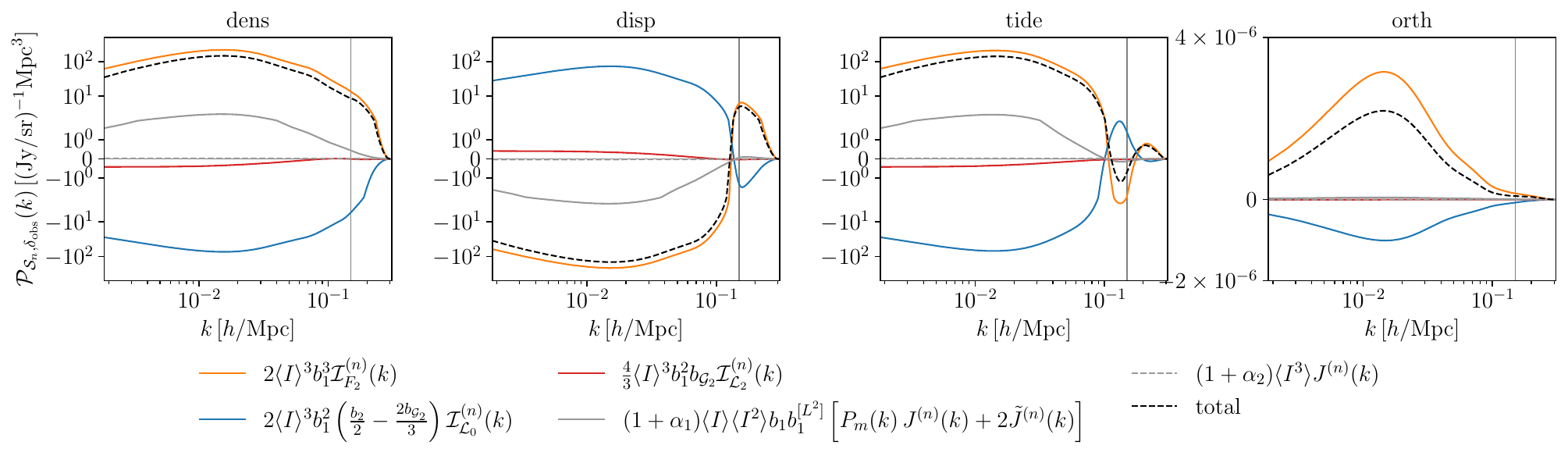}
    \caption{Contributions to the observed skew spectra for the SPHEREx-like survey in the second redshift bin, centered at~$z=0.8$, as an example. For each of the~$\{{\rm dens},{\rm disp},{\rm tide},{\rm orth}\}$ kernels, we show the clustering terms of Eq.~\eqref{eq:skew_clust}, the shot-noise terms of Eq.~\eqref{eq:skew_shot}, and the total. The vertical line marks~$k_{\rm max}=0.15\,h/{\rm Mpc}$.}
    \label{fig:contributions_skew}
\end{figure*}
 
\subsection{Data vectors and Fisher matrices}
\label{sec:fisher_matrices}

In each redshift bin, we consider five data vectors: the bispectrum; the skew spectra of the quadratic fields built with the kernels of Eq.~\eqref{eq:kernels}; the skew spectra of the same set augmented with the shot-sector kernel of Eq.~\eqref{eq:orth_kernel},
\begin{equation}
\label{eq:kernel_sets}
    \mathbf{S}^{(3)} = \left\{ \cS_{\rm dens}, \, \cS_{\rm disp}, \, \cS_{\rm tide} \right\} \, , \qquad
    \mathbf{S}^{(4)} = \mathbf{S}^{(3)} \cup \left\{ \cS_{\rm orth} \right\} \, ;
\end{equation}
the power spectrum; and the combination of the power spectrum with~$\mathbf{S}^{(4)}$.
Since~$\cS_{\rm orth}$ is a linear combination, with constant coefficients, of~$\cS_{\rm shot}$ and the fields in~$\mathbf{S}^{(3)}$, the~$\mathbf{S}^{(4)}$ forecasts are unchanged if~$\cS_{\rm orth}$ is replaced by~$\cS_{\rm shot}$.
 
The skew-spectrum data vector collects the~$N_c$ skew spectra of a kernel set, with~$N_c = 3$ for~$\mathbf{S}^{(3)}$ and~$N_c = 4$ for~$\mathbf{S}^{(4)}$, in~$N_k$ wavenumber bins,
\begin{equation}
\label{eq:data_vector}
       \mathbf{d} = \left\{ \cP_{\cS_n,\delta_{\rm obs}}(k_j) \right\} \, , \qquad \cS_n \in \mathbf{S}^{(3)} \text{ or } \mathbf{S}^{(4)} \, , \quad j = 1, \ldots, N_k \, ,
\end{equation}
and its Fisher matrix in the redshift bin~$z_i$ is
\begin{equation}
\label{eq:fisher_skew}
    F_{\alpha\beta}(z_i) = \sum_{a,b=1}^{N_c N_k} \frac{\partial d_a}{\partial \theta_{\alpha}} \left( {\rm Cov}^{-1} \right)_{ab} \frac{\partial d_b}{\partial \theta_{\beta}} \, ,
\end{equation}
where the covariance includes the auto- and cross-covariances of all skew spectra in the data vector, Eq.~\eqref{eq:skew_cov}, evaluated with the power spectrum~$P_{\rm obs}$ of the observed field.
The numerical evaluation of the skew spectra and the regularization of their covariance are described in Appendices~\ref{app:fftlog} and~\ref{app:covreg}.
 
For the bispectrum, the Fisher matrix is
\begin{equation}
\label{eq:fisher_bispectrum}
    F_{\alpha\beta}(z_i) = \sum_{\rm triangles} \frac{1}{{\rm Var}\left[ B_{\rm obs} \right]} \, \frac{\partial B_{\rm obs}}{\partial\theta_{\alpha}} \, \frac{\partial B_{\rm obs}}{\partial\theta_{\beta}} \, .
\end{equation}
Here, $B_{\rm obs} \equiv B_{\rm obs}(k_1,k_2,k_3)$, the sum runs over closed triangles, $k_3 \leq k_1 + k_2$, with~$k_{\rm min} \leq k_1 \leq k_2 \leq k_3 \leq k_{\rm max}$, and the Gaussian variance is
\begin{equation}
\label{eq:var_bispectrum}
    {\rm Var}\left[ B_{\rm obs} \right] = \frac{s_{123}\, (2\pi)^3 k_{\rm f}^3}{V_{123}} \prod_{j=1}^{3} P_{\rm obs}(k_j) \, ,
\end{equation}
where~$k_{\rm f} = 2\pi/V_i^{1/3}$ is the fundamental mode of the $i$th redshift bin, of volume~$V_i$, $s_{123}=6,2,1$ for equilateral, isosceles, and scalene triangles, and~$V_{123} = 8\pi^2 k_1 k_2 k_3\, \Delta k^3$.
We evaluate the sum in the continuum limit by adaptive integration.

For the power spectrum,
\begin{equation}
\label{eq:fisher_power}
    F_{\alpha\beta}(z_i) = \sum_{k = k_{\rm min}}^{k_{\rm max}} \frac{1}{{\rm Var}\left[ P_{\rm obs} \right]} \, \frac{\partial P_{\rm obs}}{\partial\theta_{\alpha}} \, \frac{\partial P_{\rm obs}}{\partial\theta_{\beta}} \, ,
\end{equation}
where the Gaussian variance is given by
\begin{equation}
    {\rm Var}\left[ P_{\rm obs} \right] = \frac{2\, k_{\rm f}^3}{V_k} \, P_{\rm obs}^2(k) \, ,
\end{equation}
with~$V_k = 4\pi k^2 \Delta k$ the volume of the shell.

We neglect connected non-Gaussian covariance and super-sample covariance.
These approximations define the scope of the compression comparison; the absolute parameter errors should be validated with a full covariance, particularly in configurations with high signal-to-noise~\cite{Pamuk:2026ubo}.
 
In each redshift bin, the minimum wavenumber~$k_{\rm min}$ is the fundamental mode of the bin volume, the bin width is~$\Delta k = k_{\rm min}$ for the bispectrum and power spectrum analysis and~$\Delta k = 2 k_{\rm min}$ for the skew spectra, and our baseline small-scale cut is~$k_{\rm max}=0.15\,h/{\rm Mpc}$, consistent with the reach of tree-level perturbation theory at these redshifts.
Assuming independent redshift bins, the total Fisher matrix is~$F_{\alpha\beta} = \sum_i F_{\alpha\beta}(z_i)$, and we quote marginalized~$1\sigma$ uncertainties.
All data vectors are computed with the same linear power spectrum, emission model, noise model, and range of scales, so that any difference between the bispectrum and skew-spectrum constraints isolates the information lost in the compression.

\subsection{Varied parameters}
\label{sec:parameters}

As discussed in Sec.~\ref{sec:lim}, the clustering statistics of line intensity fluctuations constrain only the products of the mean intensity and the luminosity-weighted biases.
The parameter sets below differ in which of these quantities, and of their redshift evolution, are fixed by the emission model.
When the redshift evolution is fixed (Stages~1 and~2 below), departures from the fiducial model are parametrized as follows: each bias, and the shot-noise correction, is varied by a redshift-independent shift about its fiducial evolution, $b_\alpha(z) = b_{\alpha,{\rm fid}}(z) + \Delta b_\alpha$, so that the quoted uncertainties on the biases are absolute; each luminosity moment is varied by a redshift-independent factor, $\VEV{I^n}(z) = A_n \VEV{I^n}_{\rm fid}(z)$, so that the uncertainties on~$\ln\VEV{I^n}$ are fractional. This choice is not arbitrary: three physical sources of departure have different signatures. A change in the calibration of the luminosity--star-formation relation, the dominant uncertainty of the emission model, rescales the mean intensity and the higher moments but leaves the luminosity-weighted biases of Eq.~\eqref{eq:L-weighted_biases} unchanged, since their weights are normalized; a dependence of the emission on the environment of the halo at fixed mass adds a term to each bias, independent of the halo bias itself~\cite{Barreira:2021ukk}, and leaves the mean intensity unchanged if the scatter is mean-preserving; a change in the mass range or slope of~$L(M)$ changes the mean intensity and the biases in a correlated but unequal, redshift-dependent way. The parametrization above covers the first two, under the further assumption that the departures are redshift-independent. The third is explored through the emission-model variants of Sec.~\ref{sec:results_perbin}, and, as a generic test, by varying all the biases by a common factor in Sec.~\ref{sec:results_Ibarb1}, which represents an uncertainty in the effective halo mass of the emitters that is redshift-independent in fractional terms; the per-bin analysis of Sec.~\ref{sec:results_perbin} makes no assumption on any of the three. The consequences of these choices for the separation of~$\VEV{I}$ from~$b_1$ are discussed in Secs.~\ref{sec:results_moments} and~\ref{sec:results_Ibarb1}.

\textit{Stage 1: skew spectra versus bispectrum.}
We compare the bispectrum with~$\mathbf{S}^{(3)}$ and~$\mathbf{S}^{(4)}$ for a known emission model: we fix the mean intensity and the higher moments to the model of Sec.~\ref{sec:emission}, and vary the biases together with the correction to the Poisson shot noise.
The redshift evolution of each bias is fixed to that of the fiducial model, Eq.~\eqref{eq:L-weighted_biases}, and each bias is varied by the redshift-independent shift~$\Delta b_\alpha$ defined above.
The parameter vector is
\begin{equation}
\label{eq:theta_bias}
    \vec{\theta} = \left\{ b_1, \; b_2, \; b_{\cG_2}, \; b_1^{[L^2]}, \; \alpha \right\} \, ,
\end{equation}
where~$b_1^{[L^2]}$ is varied independently in our baseline, and tied to~$b_1$ through the emission model, $b_1^{[L^2]} = r_{\rm fid}\, b_1$ (Sec.~\ref{sec:shot}), where stated.
We also consider independent corrections~$\alpha_1$ and~$\alpha_2$ to the two shot-noise terms of the bispectrum.
We then extend this set to the cosmological parameters,
\begin{equation}
\label{eq:theta_cosmo}
    \vec{\theta} = \left\{ \Omega_c, \; \Omega_b, \; h, \; n_s, \; \ln(10^{10}A_s), \; b_1, \; b_2, \; b_{\cG_2}, \; b_1^{[L^2]}, \; \alpha \right\} \, ,
\end{equation}
both alone and with a Gaussian prior on the cosmological block, taken as the inverse of the \emph{Planck} 2018 parameter covariance from the baseline \texttt{base\_plikHM\_TTTEEE\_lowl\_lowE\_lensing} chains~\cite{Planck:2018vyg}, projected onto these five parameters.
The results of this stage are presented in Sec.~\ref{sec:results_grav} and Appendix~\ref{app:results_cosmo}.

\textit{Stage 2: power spectrum, skew spectra, and their combination.}
We then relax the assumptions on the emission model in two steps.
First, we keep the redshift evolution of the luminosity fixed, but vary the three moments that enter the tree-level statistics,
\begin{equation}
\label{eq:theta_moments}
    \vec{\theta} = \left\{ \ln\VEV{I}, \; \ln\VEV{I^2}, \; \ln\VEV{I^3}, \; b_1, \; b_2, \; b_{\cG_2}, \; b_1^{[L^2]} \right\} \, ,
\end{equation}
where each moment is varied by the redshift-independent factor~$A_n$ defined above, and~$\alpha$ is dropped, since it is degenerate with the moments (Sec.~\ref{sec:degeneracies}).
With this set, we compare the bispectrum and~$\mathbf{S}^{(4)}$ (Sec.~\ref{sec:results_moments}), and the power spectrum, $\mathbf{S}^{(4)}$, and their combination (Sec.~\ref{sec:results_Ibarb1}).
Second, we make no assumption on the redshift evolution of either the biases or the moments, and constrain in each bin the amplitudes that the tree-level statistics measure,
\begin{equation}
\label{eq:theta_generic}
    \vec{\theta} = \left\{ A_1, \; A_2, \; A_{\cG_2}, \; C, \; D, \; E \right\} \, ,
\end{equation}
with~$A_\alpha = \VEV{I} b_\alpha$ for~$\alpha \in \{1, 2, \cG_2\}$, $C = (1+\alpha_1)\VEV{I^2} b_1^{[L^2]}$, $D = (1+\alpha_2)\VEV{I^3}$, and~$E = (1+\alpha_{\rm P})\VEV{I^2}$.
In this set, the corrections to the mixed and constant shot-noise terms of the bispectrum, $\alpha_1$ and~$\alpha_2$, and to the power-spectrum shot noise, $\alpha_{\rm P}$, are independent.
The results for the power spectrum and its combination with~$\mathbf{S}^{(4)}$ are presented in Sec.~\ref{sec:results_perbin}.

\section{Results: SPHEREx-like survey}
\label{sec:results}

We present four sets of forecasts in this section. In all presented forecasts, we consider H$\alpha$ LIM with the SPHEREx-like survey described in Sec.~\ref{sec:survey}, and include the instrumental noise in the weights and covariances.

\subsection{Biases and shot noise}
\label{sec:results_grav}

\begin{figure}[t!]
    \centering
    \includegraphics[width=\columnwidth]{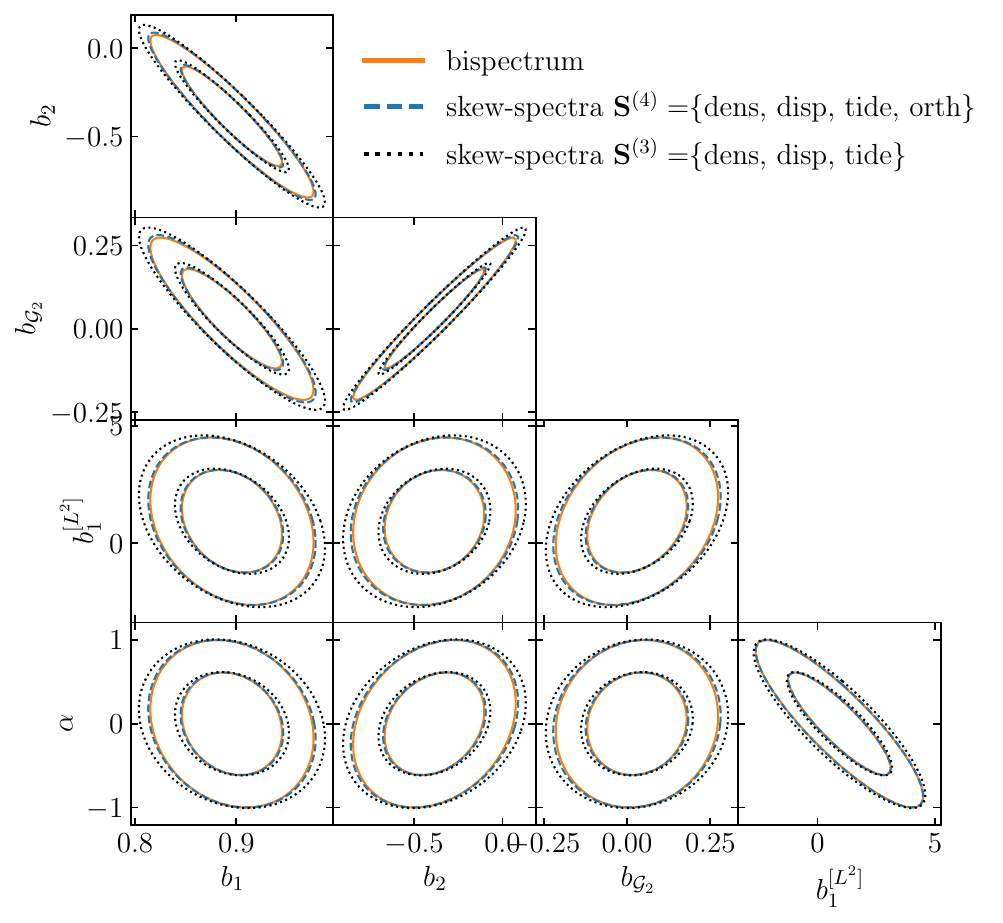}
    \caption{Marginalized constraints on the luminosity-weighted biases and the shot-noise deviation~$\alpha$ for the SPHEREx-like H$\alpha$ survey, with~$k_{\rm max}=0.15\,h/{\rm Mpc}$ and instrumental noise included: full bispectrum (solid) versus the standard three-kernel skew spectra set~$\mathbf{S}^{(3)}$ (dotted) and the extended four-kernel skew spectra set~$\mathbf{S}^{(4)}$ (dashed).}
    \label{fig:lim_constraints}
\end{figure}

Assuming a known redshift evolution for the biases and the shot-noise correction (a single redshift-independent shift for each, Sec.~\ref{sec:parameters}), and fixing the luminosity moments to the emission model of Sec.~\ref{sec:emission}, Fig.~\ref{fig:lim_constraints} compares the constraints on~$\{b_1, b_2, b_{\cG_2}, b_1^{[L^2]}, \alpha\}$ from the full bispectrum with those from the skew spectra, using either the three-kernel set~$\mathbf{S}^{(3)}$ or the extended set~$\mathbf{S}^{(4)}$; the marginalized errors are listed in Table~\ref{tab:constraints}.
The pair-weighted bias~$b_1^{[L^2]}$ is varied as an independent parameter; the case in which it is tied to~$b_1$ through the emission model is also discussed below. Since the parameters are shifts common to all bins, the contours are drawn around the fiducial values of the first bin, $z=0.4$ (Table~\ref{tab:specs}).

\begin{table*}[t]
    \centering
    \begin{tabular}{l c c c c c}
        \hline\hline
        \multicolumn{6}{c}{Bispectrum versus skew spectra compression} \\
        \hline
        & $\sigma(b_1)$ & $\sigma(b_2)$ & $\sigma(b_{\cG_2})$ & $\sigma(b_1^{[L^2]})$ & $\sigma(\alpha)$ \\
        \hline
        Bispectrum                       & $0.0325$ & $0.186$ & $0.0982$ & $1.44$ & $0.403$ \\
        skew spectra, $\mathbf{S}^{(3)}$ & $0.0371\;(+14.0\%)$ & $0.209\;(+12.4\%)$ & $0.110\;(+12.3\%)$ & $1.47\;(+2.4\%)$ & $0.405\;(+0.7\%)$ \\
        skew spectra, $\mathbf{S}^{(4)}$ & $0.0333\;(+2.6\%)$  & $0.191\;(+3.0\%)$  & $0.101\;(+3.1\%)$  & $1.44\;(-0.1\%)$ & $0.402\;(-0.1\%)$ \\
        \hline\hline
    \end{tabular}
    \caption{Marginalized~$1\sigma$ constraints for the SPHEREx-like H$\alpha$ survey with~$k_{\rm max}=0.15\,h/{\rm Mpc}$ and instrumental noise, from the full bispectrum and from the skew spectra with the kernel sets of Eq.~\eqref{eq:kernel_sets}. Percentages relative to the bispectrum row in parentheses.}
    \label{tab:constraints}
\end{table*}

The qualitative outcome of the galaxy-clustering forecasts~\cite{MoradinezhadDizgah:2019xun} carries over to intensity mapping: the four-kernel skew spectra contours track the bispectrum ones closely in both size and orientation, and the marginalized constraints agree to within~$3\%$ for the biases and~$1\%$ for~$\alpha$ (Table~\ref{tab:constraints}).
The skew spectra capture the bispectrum information on the bias and shot-noise parameters (Fig.~\ref{fig:lim_constraints}) and, non-trivially, on the cosmological parameters as well (Appendix~\ref{app:results_cosmo}) even though the kernels are built to be optimal only for amplitude-like parameters.
The new feature is that the three clustering kernels, which sufficed in the galaxy-clustering case, no longer do: a fourth skew spectrum, matched to the shot-noise templates, is required.
We show below that this requirement is set by the small volume per redshift bin of the deep-field configuration, which places the fundamental mode at~$k_{\rm f} \simeq 0.01\,h/{\rm Mpc}$ and leaves only a few~$k$-bins below~$k_{\rm max}$, not by the LIM-specific parameters. Each bin is treated as an independent volume with its own fundamental mode, since the mean intensity, biases and noise evolve across bins and modes longer than a bin's depth are not described by the fixed-parameter model; the total volume enters through the sum of the per-bin Fisher matrices.

The three-kernel set~$\{{\rm dens},{\rm disp},{\rm tide}\}$ does not recover the full bispectrum information: the errors on the three biases sit~$12$--$14\%$ above the bispectrum ones, while~$b_1^{[L^2]}$ and~$\alpha$ are recovered to~$0.5$--$2\%$. The latter is not a success of the compression but the effect of a degeneracy shared by both statistics: within a redshift bin, $b_1^{[L^2]}$ and~$\alpha$ enter only through the amplitude~$(1+\alpha)\,b_1 b_1^{[L^2]}$ of the mixed shot-clustering term and are separated only by the pure~$\VEV{I^3}$ term and by the assumed redshift evolution (see the~$\alpha$--$b_1^{[L^2]}$ panel of Fig.~\ref{fig:lim_constraints}).
Their errors are set by this degeneracy in the bispectrum and in the skew spectra alike, which masks the information lost in the compression; the loss shows up where the parameters are well measured, on the biases.
Its origin is the response structure of Fig.~\ref{fig:responses}: within each kernel the responses to the five parameters have nearly the same shape in~$k$, so each skew spectrum effectively measures one linear combination of them, and three kernels leave one direction of the bias--shot-noise space constrained only by the small residual shape differences; marginalizing over it costs~$\sim 12\%$ on each bias.

Extending the data vector by the shot-sector kernel closes the gap: the~$\mathbf{S}^{(4)}$ contours in Fig.~\ref{fig:lim_constraints} track the bispectrum ones for all parameter pairs, with errors within~$2.6$--$3.1\%$ of the bispectrum for the biases and within~$1\%$ for~$b_1^{[L^2]}$ and~$\alpha$ (Table~\ref{tab:constraints}).
These percentages are accurate to about~$2\%$, the numerical precision of our covariances (the forecasts use the regularized covariance of Appendix~\ref{app:covreg}, through eigenvalue clipping; the~$\mathbf{S}^{(3)}$ covariance is in practice unregularized, because all its eigenvalues lie above the numerical error floor); the gap between~$\mathbf{S}^{(4)}$ and the bispectrum is therefore at the level of numerical error, and an upper bound on the information lost in the compression.
What the fourth kernel adds is therefore precision on the biases, not on the shot-noise parameters: as Fig.~\ref{fig:responses} anticipates, it does not so much measure the shot noise as supply the response direction the clustering kernels lack.
Four kernels are also enough for five parameters, because~$\alpha$ and~$b_1^{[L^2]}$ enter through the same mixed template; what matters is the number of distinct response directions, not the number of parameters.

The three clustering kernels suffice for galaxy surveys~\cite{Schmittfull:2014tca,MoradinezhadDizgah:2019xun} because the shot-noise amplitude, to which none of them is matched, can still be measured from the $k$-dependence of the skew spectra, provided the data vector extends to low enough~$k$: the DESI-like survey of~\cite{MoradinezhadDizgah:2019xun} has a fundamental mode~$k_{\rm f} \simeq 0.004\,h/{\rm Mpc}$, against~$k_{\rm f} \simeq 0.01\,h/{\rm Mpc}$ for the SPHEREx deep fields, which leaves only~$N_k = 4$--$7$ bins below~$k_{\rm max}$. Reducing the volume of a DESI-like survey until its fundamental mode matches that of the SPHEREx bins reproduces both the three-kernel deficit and its recovery by the fourth kernel.\footnote{The volume is reduced by a factor of~$40$ at the same~$k_{\rm max}$; the galaxy-clustering~$\mathbf{S}^{(3)}$ constraints degrade by~$15$--$48\%$ relative to the bispectrum, and the shot-sector kernel restores them. In both the LIM and the galaxy case, the unmarginalized constraints of~$\mathbf{S}^{(3)}$ match the bispectrum to a percent (for LIM, see the un-marginalized rows of Fig.~\ref{fig:perbin_compression_cosmo}): the loss is a property of the joint fit alone.}

Tying~$b_1^{[L^2]}$ to~$b_1$ through the emission model, $b_1^{[L^2]} = r_{\rm fid}\,b_1$ (Sec.~\ref{sec:shot}), removes one parameter and the degeneracy above: $\sigma(\alpha)$ halves, from~$0.40$ to~$0.21$, and the bias errors tighten by~$1$--$5\%$.
The comparison between the statistics is unchanged, $\mathbf{S}^{(4)}$ remaining within~$2.4$--$3.0\%$ of the bispectrum for the biases and~$1.6\%$ for~$\alpha$, but the pattern of the three-kernel deficit changes: with~$\alpha$ now constrained through the mixed term and thereby coupled to~$b_1$, the degradation of~$\mathbf{S}^{(3)}$ spreads to all four parameters, at~$10$--$12\%$, with~$\alpha$ the least affected. The constraint on~$b_1^{[L^2]}$ itself is weak in either statistic ($\sigma \simeq 1.4$, comparable to its value); it improves in the joint analysis with the power spectrum, which pins the shot-noise plateau, and the tied convention is what allows that analysis to break the~$\VEV{I}$--$b_1$ degeneracy within a bin (Sec.~\ref{sec:results_Ibarb1}).

This conclusion would hold even if we were to distinguish between the shot-noise parameters~$\alpha_1 \neq \alpha_2$ in Eq.~\eqref{eq:skew_shot}: the gap between~$\mathbf{S}^{(3)}$ and bispectrum widens with the extra parameter to~$16$--$22\%$, but the four-kernel compression captures the additional degree of freedom. The bias constraints are mildly degraded (for all three statistics,~$b_1$ degrades by~$21$--$25\%$, $b_2$ by~$16$--$20\%$, $b_{\cG_2}$ by~$13$--$17\%$, and~$b_1^{[L^2]}$ by~$8$--$10\%$).
Compared to the single~$\alpha$, the error on~$\alpha_1$ more than doubles, while the pure shot term controlled by~$\alpha_2$ is essentially unconstrained.
When keeping the two shot-noise parameters distinct,~$b_1^{[L^2]}$ and~$\alpha_1$ are degenerate within a single bin: the assumed redshift evolution of the bias allows to separate them in the~$z$-summed constraints (we return to this assumption in Sec.~\ref{sec:results_perbin}).

\subsection{Line-luminosity moments}
\label{sec:results_moments}

Next, we relax the assumption that the luminosity moments are known and let the three moments that control the tree-level bispectrum vary independently.
Their redshift evolution is kept fixed to that of the fiducial model, Eq.~\eqref{eq:moments}, and each is varied by the redshift-independent factor~$A_n$ of Sec.~\ref{sec:parameters} (Eq.~\eqref{eq:theta_moments}); the per-bin analysis without this assumption is presented in Sec.~\ref{sec:results_perbin}. The shot-noise parameter~$\alpha$ is dropped, since a common correction to the Poisson prediction is exactly degenerate with a rescaling of~$\VEV{I^2}$ and~$\VEV{I^3}$ (Sec.~\ref{sec:degeneracies}).

The main finding of Sec.~\ref{sec:results_grav} is unchanged by the enlarged model space (Table~\ref{tab:moments_constraints} and the blue contours of Fig.~\ref{fig:joint_moments_constraints}): the four-kernel skew spectra recover the bispectrum constraints to within~$4.5\%$ for all parameters, and to within~$2.0\%$ for the moments themselves.
The three-kernel set now falls~$13$--$32\%$ short of the bispectrum, with the largest deficit on~$\VEV{I^2}$ and~$\VEV{I^3}$: the three-kernel set lacks enough~$k$-leverage to resolve the additional degrees of freedom.
The constraints themselves display a clear hierarchy.
The mean intensity is measured to~$3\%$, through the sum over redshift bins, as discussed below.
The second and third moments, by contrast, are individually unconstrained ($\sigma(\ln\VEV{I^2}) \simeq 1.9$, $\sigma(\ln\VEV{I^3}) \simeq 11$, with correlation coefficient~$-0.94$; for such large errors the Fisher estimate is only indicative): the mixed and constant shot-noise contributions have very similar scale dependence over the range~$k \lesssim 0.15\,h/{\rm Mpc}$ and cannot be separated, so that only a combination of the two is measured; in addition, $\VEV{I^2}$ is exactly degenerate with~$b_1^{[L^2]}$ within each bin (Sec.~\ref{sec:results_Ibarb1}).
In an analysis of actual data, external luminosity-function priors or the power spectrum (Sec.~\ref{sec:results_Ibarb1}) would anchor the moments.

Finally, we highlight the separate constraints on~$\VEV{I}$ and~$b_1$.
Within a single bin the two are exactly degenerate (Eq.~\eqref{eq:invariant_rescaling}).
The degeneracy is broken only by the sum over redshift bins, through the parametrization of Sec.~\ref{sec:parameters}: a rescaling of~$\VEV{I}$ changes~$\VEV{I}b_1$ by the same fraction in every bin, whereas a shift of~$b_1$ changes it by~$\Delta b_1/b_{1,{\rm fid}}(z)$, which varies by a factor of four across the survey, so the two cannot compensate in all bins at once.
Accordingly, the residual correlation between~$\ln\VEV{I}$ and~$b_1$ drops from~$r = -1$ in a single bin to~$r = -0.70$ after summing over bins, yielding the separate~$3.1\%$ and~$\sigma(b_1) = 0.057$ measurements for~$\mathbf{S}^{(4)}$.
Seen as in Sec.~\ref{sec:parameters}, the shift~$\Delta b_1$ can parametrize an approximately redshift-independent contribution from environmental dependence of the emission to the luminosity-weighted bias. To give an idea of the size involved, in IllustrisTNG the star-formation-rate-weighted field picks up a contribution of order~$\Delta b_1 \simeq 0.03$ at~$z \simeq 1.5$~\cite{Jun:2025knq}: too small to detect with the skew spectra alone, but within reach of the joint analysis of Sec.~\ref{sec:results_Ibarb1}. If the biases are instead varied by a common factor, as one would do when the effective halo mass of the emitters is uncertain, the separation is lost: $\VEV{I}$ and~$b_1$ stay degenerate in the sum and only the combination with the power spectrum keeps a handle on~$\VEV{I}$, through Eq.~\eqref{eq:joint_identity}.

\subsection{Combination with power spectrum: breaking the mean intensity-bias degeneracy}
\label{sec:results_Ibarb1}

The first moment of luminosity and the linear bias stay degenerate within a single-tracer analysis in real space, unless the redshift evolution of at least one of them is perfectly known.
As shown in Sec.~\ref{sec:results_moments}, the higher-order statistics alone break it only through the assumed form of the departures from the fiducial evolution, not within an individual bin. 
Combining power spectrum and bispectrum or skew spectra can instead do so bin by bin: the ratio of Eq.~\eqref{eq:joint_identity} measures~$r/\VEV{I}$, and hence the mean intensity, \emph{if the ratio~$ r = b_1^{[L^2]} / b_1$ is known} and without redshift-space information.
This requires resolving the mixed shot-clustering amplitude of the bispectrum in each bin, which the SPHEREx configurations considered here do not (Sec.~\ref{sec:results_perbin}).
We therefore present the joint constraints summed over redshift bins, where this mechanism acts on top of the cross-bin leverage of Sec.~\ref{sec:results_moments}.

We combine the four-kernel skew spectra with the tree-level power-spectrum in the same redshift bins and with the same small-scale cut~$k_{\rm max} = 0.15\,h/{\rm Mpc}$, so that the comparison between statistics does not depend on the choice of scales; a 1-loop power spectrum would allow a higher cut (Sec.~\ref{sec:results_perbin}).

\begin{figure}[t]
    \centering
    \includegraphics[width=\columnwidth]{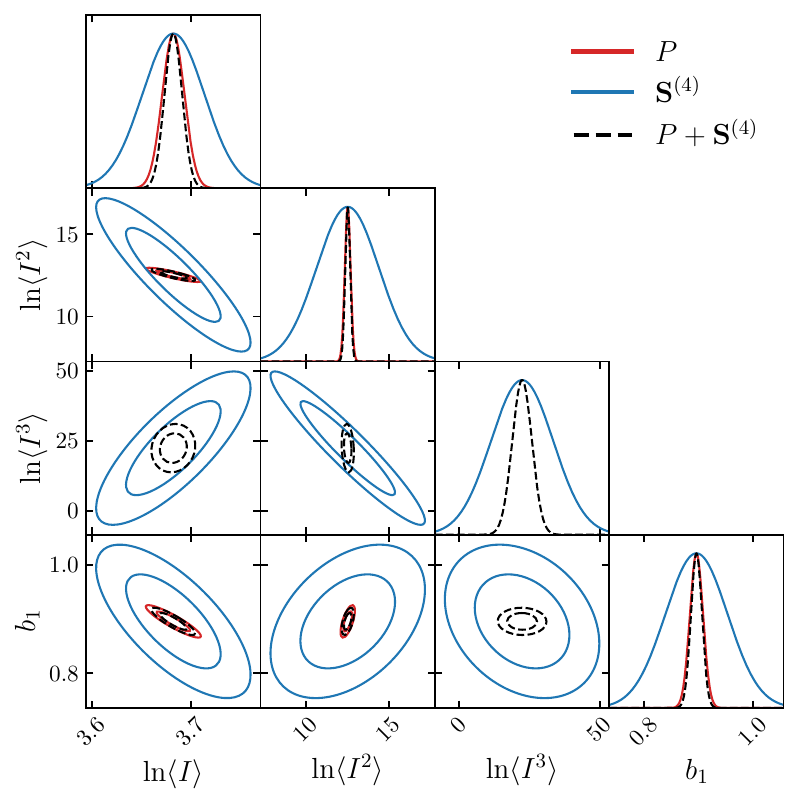}
    \caption{Marginalized constraints from the joint power spectrum and skew spectra analysis on the luminosity moments and the luminosity-weighted biases for the SPHEREx-like H$\alpha$ survey, with~$k_{\rm max}=0.15\,h/{\rm Mpc}$, four-kernel skew spectra~$\mathbf{S}^{(4)}$ (blue), tree-level power spectrum (orange), and combined (black). Only a subset of the parameters is shown, the remaining biases are marginalized over. Parametrization as in Table~\ref{tab:moments_constraints}; the separation of~$\VEV{I}$ and~$b_1$ by~$P$ alone rests on it (Sec.~\ref{sec:results_moments}).}
    \label{fig:joint_moments_constraints}
\end{figure}

Figure~\ref{fig:joint_moments_constraints} and Table~\ref{tab:moments_constraints} report the results of the joint analysis, with~$b_1^{[L^2]}$ free and the parametrization of Sec.~\ref{sec:parameters}: the moments varied by a common factor, the biases by a common shift.
With this parametrization the power spectrum alone already separates~$\VEV{I}$ and~$b_1$ in the summed analysis, to~$1.1\%$ and~$\sigma(b_1) = 0.012$, by the mechanism of Sec.~\ref{sec:results_moments}: a common shift of~$b_1$ cannot mimic a common rescaling of~$\VEV{I}$ across bins. Adding the skew spectra improves both by about~$20\%$, more than the inverse-variance combination of the two separate constraints would give ($1.06\%$ on~$\VEV{I}$): the two statistics are degenerate along different directions, this is the mechanism of Eq.~\eqref{eq:joint_identity}, acting through the sum over bins. If all the biases are instead varied by a single common multiplicative factor, so that the ratio~$r$ keeps its fiducial value, the power spectrum alone constrains only the product~$\VEV{I}b_1$, and the joint analysis measures~$\VEV{I}$ to~$34\%$ in the deep fields and~$15\%$ all-sky through Eq.~\eqref{eq:joint_identity} alone.
Giving each bias its own amplitude instead lets~$r$ float, and then no combination of the two statistics separates~$\VEV{I}$ from the biases: only the products~$\VEV{I}b_\alpha$ are measured, as in the per-bin analysis of Sec.~\ref{sec:results_perbin}.
The larger gains of the combined analysis are on the parameters the power spectrum does not measure, the higher-order biases, $\VEV{I^3}$ and~$b_1^{[L^2]}$, while its shot plateau anchors~$\VEV{I^2}$, which the higher-order statistics alone left unconstrained (Table~\ref{tab:moments_constraints}).
Tying~$b_1^{[L^2]}$ to~$b_1$ changes little in the joint analysis, because the power spectrum pins~$\VEV{I^2}$ and thereby removes its per-bin degeneracy with~$b_1^{[L^2]}$, which both enter only through the mixed term; only~$\VEV{I^3}$ remains sensitive to this choice.

\begin{table*}[t]
    \centering
    \begin{tabular}{l c c c c c c c}
        \hline\hline
        & $\sigma(\ln\VEV{I})$ & $\sigma(\ln\VEV{I^2})$ & $\sigma(\ln\VEV{I^3})$ & $\sigma(b_1)$ & $\sigma(b_2)$ & $\sigma(b_{\cG_2})$ & $\sigma(b_1^{[L^2]})$ \\
        \hline
        Bispectrum            & $0.0310$ & $1.85$ & $10.9$ & $0.0561$ & $0.219$ & $0.111$ & $2.15$ \\
        skew spectra, $\mathbf{S}^{(3)}$   & $0.0355$ & $2.31$  & $14.4$ & $0.0635$ & $0.253$ & $0.129$ & $2.50$ \\
        skew spectra, $\mathbf{S}^{(4)}$   & $0.0314$ & $1.88$ & $11.1$ & $0.0571$ & $0.228$ & $0.116$ & $2.16$ \\
        \hline
        $P(k)$                & $0.0112$            & $0.174$          & --              & $0.0121$            & --                & --                & -- \\
        Joint $P + \mathbf{S}^{(4)}$       & $0.0089$ & $0.149$ & $3.49$ & $0.0102$ & $0.074$ & $0.051$ & $0.87$ \\
        \hline\hline
    \end{tabular}
    \caption{Marginalized~$1\sigma$ constraints on the luminosity moments and the luminosity-weighted biases for the SPHEREx-like H$\alpha$ survey, with~$k_{\rm max}=0.15\,h/{\rm Mpc}$. The moments are varied by a common factor and the biases by a common shift about the fiducial evolution (Sec.~\ref{sec:parameters}); $b_1^{[L^2]}$ is free.}
    \label{tab:moments_constraints}
\end{table*}

This route to degeneracy breaking can be compared with the proposal of~\cite{Breysse:2022alx}, which reaches the same qualitative outcome by combining the power spectrum with the voxel intensity distribution (VID) for a COMAP-like CO survey.
Both close the contour in the~$\VEV{I}$--$b_1$ plane without redshift-space information, in principle within a single redshift bin.
While the joint~$P+\mathbf{S}^{(4)}$ analysis breaks the degeneracy through the ratio of Eq.~\eqref{eq:joint_identity}, the VID is a one-point statistic which relies on the shape of the intensity histogram to disentangle mean intensity and linear bias.
The two routes are complementary, and also carry different systematics (the VID is sensitive to the noise PDF and voxel-scale effects, the skew spectra to the Fourier-space noise budget entering the weights and covariance), and they could benefit from a combined analysis.

In absolute terms, the summed joint constraints on~$\VEV{I}$ and~$b_1$ are at the percent level relative to the fiducial values of Table~\ref{tab:specs}, an order of magnitude below the~$\sim 20$--$30\%$ systematic uncertainty of the H$\alpha$ luminosity calibration.
They rest on the assumed redshift evolution and are summed over bins, so they are not directly comparable to per-bin, RSD-based power-spectrum forecasts for CO/[CII] surveys~\cite{Bernal:2019jdo}.
At this level the analysis is not limited by statistics, but  by the modeling: the assumed redshift evolution and parametrization of mean intensity and biases, and the absence of interloper contamination and continuum residuals.

\subsection{Redshift evolution of the higher-order biases}
\label{sec:results_perbin}

All the results presented so far assume knowledge of the redshift evolution of the biases and the moments of intensity: the constraints are placed on common shifts of the biases and common rescalings of the moments about a fiducial evolution fixed by the emission model.
We now drop this assumption, and analyze each redshift bin on its own, with the amplitude set of Sec.~\ref{sec:parameters}, $\{A_1, A_2, A_{\cG_2}, C, D, E\}$: these six products of biases, moments, and corrections to Poissonian shot-noise are the most generic set of parameter combinations, i.e.,~the clustering and shot-noise amplitudes that power spectrum and skew spectra respond to.

\begin{figure*}[t]
    \centering
    \begin{subfigure}[b]{\textwidth}
    \includegraphics[width=0.9\textwidth]{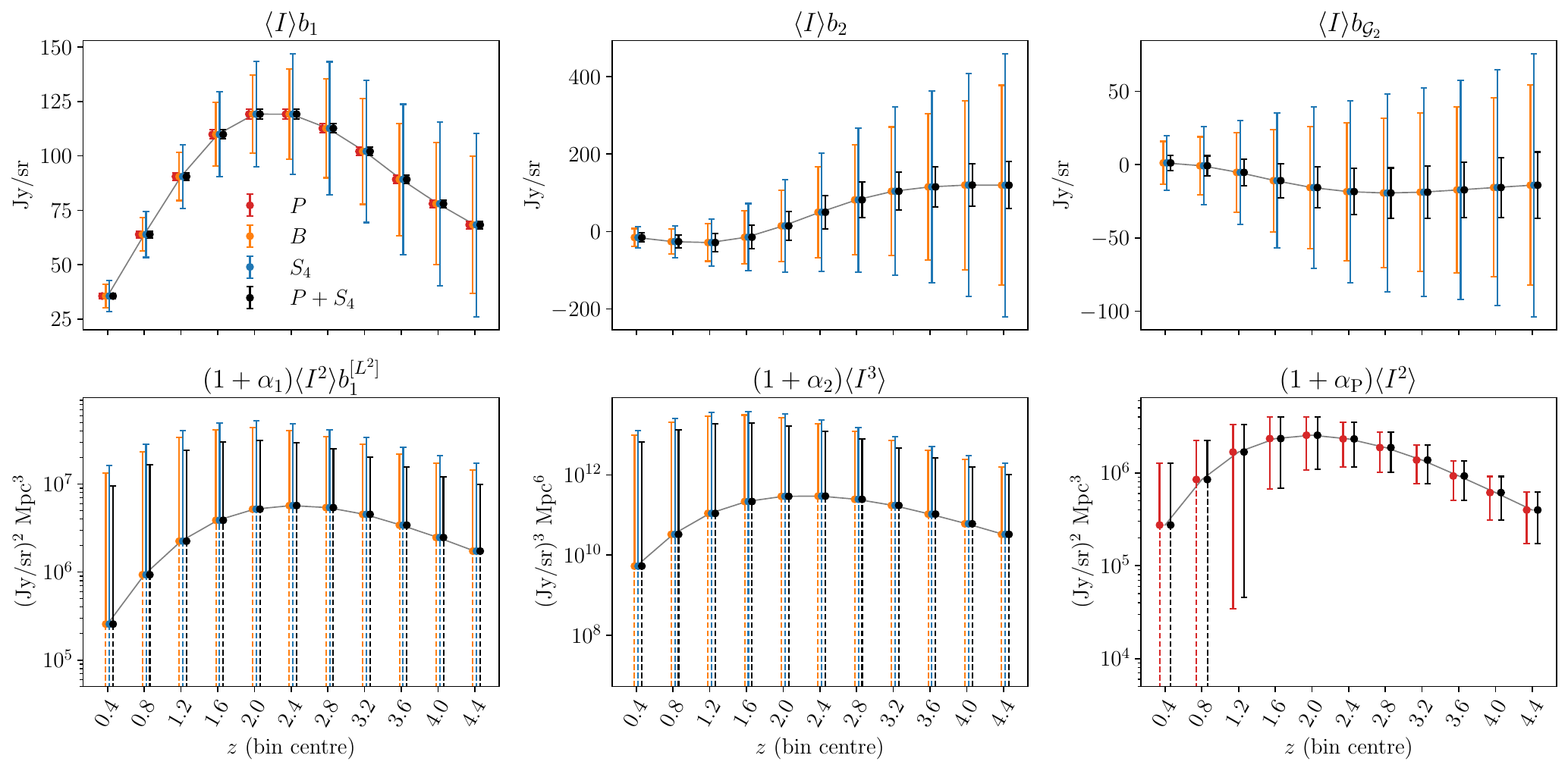}
    \caption{SPHEREx deep-field}
    \vspace{.5cm}
    \end{subfigure}
    \begin{subfigure}[b]{\textwidth}
    \includegraphics[width=0.9\textwidth]{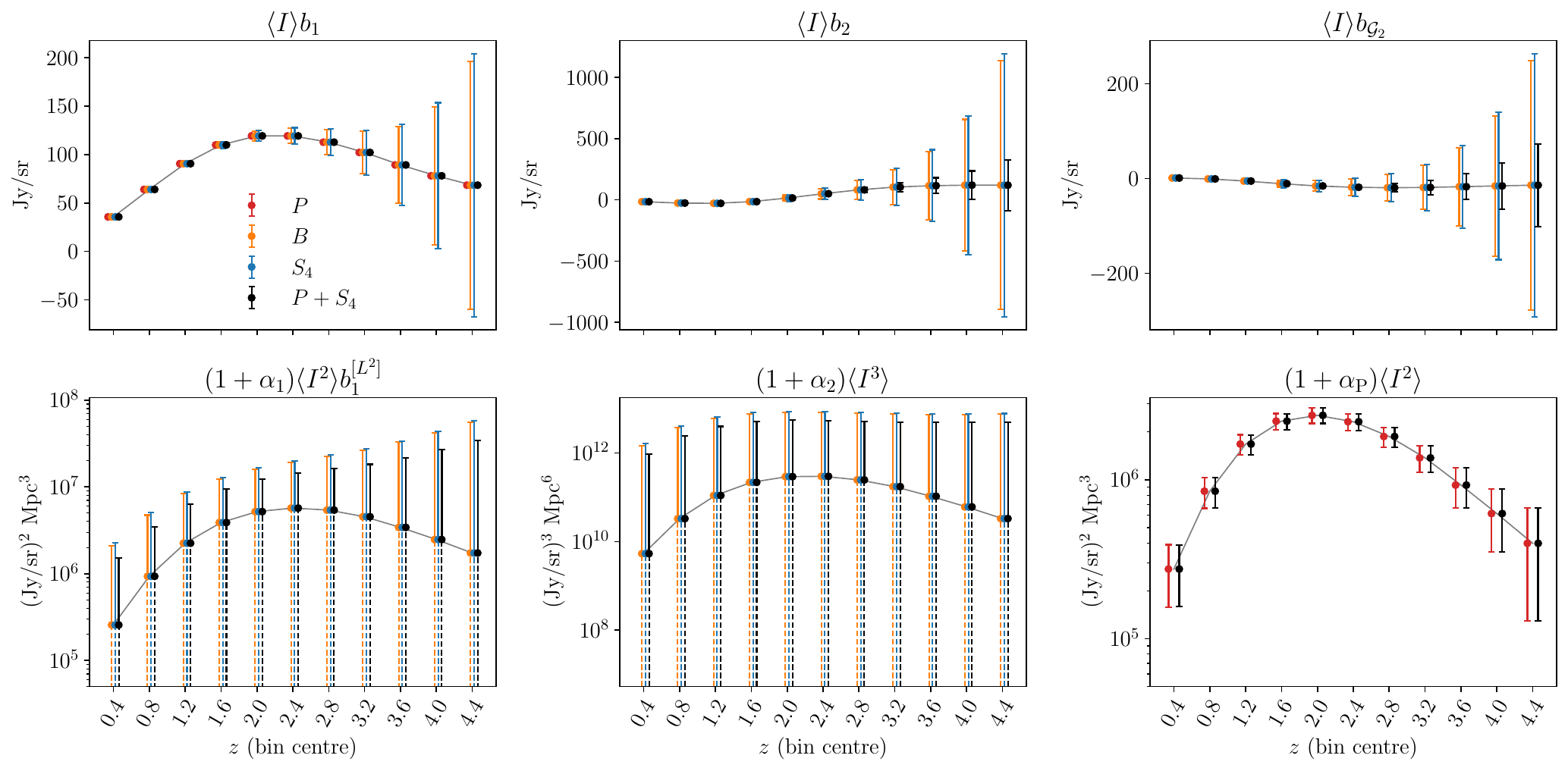}
    \caption{SPHEREx all-sky}
    \end{subfigure}
    \caption{Redshift evolution of the clustering amplitude combinations~$A_1 = \VEV{I} b_1$,~$A_2 = \VEV{I} b_2$,~$A_{\cG_2} = \VEV{I} b_{\cG_2}$, and the shot-noise amplitudes~$C = (1+\alpha_1) \VEV{I^2} b_1^{[L^2]}$,~$D = (1+\alpha_2) \VEV{I^3}$,~$E = (1+\alpha_{\rm P}) \VEV{I^2}$, from a joint analysis of power spectrum and skew spectra, for the SPHEREx-like survey of the previous sections (\emph{upper panels}:~$200 \, {\rm deg}^2$) and for an all-sky version (\emph{lower panels}:~$30\,000 \, {\rm deg}^2$). The error bars are the marginalized~$1\sigma$ from the joint~$P+\mathbf{S}^{(4)}$ analysis (dashed bars would extend below zero and are clipped).}
    \label{fig:redshift_evolution_constraints}
\end{figure*}

Figure~\ref{fig:redshift_evolution_constraints} shows the marginalized~$1\sigma$ errors from the joint~$P+\mathbf{S}^{(4)}$ analysis, for the SPHEREx-like survey in the deep-field configuration (covering~$200 \, {\rm deg}^2$), and for an all-sky scenario (covering~$30\,000 \, {\rm deg}^2$, but with enhanced instrumental noise).
The clustering amplitude~$A_1 = \VEV{I} b_1$ is measured to~$2\%$ per bin in the deep fields at all redshifts, and to~$0.3$--$0.5\%$ all-sky up to~$z \simeq 3$.
The power spectrum alone provides this, and by pinning~$A_1$, it allows the skew spectra to resolve the higher-order biases.
The higher-order amplitudes are measurable per bin, albeit weakly: $A_2 = \VEV{I} b_2$ to~$45$--$85\%$ for~$z \gtrsim 2.4$ in the deep fields ($10$--$35\%$ all-sky), and $A_{\cG_2} = \VEV{I} b_{\cG_2}$ to~$85$--$110\%$ at~$2 \lesssim z \lesssim 3.6$ ($22$--$30\%$ all-sky at~$1.2 \lesssim z \lesssim 2.4$).
The shot-noise amplitudes are poorly measured per bin: the intensity map never resolves~$C$ or~$D$, and reaches~$E$ only at the tens-of-percent level in the deep fields and~$11\%$ all-sky around~$z \simeq 2$.
Since~$C$ is unresolved, the per-bin determination of~$\VEV{I}$ through the ratio of Eq.~\eqref{eq:joint_identity} is out of reach in either configuration; the mechanism contributes only summed over bins (Sec.~\ref{sec:results_Ibarb1}).

The one shot-noise amplitude that is measured per bin, $E = (1+\alpha_{\rm P})\VEV{I^2}$, is a power-spectrum measurement, to which the skew spectra add nothing, but it is worth noting what it carries.
$\VEV{I^2}$ is the second moment of the luminosity function and, of all the quantities in Eq.~\eqref{eq:amplitudes}, the most sensitive to the scatter of the luminosity at fixed halo mass: for a lognormal scatter of~$\sigma$ dex at fixed mean, $\VEV{I^2}$ grows as~$\exp(\sigma^2 \ln^2 10)$, so the~$\sigma = 0.37$ of our model already doubles the shot noise relative to a deterministic~$L(M)$, and the larger scatter expected from bursty star formation~\cite{Kovetz:2026xfs} would raise it further, by a factor~$\sim 7$ for~$\sigma = 0.6$.
A measurement of~$E$ to~$10\%$ thus constrains~$\sigma$ to~$\simeq 0.03$ dex at fixed~$L(M)$, bin by bin, and the redshift evolution of~$E/\VEV{I}^2$ follows that of the burstiness.
As shown in Fig.~\ref{fig:redshift_evolution_constraints}, in the deep-field configuration~$E$ is measured to no better than~$45\%$ per bin, at~$z \simeq 3$, and the first two bins are unconstrained; on the other hand, the all-sky configuration measures it to~$11$--$12\%$ around~$z \simeq 2$, with precision degrading at higher redshift where the instrumental noise overtakes the shot plateau; the ten-per-cent benchmark above is thus reached only all-sky near~$z \simeq 2$.
What~$E$ alone cannot do is separate the scatter from the bright end of~$L(M)$, which enters the same integral, or from a non-Poissonian correction~$\alpha_{\rm P}$ of the kind seen in simulated intensity maps~\cite{MoradinezhadDizgah:2021dei}, to which correlated star formation of centrals and satellites also contributes~\cite{Jun:2025knq}.
That requires the third moment, which neither configuration resolves, or an external measurement of the luminosity function, for instance from the voxel intensity distribution~\cite{Ihle:2018dky,Breysse:2022alx}.

We emphasize that within the tree-level model, the power spectrum carries no information on the higher-order biases. A 1-loop treatment would, and would allow a higher~$k_{\rm max}$, but is known to constrain the higher-order biases far more weakly than the bispectrum (see e.g.,~\cite{Euclid:2026glh}); we leave this extension to a dedicated study.

\begin{figure*}[t]
    \centering
    \begin{subfigure}[b]{\textwidth}
    \includegraphics[width=0.85\textwidth]{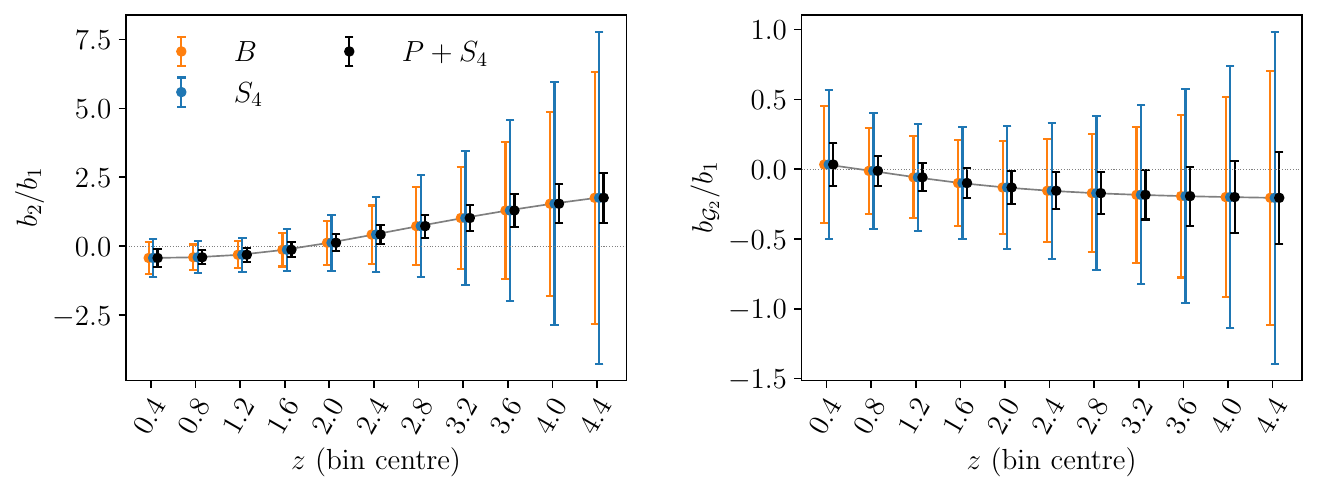}
    \caption{SPHEREx deep-field}
    \vspace{.5cm}
    \end{subfigure}
    \begin{subfigure}[b]{\textwidth}    \includegraphics[width=0.85\textwidth]{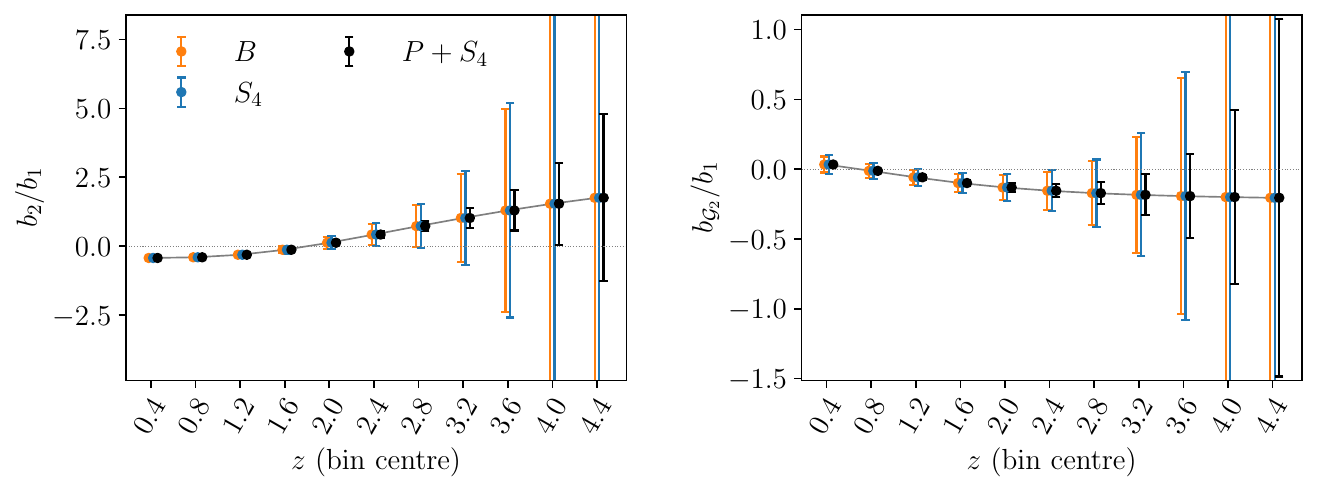}
    \caption{SPHEREx all-sky}
    \end{subfigure}
    \caption{Redshift evolution of the ratios~$A_2/A_1 = b_2/b_1$ and~$A_{\cG_2}/A_1 = b_{\cG_2}/b_1$, from the higher-order statistics alone and from a joint analysis with the power spectrum, for the SPHEREx-like survey of the previous sections (\emph{upper panels}:~$200 \, {\rm deg}^2$) and for an all-sky version (\emph{lower panels}:~$30\,000 \, {\rm deg}^2$). Axes are shared within columns to make the comparison easier between the two configurations; the higher-$z$ all-sky bins suffer from instrumental noise.}
    \label{fig:redshift_evolution_ratios}
\end{figure*}

\begin{figure*}[t]
    \centering
    \includegraphics[width=\textwidth]{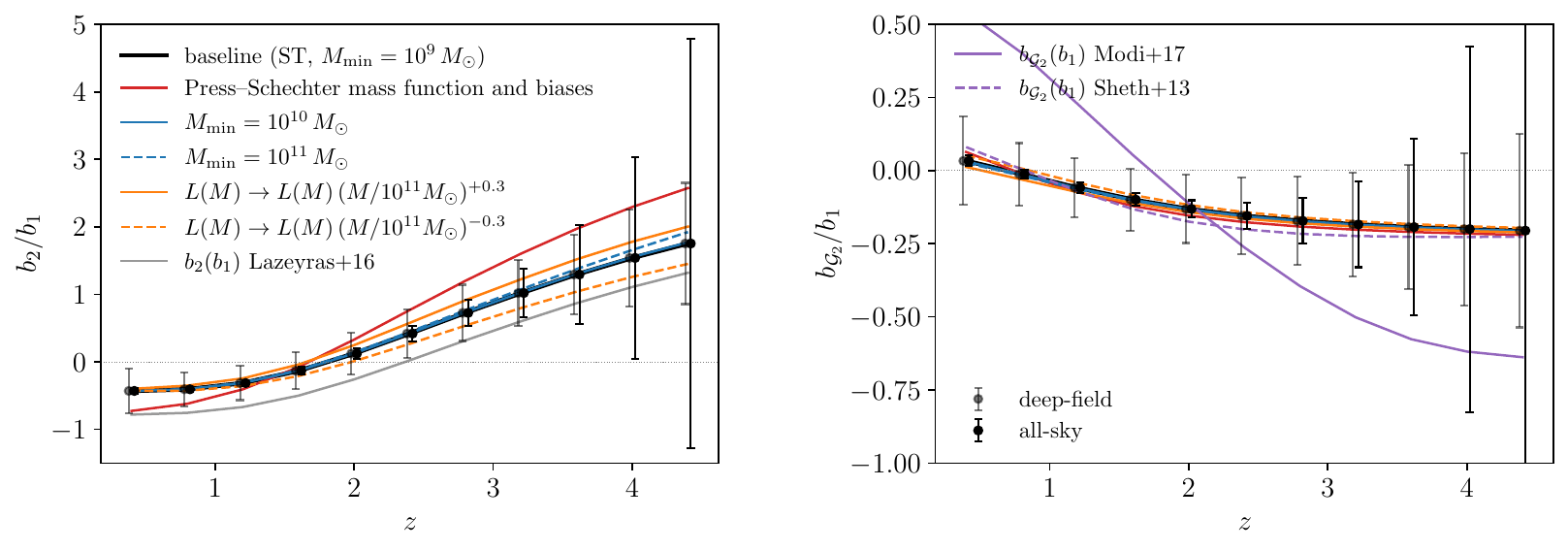}
    \caption{Variations of the ratios~$b_2/b_1$ and~$b_{\cG_2}/b_1$ within different choices of halo mass function and biases (red), lower mass cut of the emitters (blue), luminosity tilt (orange), $b_2(b_1)$ relation (light gray), presence of a Lagrangian tidal bias (purple). See main text for details. The~$1\sigma$ error bars from the joint~$P+\mathbf{S}^{(4)}$ analysis in Fig.~\ref{fig:redshift_evolution_ratios} are also shown, for the deep-field (gray) and all-sky (black) configurations.}
    \label{fig:redshift_evolution_ratios_variants}
\end{figure*}

Figure~\ref{fig:redshift_evolution_ratios} shows the constraining power on the ratios~$b_2/b_1$ and~$b_{\cG_2}/b_1$, obtained from the per-bin Fisher matrices.
These ratios are exactly invariant under the rescaling of Eq.~\eqref{eq:invariant_rescaling}: they are the quantities that a single bin measures with no reference to the mean intensity, and hence the natural per-bin observables of the bias relations.
The power spectrum contributes to them only indirectly, by pinning~$A_1$ and thereby removing the~$A_1$--$A_2$ and~$A_1$--$A_{\cG_2}$ correlations of the three-point statistics, which tightens both ratios by a factor~$2$--$3$ over the skew spectra alone.
In the deep fields the joint analysis measures~$b_2/b_1$ to~$\sigma = 0.25$--$0.35$ at~$z \lesssim 2.4$ and~$0.4$--$0.9$ at higher redshift, i.e.,~to~$45$--$80\%$ of its value wherever the fiducial is not close to zero, and~$b_{\cG_2}/b_1$ to~$\sigma = 0.10$--$0.33$, comparable to the coevolution value itself: the deep fields detect the rise of~$b_2/b_1$ with redshift but do not test the relations.
The all-sky configuration does: $b_2/b_1$ is measured to~$\sigma = 0.04$ ($10\%$) at~$z \le 1.2$ and to~$0.05$--$0.2$ ($30$--$60\%$) at~$1.6 \le z \le 2.8$, which locates the sign change of~$b_2$, predicted at~$z \simeq 1.8$ by the halo model of Appendix~\ref{app:biases}, between the~$z = 1.6$ and~$z = 2.4$ bins at more than~$3\sigma$ on either side; and~$b_{\cG_2}/b_1$ is measured to~$\sigma = 0.02$--$0.05$ at~$1.2 \le z \le 2.8$, $20$--$45\%$ of the coevolution prediction~$-\frac{2}{7}(1 - 1/b_1)$, distinguishing it from zero at~$2$--$4\sigma$ in each bin.
Beyond~$z \simeq 3.2$ the growth of the instrumental noise power degrades the all-sky ratios faster than the deep-field ones.
Per bin, the skew spectra alone lose~$15$--$20\%$ ($5$--$8\%$) on the ratios with respect to the bispectrum in the deep-field (all-sky) configuration, more than in the summed analysis of Sec.~\ref{sec:results_grav}: this is expected, since a single~$0.4$-wide redshift slice contains few~$k$-bins per kernel.

The ratios are moderately sensitive to cosmology: when marginalizing over cosmological parameters, the constraints degrade by a factor~$<2$ for~$b_2/b_1$ and~$<3$ for~$b_{\mathcal{G}_2}/b_1$.

Figure~\ref{fig:redshift_evolution_ratios_variants} puts the constraints into context, by showing how the fiducial ratios~$b_2/b_1$ and~$b_{\cG_2}/b_1$ may change, under alternative assumptions on the underlying dark matter halos and on the astrophysics of line emission.
The first class of assumptions regards the population of the emitters: the lower mass cut of the emitters ($M_{\rm min} = 10^{10}$ and~$10^{11} \, M_\odot$ against the default~$10^9 \, M_\odot$) and the slope of the luminosity-to-halo-mass relation, which we tilt with respect to Behroozi's fit by a factor~$(M/10^{11} M_\odot)^{\pm 0.3}$.
The second class concerns the halo bias relations themselves: a Press--Schechter mass function with its peak-background-split biases instead of Sheth--Tormen; the~$b_2(b_1)$ relation of the separate-universe fit of Lazeyras \textit{et al.}~\cite{Lazeyras:2015lgp},
\begin{equation} \label{eq:lazeyras}
    b_{2,h}^{\rm SU} = 0.412 - 2.143 \ b_{1,h} + 0.929 \ b_{1,h}^2 + 0.008 \ b_{1,h}^3 \, ,
\end{equation}
with~$b_{2,h}=b_{2,h}^{\rm SU}+4b_{\cG_2,h}/3$ in our basis; and a Lagrangian tidal bias which breaks the coevolution relation.
Simulations find a non-zero Lagrangian tidal bias whose sign and amplitude depend on halo mass and on the measurement method~\cite{Modi:2016dah,Abidi:2018eyd}: we show the quadratic fit to the excursion set prediction of Sheth \emph{et al.}~\cite{Sheth:2012fc,Eggemeier:2020umu,Pezzotta:2021vfn}
\begin{equation}
\label{eq:ES}
    b_{\cG_2,h} = 0.524 -0.547 \ b_{1,h} + 0.046 \ b_{1,h}^2 \, ,
\end{equation}
and the cubic fit of Modi \emph{et al.}~\cite{Modi:2016dah}
\begin{equation}
\label{eq:modi}
    b_{\cG_2,h} = 1.03 - 0.615 \ b_{1,h} + 0.118 \ b_{1,h}^2 - 0.072 \ b_{1,h}^3 \, ,
\end{equation}
which was tested in the range~$1 \lesssim b_{1,h} \lesssim 3.5$, outside of which we continue with the coevolution relation slope.

In the deep-field configuration, no variant moves the ratios by more than~$1.4\sigma$ in any bin, except for the tidal-bias relation of Modi \emph{et al.}~\cite{Modi:2016dah}.
The deep-field errors on the ratios are otherwise not informative about the bias relations.

Below~$z \lesssim 3$, the all-sky configuration is sensitive to the bias-relation variants: the two~$b_2(b_1)$ relations are distinguished at~$7$--$9\sigma$ in the lowest redshift bins, and the presence of a Lagrangian tidal bias would be detected at~$3\sigma$ combined over bins.
The quadratic tidal-bias fit shifts~$b_{\cG_2}/b_1$ by~$\gtrsim 3\sigma$ over the full range; the cubic fit by~$\sim 28\sigma$ all-sky and~$\sim 4\sigma$ deep-field. The two fits are obtained with different methods and their discrepancy is well known in the literature, see e.g.~\cite{Lazeyras:2017hxw}.
Of particular interest for the H$\alpha$ line is the possibility of probing the region~$b_{1,h}=1$, where the two fits for the tidal bias differ significantly, and which is at the edge of the range tested in~\cite{Modi:2016dah}: the per-bin ratio is a sharp discriminant in this regime.

The emission-model variants have little effect: the different mass cut~$M_{\rm min}$ is not resolved, and the tilt of~$L(M)$ moves~$b_2/b_1$ by less than~$2\sigma$ per-bin, and~$1.2\sigma$ on~$b_{\cG_2}/b_1$, concentrated at~$z \lesssim 1$.
The behavior with redshift separates the two assumptions: the luminosity tilt affects the tidal bias at low redshifts, and shifts it with the same sign across the whole redshift range, while the two fitting formulae lie above the coevolution relation at low redshifts and below it at high redshifts, so that the two effects are not degenerate across the~$0.4 \lesssim z \lesssim 2.8$ range where the all-sky map is most precise.

It is worth saying what such a departure would mean.
The luminosity-weighted biases of Eq.~\eqref{eq:L-weighted_biases} are averages of the halo biases over the luminosity--mass relation, so if the line emission depended on halo mass alone, the ratios would follow the halo relations up to a reweighting of the halo biases; the $M_{\rm min}$ and $L(M)$ variants of Fig.~\ref{fig:redshift_evolution_ratios_variants} show that this reweighting moves them by at most~$\sim 2\sigma$ per bin, even all-sky.
The halo relations themselves are known from simulations: $b_{2,h}(b_{1,h})$ to a few percent~\cite{Lazeyras:2015lgp}, better than the spread between the prescriptions of Fig.~\ref{fig:redshift_evolution_ratios_variants}, while $b_{\cG_2,h}(b_{1,h})$ lies close to the coevolution prediction, with a small negative Lagrangian tidal bias whose amplitude is still uncertain at high bias~\cite{Lazeyras:2017hxw,Abidi:2018eyd,Stucker:2024tll}.\footnote{At the all-sky precision, this Lagrangian correction is comparable to the per-bin errors on $b_{\cG_2}/b_1$. The larger Lagrangian tidal bias of Ref.~\cite{Modi:2016dah}, shown in Fig.~\ref{fig:redshift_evolution_ratios_variants}, would be distinguished at high significance.}
A larger departure would therefore test the emitters rather than the halos.
It would mean that the luminosity at fixed halo mass depends on secondary halo properties that themselves correlate with the large-scale density or tidal field, such as formation time or concentration, satellite content, or the local environment, so that the luminosity-weighted field acquires a response to the long-wavelength modes beyond that inherited from its host halos~\cite{Barreira:2021ukk}.
At the redshifts probed here, the relevant case is an assembly bias of the relation between star formation and halo mass, as discussed below; at higher redshift and lower halo mass, environmentally triggered bursts of star formation would act in the same way~\cite{Kovetz:2026xfs}.
Scatter that is independent of the environment, such as that of Sec.~\ref{sec:emission}, produces no such response: it drops out of the luminosity-weighted biases and raises only the shot noise.

For lines tracing star formation, some dependence of this kind is expected.
In hydrodynamical simulations, the occupation and star-formation rate of halos at fixed mass depend on concentration, satellite content and local density~\cite{Hadzhiyska:2020iri,Xu:2020mhz,Jun:2025knq}, enough to raise the large-scale H$\alpha$ power spectrum by~$\sim 5\%$ over a mass-only model~\cite{Jun:2025knq}.
Selecting halos by such secondary properties leaves~$b_2(b_1)$ roughly unchanged but modifies the tidal relation at order unity~\cite{Lazeyras:2021dar,Stucker:2024tll}, making~$b_{\cG_2}/b_1$ the more sensitive of the two ratios.
For galaxies in IllustrisTNG, the bias relations are broadly preserved, with tentative departures for some selections~\cite{Barreira:2021ukk} that would be several times the all-sky errors of Fig.~\ref{fig:redshift_evolution_ratios}.
The per-bin ratios can test this halo--emission connection where the two-point statistics cannot, since the power spectrum measures only~$\VEV{I}b_1$ and any change of~$b_1$ can be absorbed by~$\VEV{I}$.
Our fiducial model has no such dependence, so the forecast tells us what size of departure SPHEREx could detect, not what size to expect.
Predicting the corresponding shifts for a luminosity-weighted field requires a model of how the line luminosity correlates with secondary halo properties, for which forward models of line emission trained on hydrodynamical simulations~\cite{Gao:2026pp1} are a natural starting point.

Comparing lines sharpens the test, because lines differ in their sensitivity to secondary properties.
H$\alpha$ traces the star-formation rate with modest dependence on metallicity~\cite{Kennicutt:2012ea}, whereas the [CII] and CO luminosities at fixed star-formation rate depend strongly on it~\cite{Vallini:2015hac,Lagache:2017vxj,Bolatto:2013ks}, and on the gas content, i.e.,~on the gas depletion time or gas fraction~\cite{Liang:2023sxx,Zanella+18,Ferrara+19,Tacconi:2020wdz}.
The oxygen lines [OII] and [OIII], which SPHEREx observes in the same volume as H$\alpha$, lie in between: they arise in the same HII regions, but their luminosity at fixed star-formation rate depends non-monotonically on the gas-phase metallicity and on the ionization parameter, which correlates with the specific star-formation rate~\cite{Kewley:2002pp,Kewley:2008pp,Kewley:2019tjv}.
The bias ratios of different lines should therefore depart from the halo relations by different amounts, and evolve differently with redshift as the metallicity and ionization conditions of the emitters do; their comparison in the same volume isolates the line-specific dependence from the common halo part, with the sample variance of the shared density field largely canceling, which is one motivation for the multi-line analysis of~\cite{Vanzan:2026pp1}.

\begin{figure}[t]
    \centering
    \includegraphics[width=\columnwidth]{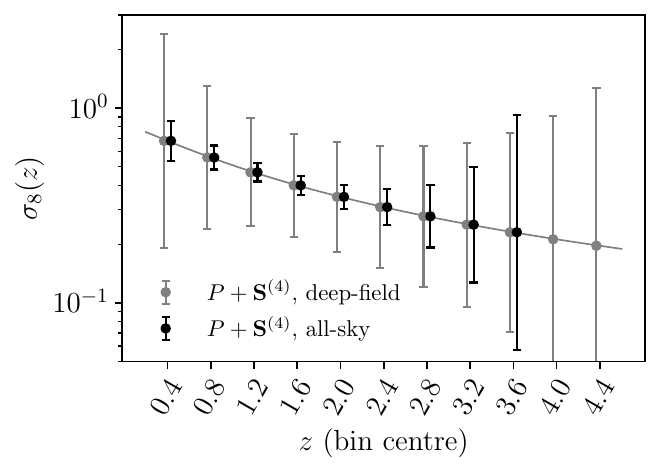}
    \caption{Marginalized~$1\sigma$ constraints on~$\ln\sigma_8(z)$ for the SPHEREx-like survey of the previous sections from a joint power spectrum and four-kernel skew spectra analysis, with~$k_{\rm max}=0.15\,h/{\rm Mpc}$.}
    \label{fig:sigma8}
\end{figure}

Finally, when the cosmological parameters are varied as well, the joint analysis of power spectrum and skew spectra separates the clustering amplitude~$A_1$ from the amplitude of the matter power spectrum in each bin, through the different powers of~$P_m(k)$ carried by the clustering terms of the two statistics; with the shape of~$P_m$ fixed by the cosmological parameters shared across bins, that amplitude is~$\sigma_8(z)$.
This is a real-space measurement of~$\sigma_8(z)$, bin by bin, with no model for the mean intensity, the biases, the shot noise, their redshift evolution.
Figure~\ref{fig:sigma8} shows the~$1\sigma$ constraints in the deep-field configuration.
In the deep-field configuration, $\sigma_8(z)$ is determined only to within a factor of about two in the best bins; the all-sky configuration measures it to~$11\%$ at~$z = 1.2$ and~$1.6$ and to better than~$25\%$ over the range~$0.4 \le z \le 2.4$, while above~$z \simeq 3$ the instrumental noise overtakes the signal and the deep-field configuration becomes the better of the two. This is not yet competitive with the precision of current redshift-space analyses at redshifts probed by Stage-IV spectroscopic galaxy surveys ($z\leq2$).
Nevertheless, this is a proof of principle that intensity maps provide a handle on the growth of structure across a wide redshift range beyond what is accessible by current galaxy surveys without any astrophysical model, and in a form complementary to RSD: combined with a measurement of~$f\sigma_8(z)$, from independent galaxy surveys or from the intensity maps themselves, a real-space~$\sigma_8(z)$ isolates the growth rate~$f(z)$.
In a full analysis including RSD, the real-space power spectrum is not directly accessible; however, $f\sigma_8$ would be obtained by measurements of the anisotropic power spectrum, and~$\sigma_8$ from combinations of redshift-space multipoles (the~$Q_0$ estimator of Ref.~\cite{Ivanov:2021fbu}), hence also giving~$f$ separately.

The Alcock--Paczy\'nski effect~\cite{Alcock:1979mp,Ballinger:1996cd}, i.e.,~the geometric distortion that arises when the fiducial cosmology used to convert angles and redshifts into comoving coordinates differs from the true one, would extend this analysis to the expansion history.
Its isotropic part can be included in a real-space analysis and, in each bin, constrains the volume-averaged distance~$D_V(z) \equiv [c\,z\,\chi^2(z)/H(z)]^{1/3}$ relative to the sound horizon.
Its anisotropic part, which separates~$H(z)$ from~$\chi(z)$, requires the joint modeling of RSD; since it relies on radial modes, its precision would be limited by the low spectral resolution of a SPHEREx-like survey, motivating higher spectral resolution in future intensity-mapping experiments.
Combined with the growth information in~$\sigma_8(z)$, these measurements would test the expansion history across a wide redshift range, extending beyond the reach of galaxy surveys, and, in combination with low-redshift probes, constrain dark energy, including models in which it is not negligible during matter domination.
Both the growth and the geometric constraints would also be somewhat degraded by the loss of modes with small~$k_\parallel$ due to continuum-foreground removal, which we have not modeled here.
We leave these extensions to future work.

\section{Conclusions}
\label{sec:conclusions}

We presented the first application of optimally weighted skew spectra to line-intensity mapping of galaxy-hosted emission, extending the galaxy-clustering formalism to luminosity-weighted biases, LIM shot noise, and instrumental noise in the optimal weights.
For a SPHEREx-like H$\alpha$ survey, four skew spectra recover the full-bispectrum constraints on the luminosity-weighted biases to within~$3\%$ and on the shot-noise parameters to within~$1\%$, and, although their kernels are built to be optimal only for amplitude-like parameters, those on the cosmological parameters to better than~$2\%$. The three clustering kernels of galaxy-clustering analyses are not enough on their own: a fourth kernel, matched to the shot-noise templates, is needed. The reason is the volume, not the physics of line emission. With a fundamental mode of~$k_{\rm f} \simeq 0.01\,h/{\rm Mpc}$, a single redshift bin of the deep fields contains too few~$k$-bins below~$k_{\rm max}$ to recover the shot-noise amplitude from the $k$-dependence of the skew spectra, as the larger volumes of galaxy surveys allow; reducing the volume of a galaxy survey to the same fundamental mode produces the same deficit. The kernel set is therefore not universal, and should be chosen for the survey at hand.

The compression survives when the luminosity moments are left free.
In our forecast, we adopt a different treatment of the bias parameters and the moments of luminosity: we constrain a redshift-independent shift of the biases, which is what a dependence of the emission on the environment of the halo at fixed mass would produce, against a redshift-independent rescaling of the moments, as a calibration error would.
Read this way, the shift is the amplitude of such an environment dependence, and the joint analysis would resolve one of the size found for the star-formation-rate-weighted field of IllustrisTNG.
With this setup, the higher-order statistics measure the mean intensity to~$3\%$, an order of magnitude below the uncertainty of the H$\alpha$ luminosity calibration.
Instead, if the biases are rescaled by a common multiplicative factor, without RSD, the mean intensity is degenerate with the biases and neither the power spectrum nor the skew spectra alone can measure it, and the combined analysis can do so only weakly at~$34\% \, (15\%)$ in the deep-field (all-sky) configuration.

Combined with the tree-level power spectrum, the skew spectra can in principle separate~$\VEV{I}$ from~$b_1$ in each redshift bin and without redshift-space information, through the mixed shot-clustering term of the bispectrum, once the emission model fixes the ratio of the two luminosity-weighted linear biases. For SPHEREx this requires resolving that term bin by bin, which neither the deep fields nor an all-sky survey do. Summed over bins, and under the same assumptions on the redshift evolution, the power spectrum alone already separates the two, to~$1.1\%$ on~$\VEV{I}$ and~$\sigma(b_1) = 0.012$; adding the skew spectra improves both by about~$20\%$, more than a naive combination of the two statistics would give, and brings what the power spectrum does not measure at tree level: the higher-order biases, tighter by factors of~$2$--$3$, the third moment, and the pair-weighted bias of the shot noise.

The results that depend on no assumption about the redshift evolution of the emitters are those of the per-bin analysis, and they carry the astrophysical payoff.
In a single redshift bin, the joint analysis measures the clustering amplitude~$\VEV{I}b_1$ to~$2\%$ in the deep fields and to a few tenths of a percent all-sky, and the ratios~$b_2/b_1$ and~$b_{\cG_2}/b_1$, which are independent of the mean intensity by construction.
The deep fields detect the rise of~$b_2/b_1$ with redshift but do not test the bias relations; an all-sky survey does, measuring~$b_2/b_1$ to~$10\%$ at~$z \le 1.2$, locating the sign change of~$b_2$ in redshift, distinguishing the tidal bias from zero in each bin, and telling apart the published relations between the halo biases.
Since those relations are known from simulations to a few percent, a measured departure would test the emitters rather than the halos: it would mean that the line luminosity at fixed halo mass depends on the assembly history or environment of the halo, an effect simulations find at a level several times the all-sky errors, and one that the two-point statistics cannot see because any change of~$b_1$ is absorbed by the mean intensity.
Lines that depend differently on metallicity and ionization conditions, such as H$\alpha$, [OII], [OIII] and [CII], would depart by different amounts, and their comparison in the same volume isolates the line-specific dependence from the common halo part.
The power-spectrum shot noise, measured per bin, adds the second moment of the luminosity function, the quantity most sensitive to the scatter of the luminosity at fixed mass: its evolution with redshift tests the predicted growth of bursty star formation toward high redshift, while the bias ratios respond only to the part of the scatter that is correlated with the halo environment. The all-sky configuration constrains the shot noise plateau to~$11\%$ around~$z \simeq 2$, quickly degrading at higher redshifts where the instrumental noise takes over.
The same per-bin analysis, with the cosmological parameters varied, separates the clustering amplitude~$\VEV{I}b_1$ from the amplitude of the matter power spectrum through the different powers of~$P_m$ in the two statistics, giving a real-space~$\sigma_8(z)$ bin by bin with no model for the emitters.
It is determined to about ten percent around~$z \sim 1.5$ in the all-sky configuration, and it is a measurement of the growth of structure of a kind no redshift-space analysis provides; combined with~$f\sigma_8(z)$, it would isolate the growth rate.

We work within a set of assumptions: tree level in real space, single line, and a specific halo-model emission prescription; cosmology is fixed in the main results, but Appendix~\ref{app:results_cosmo} shows that the compression holds when it is varied.
Our fiducial emission model contains no dependence on halo properties other than mass and no correlated scatter, so the forecasts of Sec.~\ref{sec:results_perbin} tell us what size of departure from the halo relations could be detected, not what size to expect; predicting it requires a model of how the line luminosity correlates with the environment of its halo, which forward models trained on hydrodynamical simulations can provide.
The power-spectrum shot-noise measurement assumes that the instrumental-noise power is known; in practice, uncertainties on~$P_{\rm N}$ would be marginalized alongside~$\VEV{I^2}$. Interlopers and continuum residuals are not modeled.

RSD measurements provide additional velocity information that can break the~$\VEV{I}$--$b_1$ degeneracy within a single statistic, but at the cost of a substantially enlarged kernel set for the skew spectra.

The power spectrum analysis can be extended further into the mildly nonlinear regime by including 1-loop corrections.
This would strengthen the joint analysis with the higher-order statistics, at the price of additional bias and counterterm parameters.

Cross-correlations between multiple lines observed in the same volume would probe cosmic-variance-cancelling combinations of the luminosity-weighted biases~\cite{Vanzan:2026pp1}, and, as argued above, the comparison of their bias ratios is where the dependence of line emission on the halo environment would show most clearly. \\

\begin{acknowledgments}
We thank the Galileo Galilei Institute for Theoretical Physics, where this work was initiated, for its hospitality.
We are grateful to Guochao Sun for valuable comments on the manuscript, and to Hovav Lazare and Sarah Libanore for helpful discussions.
EV is supported by an Azrieli International Postdoctoral Fellowship.
AMD acknowledges support from the Agence Nationale de la Recherche (ANR) under grant No.~ANR-23-CPJ1-0160-01.
EDK acknowledges support from the U.S.--Israel Binational Science Foundation (NSF-BSF grant 2022743 and BSF grant 2024193) and the Israel National Science Foundation (ISF grant 3135/25), as well as from the joint Israel--China program (ISF--NSFC grant 3156/23).
\end{acknowledgments}

\bibliography{bibliography}

\onecolumngrid

\appendix
\section{Skew-spectra formalism}
\label{app:formalism}

The formalism follows the same logic as in the galaxy-clustering case~\cite{Schmittfull:2014tca,MoradinezhadDizgah:2019xun}, keeping in mind that the amplitude parameters to be constrained are not the bare biases, but luminosity-weighted ones, multiplied by the mean intensity, and that the shot noise is set by the moments of the luminosity rather than by number counts.

Given a bispectrum
\begin{equation}
    \VEV{\delta_{\rm obs}(\vec{k}_1) \delta_{\rm obs}(\vec{k}_2) \delta_{\rm obs}(\vec{k}_3)} = (2\pi)^3 \delta^{(3)}_D(\vec{k}_1+\vec{k}_2+\vec{k}_3) B_{\rm obs}(\vec{k}_1,\vec{k}_2,\vec{k}_3) \, ,
\end{equation}
the skew spectra are defined as the filtered bispectrum, with a suitable kernel,
\begin{equation}
    \VEV{ \cS_n(\bk)\, \delta_{\rm obs}(\bk') } = (2\pi)^3 \delta_D^{(3)}(\vec{k}+\vec{k}') \cP_{\cS_n,\delta_{\rm obs}}(k) = \int \frac{d^3\vec{q}}{(2\pi)^3} D_2^{(n)}(\vec{q},\vec{k}-\vec{q}) \VEV{ \delta_{\rm obs}(\vec{q}) \delta_{\rm obs}(\vec{k}-\vec{q}) \delta_{\rm obs}(\vec{k}') } \, ,
\end{equation}
where the kernels~$D_2^{(n)}(\vec{q},\vec{k}-\vec{q})$ are taken to be symmetric in their arguments.

The bias expansion of Eq.~\eqref{eq:bias_expansion}, in terms of the Galileon operator~$\mathcal{G}_2$ of Eq.~\eqref{eq:G2}, follows the convention of~\cite{Schmittfull:2020hoi}, as opposed to that of~\cite{Schmittfull:2014tca,MoradinezhadDizgah:2019xun}, which expands the overdensity as
\begin{equation}
    \delta(\vec{x}) = b_1 \delta_m(\vec{x}) + b_{\delta^2} \delta_m^2(\vec{x}) + b_{K^2} K^2(\vec{x}) \, ,
\end{equation}
where~$K^2=K_{ij}K^{ij}$ is the squared trace-free tidal operator
\begin{equation}
    K_{ij}(\vec{x}) = \left( \frac{\partial_i\partial_j}{\partial^2}-\frac{1}{3}\delta_{ij} \right) \delta_m(\vec{x}) \, .
\end{equation}
This latter convention has the advantage that the tidal operator is directly linked to the Legendre polynomial~$\mathcal{L}_2$.
Here we choose the expansion in terms of the Galileon operator instead, whose large-scale mode vanishes.
The two bases are related by
\begin{equation}
    \mathcal{G}_2 = K^2 - \frac{2}{3} \delta^2 \, , \qquad\qquad b_{K^2} = b_{\mathcal{G}_2} \, , \qquad b_{\delta^2} = \frac{b_2}{2} - \frac{2 b_{\mathcal{G}_2}}{3} \, .
\end{equation}

The tree-level single-tracer bispectrum is
\begin{equation}
\label{eq:bis}
    B(\vec{k}_1,\vec{k}_2,\vec{k}_3) = 2 b_1^2 P_m(k_1) P_m(k_2) \left[ b_1 F_2(\vec{k}_1,\vec{k}_2) + \frac{b_2}{2} + b_{\mathcal{G}_2} \left( \mu_{12}^2-1 \right) \right] + 2~{\rm perm.} \, ,
\end{equation}
where~$P_m$ is the linear matter power spectrum,~$\mu_{ij} = \vhat{k}_i \cdot \vhat{k}_j$, and~$F_2$ is the matter mode-coupling kernel
\begin{equation}
    F_2(\vec{k}_1,\vec{k}_2) = \frac{5}{7} + \frac{1}{2}\left(\frac{k_1}{k_2}+\frac{k_2}{k_1}\right) \mu_{12} + \frac{2}{7} \mu_{12}^2 \, .
\end{equation}
The filtering kernels are designed to pick out contributions from the three quadratic fields~$\delta_m^2$,~$\Psi_m^i \partial_i \delta_m$, and~$\left(K_{ij}\right)_m^2$ respectively, corresponding to a decomposition of the tree-level matter bispectrum into Legendre polynomials~\cite{Baldauf:2012hs}.
Namely,~$\mathbf{D}_2(\bk_1,\bk_2) = \left\{ \Legendre{0}(\mu_{12}), \; F_2^1(k_1,k_2)\, \Legendre{1}(\mu_{12}), \; \Legendre{2}(\mu_{12}) \right\}$, with
\begin{equation}
    F_2^1(k_1,k_2) = \frac{1}{2} \left( \frac{k_1}{k_2} + \frac{k_2}{k_1} \right) \, .
\end{equation}
In the main text, the kernels are referred to as~$\{ {\rm dens}, {\rm disp}, {\rm tide} \}$, after the quadratic fields~$\delta_m^2$,~$\Psi_m^i\partial_i\delta_m$, and~$\left(K_{ij}\right)_m^2$ whose contributions they are designed to extract; in Sec.~\ref{sec:shot_kernels} an additional kernel is introduced, matched to the shot-noise template.
We note that, in the Galileon operator basis, the constant piece of the Galileon operator will contribute to the Legendre polynomial~$\mathcal{L}_0$ as well, since the bispectrum decomposes as
\begin{equation}
\begin{split}
    B(\vec{k}_1,\vec{k}_2,\vec{k}_3) =& \left[ \left( \frac{17}{21} b_1 + \left(\frac{b_2}{2}-\frac{2}{3}b_{\mathcal{G}_2}\right) \right) \mathcal{L}_0(\mu_{12}) + b_1 \frac{1}{2} \left( \frac{k_1}{k_2}+\frac{k_2}{k_1} \right) \mathcal{L}_1(\mu_{12}) + \left( \frac{4}{21} b_1 + \frac{2}{3} b_{\mathcal{G}_2} \right) \mathcal{L}_2(\mu_{12}) \right] 2 b_1^2 P_m(k_1) P_m(k_2) \\
    & + 2~{\rm perm.}
\end{split}
\end{equation}

The kernels must inherit the inverse-variance weighting of the optimal bispectrum estimator for the skew spectra to preserve its optimality for amplitude-like parameters, Eq.~\eqref{eq:weighted_kernel}; in what follows~$\Tilde{D}_2^{(n)}$ denotes the weighted kernel, and the smoothing window~$W$ of Eq.~\eqref{eq:tanh_window} multiplies each internal leg in the signal, the estimator and the covariance alike.

Following~\cite{MoradinezhadDizgah:2019xun}, we define
\begin{align}
    I^{(n)}_Y(k) &= \int \frac{d^3\vec{q}}{(2\pi)^3} \Tilde{D}_2^{(n)}(\vec{q},\vec{k}-\vec{q}) Y(\vec{q},\vec{k}-\vec{q}) P_m(q) P_m(|\vec{k}-\vec{q}|) \, , \label{eq:I_DE} \\
    I^{(n),\rm bare}_Y(k) &= \int \frac{d^3\vec{q}}{(2\pi)^3} \Tilde{D}_2^{(n)}(\vec{q},\vec{k}-\vec{q}) Y(\vec{q},-\vec{k}) P_m(q) P_m(k) \label{eq:I_DE_bare} \, ,
\end{align}
where the latter is not symmetric under~$\Tilde{D}_2^{(n)} \longleftrightarrow Y$.
For each kernel, the skew spectra read
\begin{equation}
\begin{split}
    \cP^{\rm clust}_{\cS_n, \delta_{\rm obs}}(k) = & 2 b_1^3 \left[ I^{(n)}_{F_2}(k) + 2 I^{(n),{\rm bare}}_{F_2}(k) \right] + 2 b_1^2 \left(\frac{b_2}{2}-\frac{2b_{\mathcal{G}_2}}{3}\right) \left[ I^{(n)}_{\Legendre{0}}(k) + 2 I^{(n),{\rm bare}}_{\Legendre{0}}(k) \right] \\
    & + \frac{4}{3} b_1^2 b_{\mathcal{G}_2} \left[ I^{(n)}_{\Legendre{2}}(k) + 2 I^{(n),{\rm bare}}_{\Legendre{2}}(k) \right] \, .
\end{split}
\end{equation}
For the intensity field, the clustering contribution above is multiplied by~$\VEV{I}^3$ and supplemented by the shot-noise terms of Eq.~\eqref{eq:skew_shot}, and the shot-sector kernel of Sec.~\ref{sec:shot_kernels} enters Eqs.~\eqref{eq:I_DE}--\eqref{eq:J_nm} on the same footing as the three clustering kernels.

\begin{figure}
    \centering
    \includegraphics[height=3.3cm]{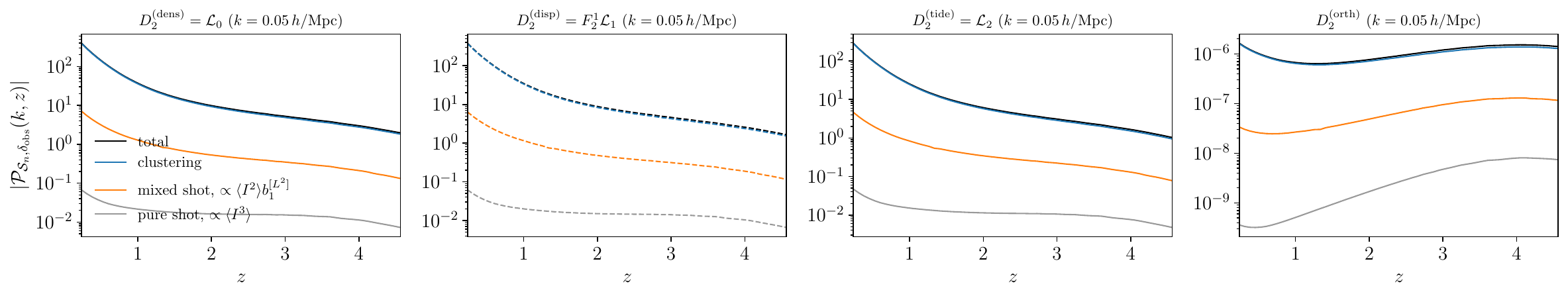}
    \includegraphics[height=3.3cm]{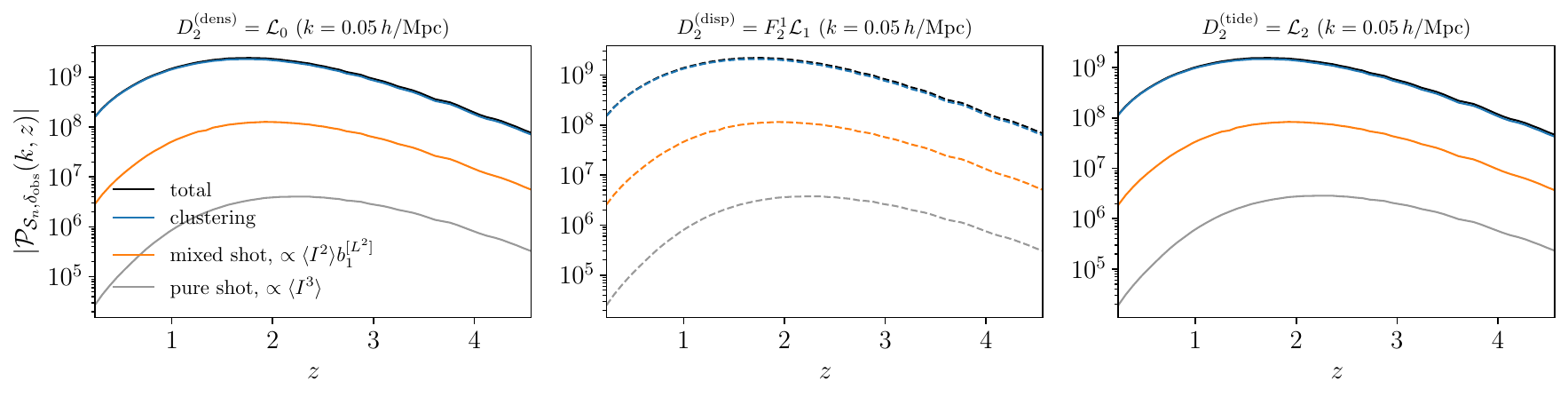}
    \caption{Weighted (\emph{upper row}) and unweighted (\emph{lower row}) skew spectra as a function of redshift, at fixed~$k=0.05\,h/{\rm Mpc}$, for the SPHEREx-like deep-field survey of the main text. Dashed lines indicate negative values. In the unweighted case, only the clustering kernels are shown, since the definition of the shot noise ones relies on the weights.}
    \label{fig:contributions_skew_z}
\end{figure}

Figure~\ref{fig:contributions_skew_z} shows the skew spectra as a function of redshift, at fixed~$k=0.05\, h/{\rm Mpc}$.
As mentioned in the main text, the amplitude of the signal does not closely track the star formation rate, contrary to the power spectrum (Fig.~\ref{fig:contributions_pk}): because of the inverse-variance weights, the amplitude in the clustering-dominated regime scales as~$[\VEV{I}b_1]^{-1}$.
On the other hand, the unweighted skew spectra do not decrease monotonically with redshift, and their peak occurs around~$z \simeq 2$, at slightly higher redshift than the power spectrum's.

We briefly recall the expressions for the estimators and covariance for the skew spectra, and refer the reader to~\cite{Schmittfull:2014tca,MoradinezhadDizgah:2019xun} for more details.

The estimators for the skew spectra are
\begin{equation}
    \Hat{\cP}_{\cS_n, \delta_{\rm obs}}(k) = \frac{1}{V_i V_k} \int_k d^3\vec{k}' \left[ \int^{\prime} \frac{d^3\vec{q}}{(2\pi)^3} \Tilde{D}_2^{(n)}(\vec{q},\vec{k}'-\vec{q})\, \delta_{\rm obs}(\vec{q})\, \delta_{\rm obs}(\vec{k}'-\vec{q}) \right] \delta_{\rm obs}(-\vec{k}') \, .
\end{equation}
The outer integral runs over the shell of width~$\Delta k$ around~$k$, the primed integral over internal modes with~$q, |\vec{k}'-\vec{q}| < k_{\rm max}$, a cut implemented smoothly through the window~$W$ below, and
\begin{equation}
    V_k = 4\pi k^2 \Delta k \, , \qquad V_i = \frac{4\pi f_{\rm sky}}{3} \left[ \chi^3(z_{\rm max}) - \chi^3(z_{\rm min}) \right] \, ,
\end{equation}
where~$\chi$ is the comoving distance.
The small-scale cutoff of the skew-spectrum estimators is implemented as a smoothed step in the internal momenta: each internal leg carries a window
\begin{equation}
\label{eq:tanh_window}
    W(q) = \frac{1}{2}\left[ 1 - \tanh\!\left( \beta \ln \frac{q}{k_{\rm max}} \right) \right] ,
    \qquad \beta = 50 \, ,
\end{equation}
the value adopted in Ref.~\cite{Gao:2026zaz}; as~$\beta \to \infty$ the window becomes a sharp cutoff.
Internal wavenumbers are likewise restricted from below to the fundamental mode of each redshift shell, consistently with the mode content of the estimators.

The Gaussian cross-covariance terms between skew spectra, with~$P_{\rm obs}$ the power spectrum of the observed field including shot and instrumental noise, read
\begin{equation}
\label{eq:skew_cov}
\begin{split}
    {\rm Cov}\left[ \Hat{\cP}_{\cS_n, \delta_{\rm obs}}(k), \Hat{\cP}_{\cS_m, \delta_{\rm obs}}(k') \right] =& \frac{(2\pi)^3}{V_i V_k V_{k'}} \int_k d^3\vec{p}\, 2 P_{\rm obs}(p) I^{(n,m)}(p) \int_{k'} d^3\vec{p}' \delta^{(3)}_D(\vec{p}+\vec{p}') \\
    & + \frac{4}{V_i V_k V_{k'}} \int_k d^3\vec{p} \int_{k'} d^3\vec{p}'  P_{\rm obs}(p) P_{\rm obs}(p') P_{\rm obs}(|\vec{p}-\vec{p}'|)  \Tilde{D}_2^{(n)}(\vec{p}',\vec{p}-\vec{p}') \Tilde{D}_2^{(m)}(\vec{p},\vec{p}'-\vec{p}) \\
    =& \frac{2 (2\pi)^3}{V_i V_k} P_{\rm obs}(k) I^{(n,m)}(k) \delta^K_{kk'} + \frac{4}{V_i} P_{\rm obs}(k) P_{\rm obs}(k') J^{(n,m)}(k,k') \, ,
\end{split}
\end{equation}
where\footnote{This agrees with Eq. (B2) of~\cite{Schmittfull:2014tca}; note that in Eq. (3.44) of~\cite{MoradinezhadDizgah:2019xun} the roles of~$D_2^{(n)}$ and~$D_2^{(m)}$ in~$J^{(n,m)}$ (there denoted as~$J_{\Tilde{D}\Tilde{E}}$) are interchanged.
The difference affects only the~$n \neq m$ cross-covariance blocks and is invisible in any single-kernel comparison.}
\begin{align}
    I^{(n,m)}(k) &= \int \frac{d^3\vec{q}}{(2\pi)^3} \Tilde{D}_2^{(n)}(\vec{q},\vec{k}-\vec{q}) \Tilde{D}_2^{(m)}(\vec{q},\vec{k}-\vec{q}) P_{\rm obs}(q) P_{\rm obs}(|\vec{k}-\vec{q}|) \label{eq:I_nm} \, , \\
    J^{(n,m)}(k,k') &= \frac{1}{2} \int_{-1}^{+1} d\mu\, \Tilde{D}_2^{(n)}(\vec{k}',\vec{k}-\vec{k}') \Tilde{D}_2^{(m)}(\vec{k},\vec{k}'-\vec{k}) P_{\rm obs}(|\vec{k}-\vec{k}'|) \label{eq:J_nm} \, .
\end{align}

\section{Bispectrum shot noise}
\label{app:shot}

We briefly sketch the derivation of the bispectrum shot noise for LIM, following the notation of~\cite{Chan:2016ehg}.
The calculation is analogous to the galaxy clustering case~\cite{Chan:2016ehg}.
The volume is divided into infinitesimal cells, each containing~$N_i = \{0,1\}$ emitters~\cite{Sato-Polito:2022fkd}, and the intensity of cell~$i$ is proportional to the luminosity of the emitter it contains,~$I_i \propto N_i L_i$.
The emitters are identified with halos, as in Eq.~\eqref{eq:L-weighted_biases}: their luminosities~$L_i = L(M_i)$ follow the halo mass function and each carries the linear bias~$b_{1,h}(M_i)$ of its halo, so that the averages over emitters below produce the moments~$\VEV{I^n}$ of Eq.~\eqref{eq:moments} and the luminosity-weighted biases.
The emitters are Poisson sampled; the corrections~$\alpha_{\rm P}$,~$\alpha_1$,~$\alpha_2$ of Secs.~\ref{sec:lim} and~\ref{sec:shot} parametrize departures from this assumption and are added to the final expressions.
The fluctuation is defined as~$I(\vec{x})-\VEV{I}$.

Each emitter enters with its luminosity, so that when two of the fields in a correlator are evaluated at the same emitter the pair carries~$L^2$, and the bias associated with that pair is~$L^2$-weighted.
Explicitly,
\begin{align}
    & \VEV{\sum_{i \neq j} I_i I_j \, \delta_D^{(3)}(\vec{x}_1-\vec{x}_i) \, \delta_D^{(3)}(\vec{x}_2-\vec{x}_j)} = \VEV{I}^2 \left[ 1 + \left(b_1^{[L]}\right)^2 \xi_m(\vec{x}_1-\vec{x}_2) \right] \, , \\
    & \VEV{\sum_{i \neq j} I_i I_j^2 \, \delta_D^{(3)}(\vec{x}_1-\vec{x}_i) \, \delta_D^{(3)}(\vec{x}_2-\vec{x}_j)} = \VEV{I} \VEV{I^2} \left[ 1 + b_1^{[L]} b_1^{[L^2]} \xi_m(\vec{x}_1-\vec{x}_2) \right] \, ,
\end{align}
where~$\xi_m$ is the matter two-point function,~$b_1^{[L]} = b_1$ is the luminosity-weighted bias of Eq.~\eqref{eq:L-weighted_biases}, and
\begin{equation}
\label{eq:b1L2_app}
    b_1^{[L^2]}(z) = \frac{\int dM \, \frac{dn}{dM} \, b_{1,h}(M,z) \, L^2(M,z)}{\int dM \, \frac{dn}{dM} \, L^2(M,z)}
\end{equation}
is weighted by two powers of luminosity.
The second correlator above is the cross-correlation of the intensity field with the composite~$L^2$-weighted field, and it is the origin of the mixed shot-clustering term of the bispectrum derived below.
The Dirac deltas carry the inverse volume, so that~$\VEV{I^2}$ and~$\VEV{I^3}$ below are the dimensionful moments of Eq.~\eqref{eq:moments}, with the abuse of notation noted there.
A scatter of the luminosity at fixed halo mass that is independent of mass multiplies~$\VEV{I^2}$ and~$\VEV{I^3}$ by the factors~$p_{n,\sigma}$ of Eq.~\eqref{eq:moments} and leaves Eq.~\eqref{eq:b1L2_app} unchanged, since the factor cancels between numerator and denominator; a mass-dependent scatter would instead reweight the integrand of Eq.~\eqref{eq:b1L2_app}.

The single-tracer two-point correlator is
\begin{equation}
\begin{split}
    \VEV{I(\vec{x}_1)I(\vec{x}_2)} =& \VEV{ \sum_i I_i^2 \delta_D^{(3)}(\vec{x}_1-\vec{x}_i) \delta_D^{(3)}(\vec{x}_2-\vec{x}_i)} + \VEV{\sum_{i \neq j} I_i I_j \delta_D^{(3)}(\vec{x}_1-\vec{x}_i) \delta_D^{(3)}(\vec{x}_2-\vec{x}_j)} \\
    =& \VEV{I^2} \delta_D^{(3)}(\vec{x}_1-\vec{x}_2) + \VEV{I}^2 \left[ 1 + \left(b_1^{[L]}\right)^2 \xi_m(\vec{x}_1-\vec{x}_2) \right] \, .
\end{split}
\end{equation}
The observed connected two-point function contains a shot noise component and a clustering component,
\begin{equation}
    \xi_{\rm obs}(\vec{x}_1,\vec{x}_2) = \VEV{I(\vec{x}_1)I(\vec{x}_2)}- \VEV{I}^2 = \VEV{I^2} \delta_D^{(3)}(\vec{x}_1-\vec{x}_2) +\VEV{I}^2 \left(b_1^{[L]}\right)^2 \xi_m(\vec{x}_1-\vec{x}_2) \, .
\end{equation}
Upon Fourier transforming, and including the instrumental noise,
\begin{equation}
    P_{\rm obs}(k) = \VEV{I}^2 b_1^2 P_m(k) + \VEV{I^2} + P_{\rm N} \, .
\end{equation}
The clustering component depends on the first moment of luminosity, whereas the shot noise component yields information on the second moment of luminosity.

Similarly, the observed connected three-point correlator is
\begin{equation}
\begin{split}
    \zeta_{\rm obs}(\vec{x}_1,\vec{x}_2,\vec{x}_3) =& \VEV{\left(I(\vec{x}_1)-\VEV{I}\right) \left(I(\vec{x}_2)-\VEV{I}\right) \left(I(\vec{x}_3)-\VEV{I}\right)} \\
    =& \VEV{I(\vec{x}_1) I(\vec{x}_2) I(\vec{x}_3)} - \left[ \VEV{I} \VEV{I(\vec{x}_1) I(\vec{x}_2)} + 2~{\rm perm.} \right] \\
    & + \VEV{I}^2 \VEV{I(\vec{x}_1)} + 2~{\rm perm.} - \VEV{I}^3 \, .
\end{split}
\end{equation}
The first piece receives the following contributions, according to whether some or all of the emitters are located within the same cell,
\begin{equation}
\begin{split}
    & \VEV{I(\vec{x}_1) I(\vec{x}_2) I(\vec{x}_3)} \\
    =& \VEV{ \sum_i I_i^3 \delta_D^{(3)}(\vec{x}_1-\vec{x}_i) \delta_D^{(3)}(\vec{x}_2-\vec{x}_i) \delta_D^{(3)}(\vec{x}_3-\vec{x}_i) } \\
    & + \left[ \VEV{ \sum_{i \neq j} I_i I_j^2 \delta_D^{(3)}(\vec{x}_1-\vec{x}_i) \delta_D^{(3)}(\vec{x}_2-\vec{x}_j) \delta_D^{(3)}(\vec{x}_3-\vec{x}_j) } + 2~{\rm perm.} \right] \\
    & + \VEV{\sum_{i \neq j \neq k } I_i I_j I_k \delta_D^{(3)}(\vec{x}_1-\vec{x}_i) \delta_D^{(3)}(\vec{x}_2-\vec{x}_j) \delta_D^{(3)}(\vec{x}_3-\vec{x}_k)} \\
    =& \delta_D^{(3)}(\vec{x}_1-\vec{x}_2) \delta_D^{(3)}(\vec{x}_2-\vec{x}_3) \VEV{I^3} \\
    & + \left\{ \VEV{I} \VEV{I^2} \left[ 1 + b_1^{[L]} b_1^{[L^2]} \xi_m(\vec{x}_1-\vec{x}_2) \right] \delta_D^{(3)}(\vec{x}_2-\vec{x}_3) + 2~{\rm perm.} \right\} \\
    &  + \VEV{I}^3 \left[ 1 + \left(b_1^{[L]}\right)^2 \xi_{m,12} + \left(b_1^{[L]}\right)^2 \xi_{m,13} + \left(b_1^{[L]}\right)^2 \xi_{m,23} +\zeta_{123} \right] \, ,
\end{split}
\end{equation}
where~$\zeta_{123}$ is the connected three-point function of the biased field, whose Fourier transform is the tree-level bispectrum of Eq.~\eqref{eq:bis}.
In the $i \neq j = k$ terms the coincident pair and the spectator enter with different luminosity weights (two powers for the pair, one for the spectator), which is why the pair carries~$b_1^{[L^2]}$ and the spectator~$b_1^{[L]}$.

Substituting above,
\begin{equation}
\begin{split}
    \zeta_{\rm obs}(\vec{x}_1,\vec{x}_2,\vec{x}_3) =& \VEV{I^3} \delta_D^{(3)}(\vec{x}_1-\vec{x}_2) \delta_D^{(3)}(\vec{x}_2-\vec{x}_3) \\
    & +\VEV{I} \VEV{I^2} b_1^{[L]} b_1^{[L^2]} \xi_{m,12} \delta_D^{(3)}(\vec{x}_2-\vec{x}_3) + 2~{\rm perm.} \\
    & +\VEV{I}^3 \zeta_{123} \, .
\end{split}
\end{equation}
After Fourier transforming,
\begin{equation}
\begin{split}
    B_{\rm obs}(\vec{k}_1,\vec{k}_2,\vec{k}_3) &= \VEV{I}^3 B_I(\vec{k}_1,\vec{k}_2,\vec{k}_3) + \VEV{I^2} \VEV{I} b_1^{[L]} b_1^{[L^2]} \left[ P_m(k_1) + P_m(k_2) + P_m(k_3) \right] + \VEV{I^3} \, .
\end{split}
\end{equation}
The purely shot noise contribution contains the third moment of luminosity~$\VEV{I^3}$.
With the Poisson corrections of Sec.~\ref{sec:shot}, this is Eq.~\eqref{eq:Bobs_tot} of the main text.
The result agrees with Eq.~(A.6) in~\cite{Sato-Polito:2022fkd} if one makes the additional assumption~$b_1^{[L^2]} = b_1^{[L]}$.

\section{Theoretical halo biases}
\label{app:biases}

The halo mass function and the halo bias parameters~$b_{\alpha,h}(M,z)$ are computed with the Sheth--Tormen prediction~\cite{Sheth:1999mn,Scoccimarro:2000gm}, with the tidal bias fixed by the coevolution of halos and dark matter~\cite{Baldauf:2012hs,Chan:2012jj}.
The mass function is
\begin{equation}
    \frac{dn}{dM} = A\sqrt{\frac{2q}{\pi}}\,\frac{\bar\rho_m}{M^2}\,\left[1+\left(q\nu^2\right)^{-p}\right]\nu\, e^{-q\nu^2/2}\,\left|\frac{d\ln\sigma}{d\ln M}\right| , \qquad \nu \equiv \frac{\delta_c}{\sigma(M,z)} \, ,
\end{equation}
with~$\delta_c = 1.686$,~$\sigma(M,z)$ the rms of the linear density field smoothed on the Lagrangian scale of~$M$ in the fiducial cosmology of Sec.~\ref{sec:fisher},~$A = 0.3222$,~$q = 0.707$ and~$p = 0.3$.
The Eulerian biases of the~$\delta^2/2$ convention of Eq.~\eqref{eq:bias_expansion} are
\begin{equation}
    b_{1,h} = 1+\epsilon_1+E_1 \, , \qquad b_{2,h} = 2\left(1-\frac{17}{21}\right)\left(\epsilon_1+E_1\right)+\epsilon_2+E_2 \, , \qquad b_{\mathcal{G}_2,h} = -\frac{2}{7}(b_{1,h}-1) \, ,
\end{equation}
where
\begin{equation}
\begin{gathered}
    \epsilon_1 = \frac{q\nu^2-1}{\delta_c} \, , \qquad \epsilon_2 = \frac{q\nu^2}{\delta_c^2}(q\nu^2-3) \, , \\
    E_1 = \frac{2p/\delta_c}{1+\left(q\nu^2\right)^p} \, , \qquad E_2 = E_1 \left(\frac{1+2p}{\delta_c}+2\epsilon_1\right) \, ,
\end{gathered}
\end{equation}
Here~$\epsilon_1+E_1$ and~$\epsilon_2+E_2$ are the Lagrangian biases, and the factor~$2(1-17/21) = 8/21$ is the second-order term of the spherical-collapse mapping to Eulerian space that goes together with the coevolution value of~$b_{\mathcal{G}_2,h}$.
The alternative prescriptions explored in Sec.~\ref{sec:results_perbin} (Press--Schechter mass function and biases, the~$b_{2,h}(b_{1,h})$ fit of Ref.~\cite{Lazeyras:2015lgp}, the excursion set prediction, and a Lagrangian tidal bias) are given in Eqs.~\eqref{eq:lazeyras}--\eqref{eq:modi}.

\section{Survey specification details}
\label{app:survey_specs}

Table~\ref{tab:specs} reports the per-bin properties resulting from the survey specifications of Sec.~\ref{sec:fisher}: the volume and the corresponding fundamental Fourier mode, the mean intensity, the luminosity-weighted biases, the shot-noise plateau, the instrumental noise.

\begin{table*}[h!]
    \centering
    \begin{tabular}{c c c c c c c c c}
        \hline\hline
        $z_i$ & $V_i \, [({\rm Gpc}/h)^3]$ & $k_{\rm min} \, [h/{\rm Mpc}]$ & $\VEV{I} \, [{\rm Jy/sr}]$ & $b_1$ & $b_2$ & $b_{\mathcal{G}_2}$ & $P_{\rm shot} \, [({\rm Jy/sr})^2 ({\rm Mpc})^3]$ & $P_{\rm N} \, [({\rm Jy/sr})^2 ({\rm Mpc})^3]$ \\
        \hline
        $0.4$ & $7.0 \times 10^{-2}$ & $0.015$  & $39.8$ & $0.896$ & $-0.385$ & $0.0298$  & $2.77 \times 10^5$ & $1.52 \times 10^5$ \\
        $0.8$ & $1.7 \times 10^{-1}$ & $0.011$  & $61.2$ & $1.04$  & $-0.424$ & $-0.0127$ & $8.57 \times 10^5$ & $4.96 \times 10^5$ \\
        $1.2$ & $2.5 \times 10^{-1}$ & $0.010$  & $71.9$ & $1.26$  & $-0.394$ & $-0.0734$ & $1.69 \times 10^6$ & $8.72 \times 10^5$ \\
        $1.6$ & $2.9 \times 10^{-1}$ & $0.0094$ & $71.8$ & $1.53$  & $-0.203$ & $-0.152$  & $2.36 \times 10^6$ & $1.21 \times 10^6$ \\
        $2.0$ & $3.2 \times 10^{-1}$ & $0.0092$ & $64.4$ & $1.85$  & $0.228$  & $-0.243$  & $2.55 \times 10^6$ & $1.49 \times 10^6$ \\
        $2.4$ & $3.2 \times 10^{-1}$ & $0.0092$ & $54.8$ & $2.18$  & $0.915$  & $-0.338$  & $2.33 \times 10^6$ & $1.72 \times 10^6$ \\
        $2.8$ & $3.2 \times 10^{-1}$ & $0.0092$ & $45.2$ & $2.51$  & $1.82$   & $-0.431$  & $1.89 \times 10^6$ & $1.91 \times 10^6$ \\
        $3.2$ & $3.1 \times 10^{-1}$ & $0.0093$ & $36.2$ & $2.81$  & $2.87$   & $-0.518$  & $1.36 \times 10^6$ & $2.06 \times 10^6$ \\
        $3.6$ & $3.0 \times 10^{-1}$ & $0.0094$ & $29.3$ & $3.09$  & $4.00$   & $-0.597$  & $9.47 \times 10^5$ & $2.18 \times 10^6$ \\
        $4.0$ & $2.9 \times 10^{-1}$ & $0.0095$ & $23.4$ & $3.33$  & $5.13$   & $-0.667$  & $6.16 \times 10^5$ & $2.28 \times 10^6$ \\
        $4.4$ & $2.8 \times 10^{-1}$ & $0.0096$ & $19.2$ & $3.55$  & $6.23$   & $-0.729$  & $3.98 \times 10^5$ & $2.36 \times 10^6$ \\
        \hline\hline
    \end{tabular}
    \caption{Specifications and fiducial values for the SPHEREx-like H$\alpha$ survey in each of the 11 redshift bins: bin center, comoving volume, fundamental mode, mean intensity, luminosity-weighted biases, Poissonian shot-noise power, and instrumental-noise power.}
    \label{tab:specs}
\end{table*}

\section{Varying cosmological parameters}
\label{app:results_cosmo}

\begin{figure}[t]
    \centering
    \includegraphics[width=0.8\textwidth]{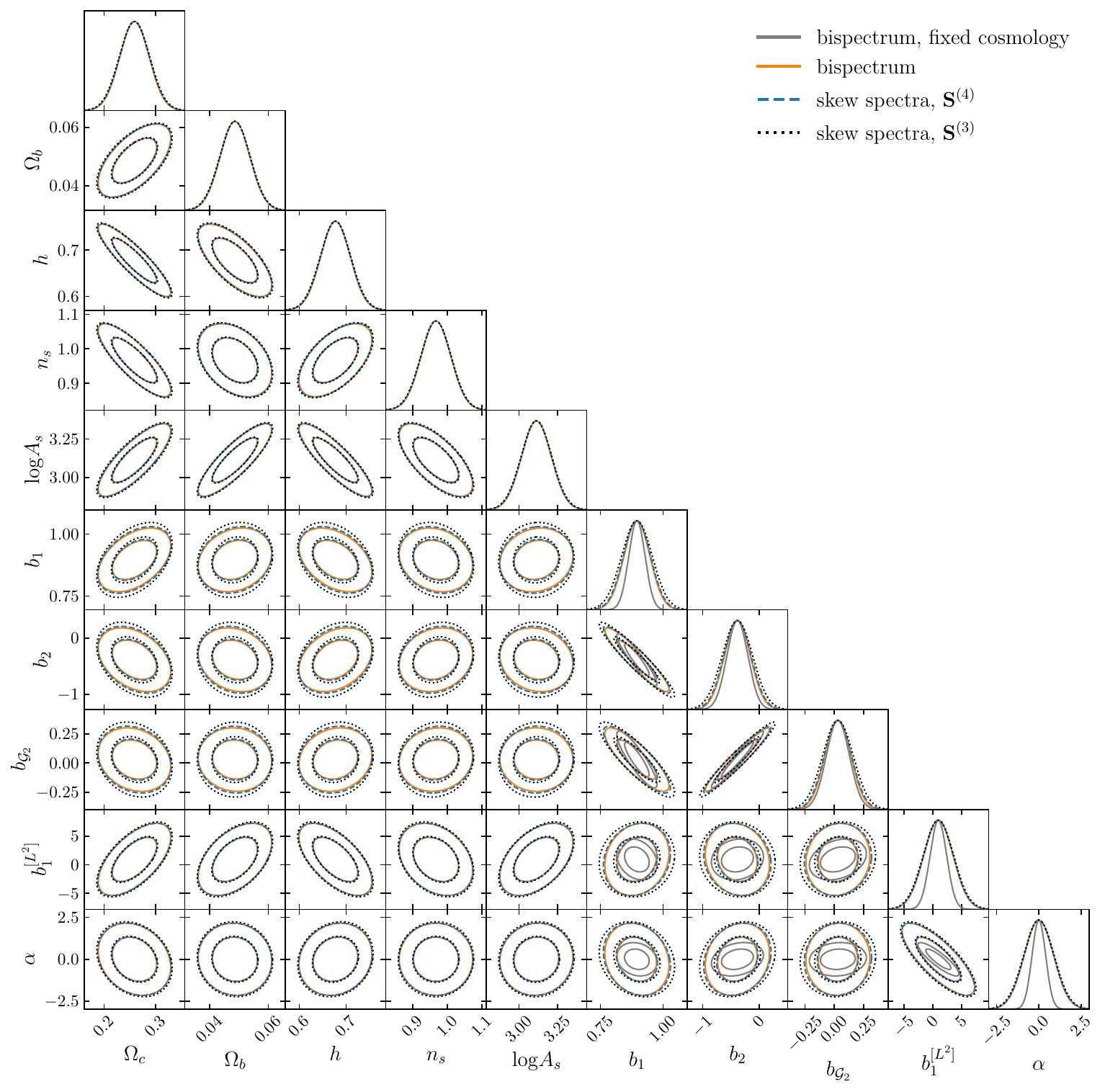}
    \caption{Marginalized constraints on~$\{\Omega_c, \Omega_b, h, n_s, \ln(10^{10}A_s), b_1, b_2, b_{\mathcal{G}_2}, b_1^{[L^2]}, \alpha\}$ for the SPHEREx-like H$\alpha$ survey,~$k_{\rm max}=0.15\,h/{\rm Mpc}$, from the full bispectrum (solid) versus the standard three-kernel skew spectra set~$\mathbf{S}^{(3)}$ (dotted) and the extended four-kernel skew spectra set~$\mathbf{S}^{(4)}$ (dashed); the gray contours show the bispectrum with cosmology fixed. The agreement between bispectrum and skew-spectra constraints in the presence of non-amplitude-like parameters is a non-trivial result that was not guaranteed by the construction of the skew-spectra estimators.}
    \label{fig:cosmo_constraints}
\end{figure}

We show that the skew spectra recover the full tree-level bispectrum constraints on~$\left\{ \Omega_c, \Omega_b, h, n_s, \ln(10^{10}A_s) \right\}$ to better than~$2\%$, with either kernel set.
This goes beyond the skew spectra's original design: the weighted kernels were built to be optimal for amplitude parameters, and there was no a priori guarantee for shape parameters, whose responses have nontrivial~$k$-dependence.
Empirically, we find that the~$k$-dependence of the skew spectra carries essentially all of the information.
The constraints themselves are weak, as expected for the real-space tree-level bispectrum of the deep fields alone (Fig.~\ref{fig:cosmo_constraints}); the point of this appendix is the compression, not the constraining power.

Figure~\ref{fig:cosmo_constraints} shows marginalized constraints on cosmological parameters, biases, and shot-noise parameter, as described in Sec.~\ref{sec:parameters}.
Both~$\mathbf{S}^{(3)}$ and~$\mathbf{S}^{(4)}$ recover all of the bispectrum information on the cosmological parameters equally well, while for the biases the fourth kernel is necessary, as in Sec.~\ref{sec:results_grav}.
With the amplitude~$\ln(10^{10}A_s)$ free, the constraints on~$\alpha$ and~$b_1^{[L^2]}$ degrade by a factor of about two with respect to the fixed-cosmology analysis, and those on~$b_1$ by~$60\%$ (Table~\ref{tab:cosmo_constraints}): the two shot-noise parameters enter only through the mixed and constant shot-noise terms, whose amplitudes relative to the clustering terms are what fixes~$A_s$ once the biases are free, so the three are now shared between~$A_s$ and the shot sector.

We have verified that freeing~$b_1^{[L^2]}$, rather than tying it to~$b_1$ through the fixed ratio~$r_{\rm fid}$, leaves the biases and the cosmological shape parameter~$n_s$ essentially unaffected, but degrades the amplitude-degenerate parameters appreciably: $\sigma(\Omega_c)$ and~$\sigma(h)$ grow by~$\simeq 25\%$, $\sigma(\Omega_b)$ and~$\sigma(\ln(10^{10}A_s))$ by~$13$--$15\%$, identically for the bispectrum and both skew spectra.
The mixed shot-clustering term carries a power of~$b_1$ different from the clustering sector, and thus contributes to separating~$b_1$ from the amplitude~$\ln(10^{10}A_s)$; marginalizing over its bias weighting removes that handle, with repercussions on cosmological amplitude parameters.
Consistently with the baseline of Sec.~\ref{sec:results_grav}, all constraints quoted in this appendix have~$b_1^{[L^2]}$ free.

We compare our results to a \emph{Planck} prior on the cosmological parameters widened by a factor~$10$, so that the prior and the intensity maps carry comparable information (Sec.~\ref{sec:parameters}).
In the combination the maps reduce the width of the widened prior by~$63$--$72\%$ on~$\Omega_c$, $\Omega_b$, $h$ and~$\ln(10^{10}A_s)$, and by~$38\%$ on~$n_s$, with the same numbers for the bispectrum and for either skew-spectra set.
With the full \emph{Planck} covariance instead, the cosmological block is prior-dominated except for the amplitude: the maps still halve the error on~$\ln(10^{10}A_s)$ and tighten~$n_s$ by~$15\%$.

\begin{table}[h!]
    \centering
    \begin{tabular}{l l c c c c c}
        \hline\hline
        & & $\sigma(b_1)$ & $\sigma(b_2)$ & $\sigma(b_{\mathcal{G}_2})$ & $\sigma(b_1^{[L^2]})$ & $\sigma(\alpha)$ \\
        \hline
        \multirow{3}{*}{Bispectrum}
            & free cosmology       & $0.0518$ & $0.229$ & $0.110$  & $2.54$ & $0.870$ \\
            & + $10\times$ \emph{Planck} prior & $0.0505$ & $0.225$ & $0.110$ & $2.28$ & $0.860$ \\
            & cosmology fixed      & $0.0325$ & $0.186$ & $0.0982$ & $1.44$ & $0.403$ \\
        \hline
        \multirow{3}{*}{Skew spectra, $\mathbf{S}^{(3)}$}
            & free cosmology       & $0.0609$ & $0.269$ & $0.130$  & $2.64$ & $0.903$ \\
            & + $10\times$ \emph{Planck} prior & $0.0596$ & $0.265$ & $0.128$ & $2.34$ & $0.892$ \\
            & cosmology fixed      & $0.0371$ & $0.209$ & $0.110$  & $1.47$ & $0.405$ \\
        \hline
        \multirow{3}{*}{Skew spectra, $\mathbf{S}^{(4)}$}
            & free cosmology       & $0.0541$ & $0.240$ & $0.116$  & $2.55$ & $0.871$ \\
            & + $10\times$ \emph{Planck} prior & $0.0530$ & $0.237$ & $0.115$ & $2.27$ & $0.861$ \\
            & cosmology fixed      & $0.0333$ & $0.191$ & $0.101$  & $1.44$ & $0.402$ \\
        \hline\hline
    \end{tabular}
    \caption{Marginalized~$1\sigma$ constraints on the bias and shot-noise parameters for the SPHEREx-like H$\alpha$ survey with~$k_{\rm max}=0.15\,h/{\rm Mpc}$, and~$b_1^{[L^2]}$ free, from the full bispectrum and the skew-spectra kernel sets of Eq.~\eqref{eq:kernel_sets}: with the cosmological parameters~$\{\Omega_c,\Omega_b,h,n_s,\ln(10^{10}A_s)\}$ free, with the \emph{Planck} 2018 prior widened by a factor~$10$ on the cosmological block, and with cosmology held fixed (the rows of Table~\ref{tab:constraints}).
    The widened prior leaves the degradation induced by the free cosmology essentially intact, $\sigma(b_1)$ remaining~$1.6$ times its fixed-cosmology value: the degradation is the~$b_1$--$\ln(10^{10}A_s)$ amplitude degeneracy, which only a prior on the amplitude at the full \emph{Planck} level breaks, recovering then about half of it.}
    \label{tab:cosmo_constraints}
\end{table}

\begin{figure}[h!]
    \centering
    \includegraphics[width=\textwidth]{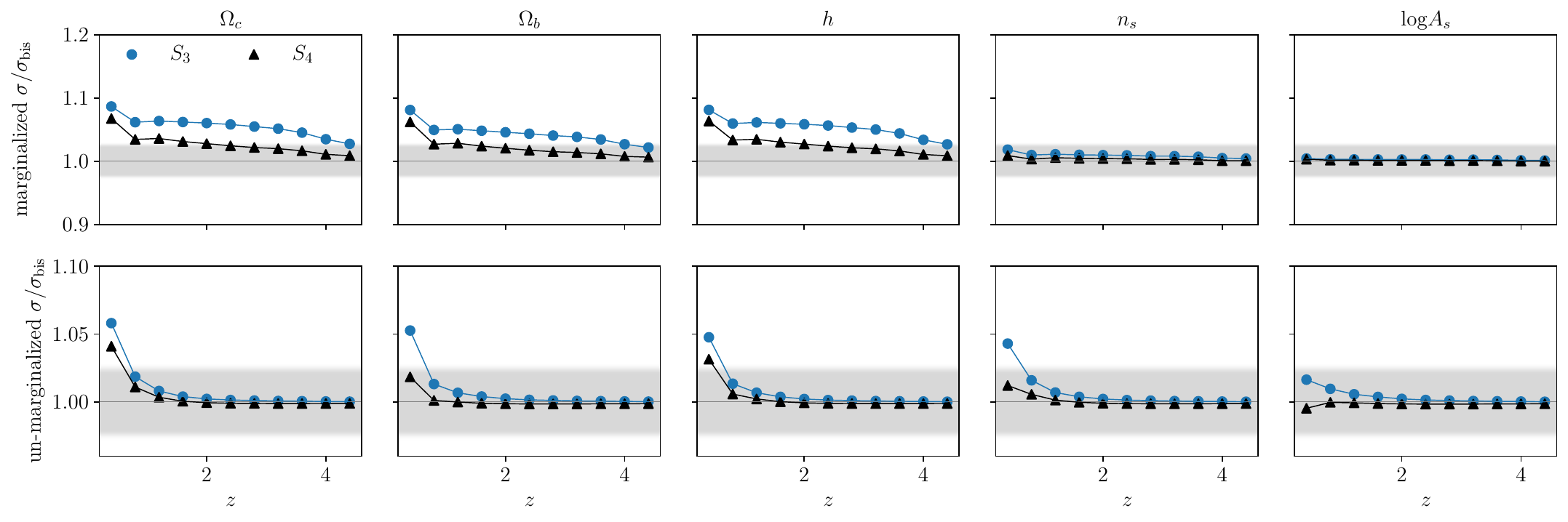}
    \includegraphics[width=\textwidth]{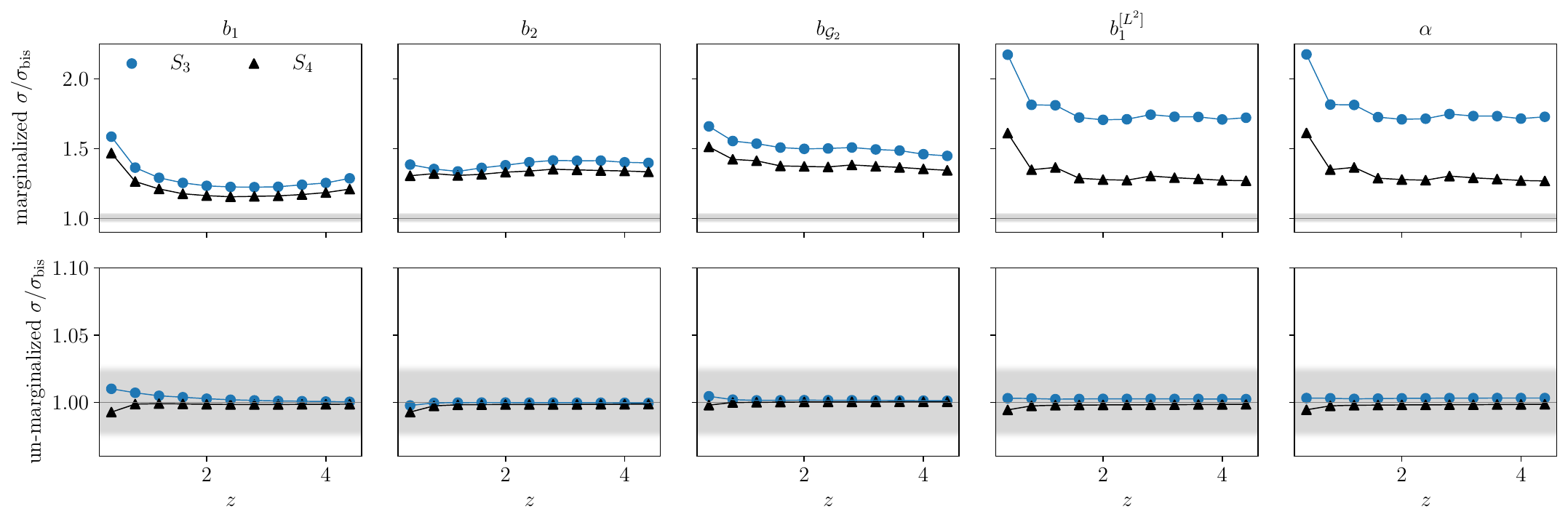}
    \caption{Per-bin constraint ratios between the compressed statistics and the full bispectrum, marginalized (\emph{top rows}) and un-marginalized (\emph{bottom rows}), when varying~$\{\Omega_c, \Omega_b, h, n_s, \ln(10^{10}A_s), b_1, b_2, b_{\mathcal{G}_2}, b_1^{[L^2]}, \alpha\}$ for the SPHEREx-like H$\alpha$ survey,~$k_{\rm max}=0.15\,h/{\rm Mpc}$.
    The~$\pm2\%$ gray band marks the numerical error floor: ratios inside the band should be read as unity. The marginalized constraints need a prior on the cosmological block to regularize the matrix inversion, since a single redshift contains highly degenerate directions: we show marginalized constraints including a \emph{Planck} prior widened by a factor~$10$; the marginalized ratios depend on the width of this prior because the per-bin parameter space is degenerate, whereas the un-marginalized row is prior-free.}
    \label{fig:perbin_compression_cosmo}
\end{figure}

It might seem surprising that the skew spectra recover the bispectrum information on shape-like parameters nearly as well as they do for amplitude-like parameters.
More insight can be gained by inspecting the per-bin constraining power, both before and after marginalization, shown in Fig.~\ref{fig:perbin_compression_cosmo}.
For amplitude-like parameters ($\ln(10^{10}A_s)$, biases, shot-noise parameter), the skew spectra recover the full bispectrum information bin by bin before marginalization, but lose part of it in the marginalization, most severely when only three filtering kernels are employed.
For the shape parameters no equivalence is expected, and the per-bin unmarginalized constraints show that bispectrum and skew spectra do not encode the same information in the low-redshift bins, which are the ones with the least $k$-leverage; the difference is at most~$5\%$ in the lowest bin and disappears beyond~$z\simeq 1.5$.
Moreover, when marginalizing, the imperfect recovery of the non-amplitude parameters penalizes the biases and~$\alpha$ as well.
The per-bin marginalized constraints shown in Fig.~\ref{fig:perbin_compression_cosmo} include a \emph{Planck} prior widened by a factor~$10$ in order to break high-degeneracy directions, which a single bin cannot resolve: the resulting constraints on cosmological parameters are prior-dominated for all statistics, so that their ratios lie within~$10\%$ of unity; it remains visible in the bias panels, however, that the skew spectra do not recover all of the bispectrum information, per bin, in the presence of non-amplitude-like parameters.

\begin{figure}[t]
    \centering
    \includegraphics[width=0.8\textwidth]{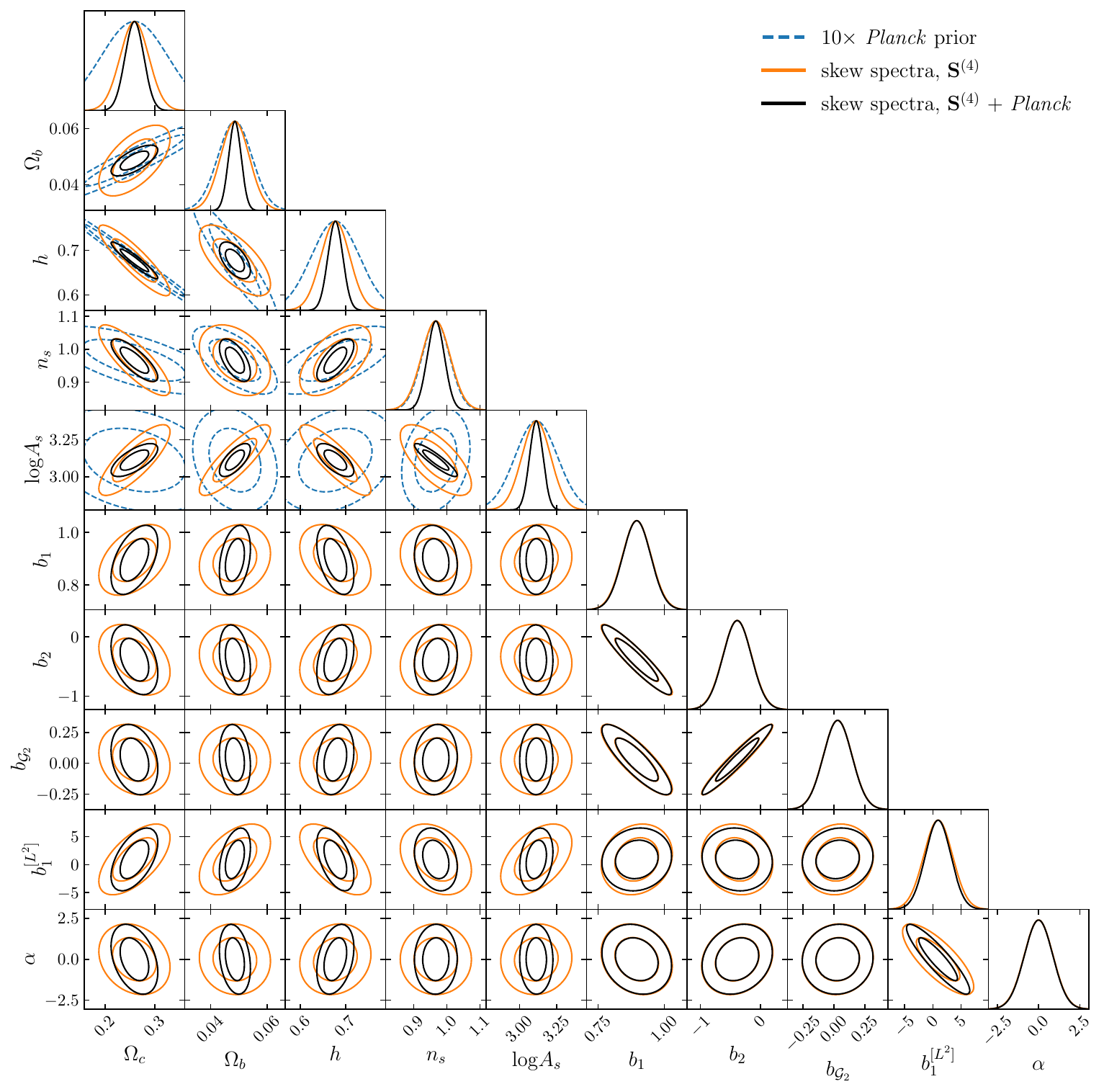}
    \caption{Marginalized constraints on the parameters of Fig.~\ref{fig:cosmo_constraints} from the four-kernel skew spectra~$\mathbf{S}^{(4)}$ alone (orange), from the \emph{Planck} 2018 prior widened by a factor~$10$ alone (blue dashed; Sec.~\ref{sec:parameters}), and from their combination (black).}
    \label{fig:cosmo_constraints_Planckprior}
\end{figure}

Figure~\ref{fig:cosmo_constraints_Planckprior} shows the effect of including a~$10 \times$ \emph{Planck} prior on the cosmological parameter block, for the four-kernel skew spectra.
The factor has been chosen so that the LIM constraints and the information from the prior are roughly comparable.

\section{FFTLog implementation}
\label{app:fftlog}

The mode-coupling integrals~$I_Y^{(n)}$ and~$I_Y^{(n),\rm bare}$ of Appendix~\ref{app:formalism}, and the shot-noise integrals~$J^{(n)}$ and~$\tilde J^{(n)}$ of Eq.~\eqref{eq:J_integrals}, are evaluated via FFTLog, following the formalism of Ref.~\cite{Simonovic:2017mhp} extended to the inverse-variance weighted kernels of Eq.~\eqref{eq:weighted_kernel}.
We validated the method against adaptive cubature and brute-force quadrature of the exact integrands (at the~$10^{-4}$--$10^{-6}$ level per component): FFTLog is one or two orders of magnitude faster than adaptive cubature (about $30\times$ faster at $k_{\rm max}=0.11\,h/{\rm Mpc}$, $80\times$ at $0.15$).
This is a necessary step towards sampling-based analyses in which the model is evaluated at every likelihood call.

\paragraph{The integrals.}
Every integral entering the skew-spectrum signal has the form
\begin{equation}
    \int \frac{d^3\vec{q}}{(2\pi)^3}\, \omega(q)\,\omega(p)\,
    D_2^{(n)}(\vec{q},\vec{k}-\vec{q})\, \mathcal{F}(\vec{q},\vec{k}) \, ,
    \qquad p \equiv |\vec{k}-\vec{q}| \, ,
\label{eq:fftlog_conv}
\end{equation}
where~$\omega(x) = W(x)\,P_m(x)/P_{\rm obs}(x)$ combines the inverse-variance weight of Eq.~\eqref{eq:weighted_kernel} with the window of Eq.~\eqref{eq:tanh_window}, so that~$\omega(q)\,\omega(p)\,D_2^{(n)} = W(q)\,W(p)\,\tilde{D}_2^{(n)}$ is the weighted kernel of Appendix~\ref{app:formalism}, and~$P_{\rm obs}$ includes the shot and the instrumental noise.
The kernel~$D_2^{(n)}$ is either one of the clustering kernels of Eq.~\eqref{eq:kernels} or one of the shot-sector kernels of Sec.~\ref{sec:shot_kernels}, $1/[P_m(q)P_m(p)]$ and~$[P_m(q)+P_m(p)]/[P_m(q)P_m(p)]$; the orthogonalized kernel is a linear combination of these.
The~$P_m(k)$ term is pulled out of the bare integrals.

The function~$\mathcal{F}$ takes four forms (see Eqs.~\eqref{eq:I_DE}--\eqref{eq:I_DE_bare} for the clustering integrals, Eq.~\eqref{eq:J_integrals} for the shot noise integrals),
\begin{equation}
    \mathcal{F} = \begin{cases}
        Y(\vec{q},\vec{k}-\vec{q})\,P_m(q)\,P_m(p) & I^{(n)}_Y \, , \\
        Y(\vec{q},-\vec{k})\,P_m(q) & I^{(n),\rm bare}_Y \, , \\
        1 & J^{(n)} \, , \\
        P_m(q) & \tilde{J}^{(n)} \, , \\
    \end{cases}
\end{equation}
with~$Y \in \{F_2, \Legendre{0}, \Legendre{2}\}$, therefore each kernel requires six clustering convolutions and two shot-noise ones.

\paragraph{Reduction to power-law pairs.}
These integrals are separable since (1) the weights factorize, $\omega(q)\,\omega(p)$, because the inverse-variance weighting is applied to each field before correlating (its functional form is irrelevant, the window and the noise are simply part of the function being decomposed), and (2) every product~$D_2^{(n)}\mathcal{F}$ is a finite sum of monomials in~$q^2$, $p^2$ and~$k^2$, for example
\begin{equation}
    F_2(q,p,\mu) = \frac{5}{14} + \frac{3}{28}\frac{k^2}{q^2} + \frac{3}{28}\frac{k^2}{p^2}
    - \frac{5}{28}\frac{q^2}{p^2} - \frac{5}{28}\frac{p^2}{q^2} + \frac{1}{14}\frac{k^4}{q^2p^2} \, ,
\end{equation}
where~$\mu = (k^2-q^2-p^2)/(2qp)$ is the cosine of the angle between~$\vec{q}$ and~$\vec{k}-\vec{q}$.
Each component is a table of terms~$c\,k^{2\gamma}\int [q^{2a} g_q(q)]\,[p^{2b} g_p(p)]$ with integer~$(a,b,\gamma)$.
For all kernels, the functions fall into three classes only: $g = \omega P_m$, $\omega$, and~$\omega/P_m$.
The integer powers~$q^{2a}$, $p^{2b}$ with~$a,b\in[-2,3]$ are absorbed into the functions before decomposition; if these powers were instead carried as shifts of the exponents in Eq.~\eqref{eq:fftlog_master}, individual terms would diverge in the infrared or ultraviolet and the analytic continuation would silently drop finite, cutoff-dependent pieces.

\paragraph{Decomposition and the separable master integral.}
Each function of~$x=q,p$ is sampled on~$N$ logarithmically spaced points covering the support~$[k_{\rm f}, k_{\rm max}]$ extended by~$n_{\rm pad}$ decades of zeros on either side, and expanded by a fast Fourier transform in~$\ln x$ as
\begin{equation}
    g(x) = \sum_{m=-N/2}^{N/2-1} c_m\, x^{\nu + i\eta_m} \, , \qquad
    \eta_m = \frac{2\pi m}{N\,\Delta\ln x} \, ,
\label{eq:fftlog_decomp}
\end{equation}
where~$\nu$ is a real bias chosen per function.
The highest-frequency coefficients are tapered over the top~$25\%$ of the mode range to suppress ringing from the cutoffs.
Every pair of terms then reduces to the master integral~\cite{Simonovic:2017mhp}
\begin{equation}
    \int \frac{d^3\vec{q}}{(2\pi)^3}\, q^{-2\nu_1}\,|\vec{k}-\vec{q}|^{-2\nu_2}
    = \frac{k^{3-2\nu_{12}}}{8\pi^{3/2}}\,
      \frac{\Gamma(\tfrac32-\nu_1)}{\Gamma(\nu_1)}\,
      \frac{\Gamma(\tfrac32-\nu_2)}{\Gamma(\nu_2)}\,
      \frac{\Gamma(\nu_{12}-\tfrac32)}{\Gamma(3-\nu_{12})} \, ,
    \qquad \nu_{1,2} = -\tfrac12\left(\nu_{q,p} + i\eta_{m,n}\right),\ \ \nu_{12}=\nu_1+\nu_2 \, ,
\label{eq:fftlog_master}
\end{equation}
which converges for~$\nu_q,\nu_p > -3$ and~$\nu_q+\nu_p < -3$.
The~$\Gamma$ ratio is written as a product of a factor depending on~$m$ alone, one on~$n$ alone, and one on~$m+n$ alone; because the frequencies~$\eta_m$ are uniformly spaced, the last factor depends only on~$s = m+n$ and the double mode sum collapses to a linear convolution,
\begin{equation}
    \mathcal{I}(k) = k^{3+\nu_q+\nu_p}\sum_{s} T_s\, k^{i\eta_s} \, , \qquad
    T_s = \frac{1}{8\pi^{3/2}}\,\frac{\Gamma(\nu_{12}(s)-\tfrac32)}{\Gamma(3-\nu_{12}(s))}\,
    \big[(c A)\star(d B)\big]_s \, ,
\label{eq:fftlog_conv_sum}
\end{equation}
with~$A_m$ and~$B_n$ the single-index~$\Gamma$ ratios and~$c_m$,~$d_n$ the coefficients of the two functions.
The convolution is one zero-padded transform of length~$2N$ per pair of functions, at cost~$\mathcal{O}(N\log N)$ and independent of~$k$; each external wavenumber then costs a single~$\mathcal{O}(N)$ sum, and no~$N\times N$ matrix is ever formed.
The bias~$\nu$ can be chosen separately for each function at no extra cost, which is what makes the next step possible.

\paragraph{Biases.}
Functions carrying the extreme absorbed powers~$x^{-4}$, present in the displacement and tidal tables, span a large dynamic range above the infrared cutoff.
At a single common bias this produces an aliasing error growing with~$k$ that reaches~$14\%$ for the displacement kernel at~$N=4096$ with~$4.5$ decades of padding, and~$23\%$ for the tidal kernel even at~$N=8192$ with six decades of padding.
Assigning~$\nu = -1.8 + 0.3\,a$ to the function carrying~$x^{2a}$ keeps every pair inside the convergence window of Eq.~\eqref{eq:fftlog_master}, since the pairs that occur have~$a+b \leq 1$ and the ultraviolet condition requires~$a+b<2$, and removes the effect entirely at~$N=2048$; the window is checked at run time for every pair actually used.

\paragraph{Validation.}
The method was validated with a three-way comparison.
For every component and wavenumber we tabulated the adaptive cubature (Divonne method in the CUBA library~\cite{Hahn:2004fe,Hahn:2005pf}, relative tolerance~$10^{-5}$), the FFTLog evaluation, and a brute-force midpoint quadrature of the exact integrand on the identical domain, cutoffs, weights, window and kernel.

With the top-hat window in~$k$ used in earlier work, both FFTLog and Divonne carry systematics at the~$10^{-3}$ level tied to the discontinuity of the integrand.
With the smoothed window of Eq.~\eqref{eq:tanh_window}, both methods agree with the quadrature to~$2.3\times10^{-4}$ of the component scale at worst and to a few~$10^{-6}$ typically.
The comparison with the bispectrum, which is binned with hard triangle bounds and no window, remains fair: the net distortion of the~$k^3$-weighted mode volume with respect to a sharp cutoff is~$3\pi^2/(8\beta^2) \simeq 1.5\times10^{-3}$ at leading order
in~$1/\beta$.

We checked that the complete Fisher forecasts run end to end with the two signal evaluations, FFTLog versus Divonne, agree on every marginalized error to~$3.6\times10^{-4}$ with four kernels and~$7.2\times10^{-4}$ with three at~$k_{\rm max}=0.15\,h/{\rm Mpc}$.

\paragraph{Settings and cost.}
The runs use~$N=2048$ points, $n_{\rm pad}=4.5$ decades, the~$25\%$ coefficient taper and the per-shift biases above.
The signal derivatives of one redshift bin drop from~$25\,{\rm s}$ to~$0.3\,{\rm s}$ with four kernels, and from~$15\,{\rm min}$ to~$4\,{\rm s}$ when five numerical cosmology derivatives multiply the number of integrals.

The covariance is evaluated once at the fiducial cosmology by adaptive integration at relative tolerance~$10^{-5}$; it is now the dominant cost of a forecast.
The diagonal term~$I^{(n,m)}(k)$ of Eq.~\eqref{eq:I_nm} has the same structure as Eq.~\eqref{eq:fftlog_conv} and can be evaluated with the same machinery, which would also allow proper~$k$-bin averaging (Appendix~\ref{app:covreg}); the off-diagonal term~$J^{(n,m)}(k,k')$ of Eq.~\eqref{eq:J_nm} is a one-dimensional angular integral and is already cheap.

It is worth mentioning one limitation of the FFTLog implementation: the weight must be isotropic.
This requirement breaks down in redshift space, where~$P_{\rm obs}$ depends on the angle to the line of sight, or in the presence of an anisotropic noise; in those cases~$\omega$ should be expanded in Legendre multipoles first, each multipole contributing a separable term.

\section{Regularization of the skew-spectrum covariance}
\label{app:covreg}

The Fisher forecasts of Sec.~\ref{sec:results} require the inverse of the~$N_c N_k \times N_c N_k$ covariance of the skew spectra data vector, Eq.~\eqref{eq:skew_cov}.
For numerical stability, the inversion is performed on the correlation matrix,~$r_{mn} = {\rm Cov}_{mn}/\sqrt{{\rm Cov}_{mm}\,{\rm Cov}_{nn}}$, and transformed back.
This step is essential and makes the analysis exactly invariant under arbitrary rescalings of individual data-vector entries: the raw covariance spans~$\mathcal{O}(10^{18})$ between the clustering and (dimensionful) shot-sector blocks, which would render a direct inversion meaningless in double precision, whereas the correlation matrix, and hence the eigenvalue structure discussed below, is independent of the normalization of the kernels.
In this appendix we show that, for the four-kernel data vector, a direct inversion is unreliable -- not because of the conditioning of the linear algebra, but because the smallest eigenvalue of the computed correlation matrix lies below the accuracy at which its entries are evaluated -- and we describe the regularization procedure adopted in the main text.
The covariance, and therefore everything in Secs.~\ref{app:covreg_floor} and~\ref{app:covreg_spectrum}, does not depend on the parameters varied; the tests of the Fisher inequalities in Sec.~\ref{app:covreg_projection} were performed for the set~$\{b_1,b_2,b_{\mathcal{G}_2},\alpha\}$ directly comparable with the galaxy-clustering results of Ref.~\cite{MoradinezhadDizgah:2019xun}.

\subsection{Measuring the error floor of the computed covariance}
\label{app:covreg_floor}

A covariance matrix is positive semi-definite by construction, so any negative eigenvalue of the computed correlation matrix is an unambiguous signature of numerical error: if the exact matrix~$r^{\rm true}$ is PSD and the computed one is~$r = r^{\rm true} + \delta r$, then the most negative eigenvalue of~$r$ bounds the error norm from below,~$\|\delta r\| \geq |\lambda_{\rm min}|$.
This is a measurement of the accuracy of the assembled matrix that requires no convergence test of the individual integrals.
For the four-kernel data vector, the computed correlation matrix has a negative eigenvalue,~$\lambda_{\rm min} \simeq -7\times10^{-3}$, in the first redshift bin, which is the coarsest ($N_k = 4$) and the one with the largest discretization error, while in the remaining bins the smallest eigenvalue is positive but small,~$\lambda_{\rm min} \simeq 10^{-4}$--$1.6\times10^{-3}$, i.e.,~consistent with zero at the accuracy set by the first bin.
We adopt
\begin{equation}
\label{eq:eps_floor}
    \varepsilon \simeq 2\times10^{-3}
\end{equation}
as the error floor of the correlation matrix over most of the redshift range, a few times below the bound from the first bin.
The clipping threshold of the pseudo-inverse, Eq.~\eqref{eq:eigclip}, is accordingly placed \emph{above} this floor, as described below.
Two error sources contribute at this level: the accuracy of the individual covariance integrals (a few parts in~$10^{4}$ of the relevant scale for the adaptive integration, accumulating over the~$\mathcal{O}(N_c^2 N_k^2)$ entries), and the evaluation of the continuous covariance expressions at the~$k$-bin centers, which at our coarse binning~($\Delta k = 2 k_{\rm f}$) introduces discretization errors of comparable size and can itself violate positive-definiteness.

Positive-definiteness is the diagnostic, not the criterion: since the inverse assigns each mode the weight~$1/\lambda$, any eigenvalue that is not well separated from zero on the scale of~$\varepsilon$ is untrusted, whatever its sign.
By this criterion the three-kernel spectrum is bounded away from zero ($\lambda_{\rm min} \gtrsim \varepsilon$) in every bin, while the four-kernel one is consistent with zero ($|\lambda_{\rm min}| \lesssim \varepsilon$) in every bin.

\subsection{Spectral structure of the four-kernel covariance}
\label{app:covreg_spectrum}

\begin{figure}[h]
    \centering
    \includegraphics[width=0.7\textwidth]{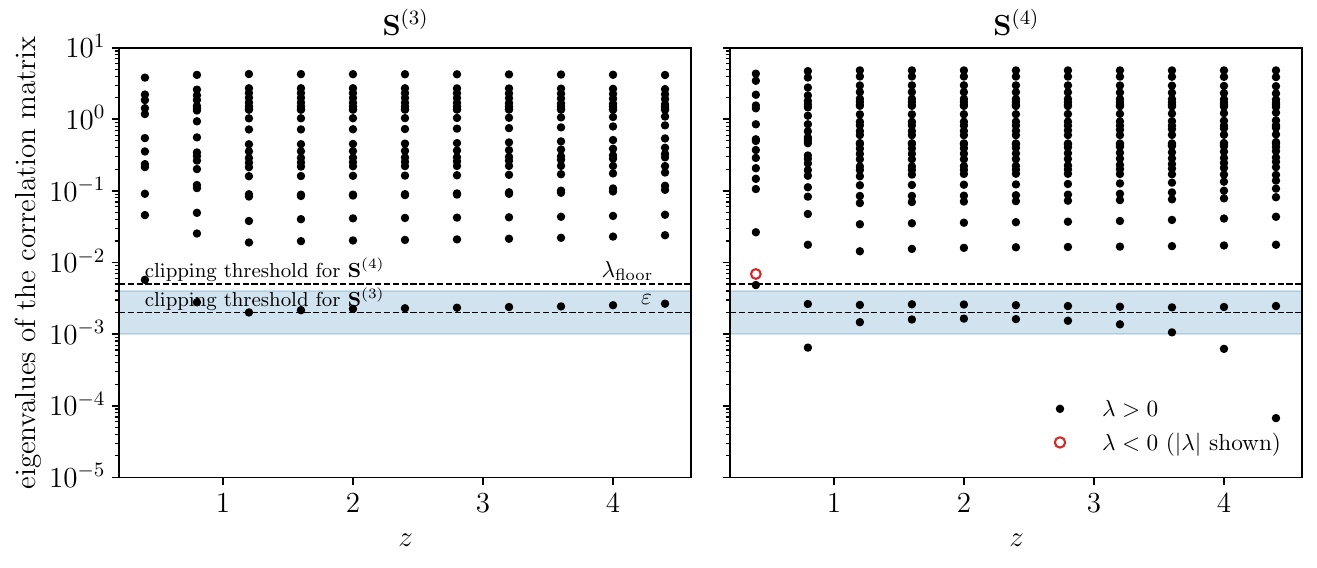}
    \caption{Eigenvalue spectrum of the computed correlation matrix of the three-kernel (\emph{left}) and four-kernel (\emph{right}) skew-spectrum data vector, per redshift bin, before regularization. Exactly one collective eigen-direction per bin lies at or below the measured error floor~$\varepsilon = 2\times10^{-3}$ of Eq.~\eqref{eq:eps_floor}, shown as a horizontal band; a second eigenvalue lies close to it; the rest of the spectrum is separated by an order of magnitude. The three-kernel spectrum does not fall below the floor in any bin. The dashed lines mark the clipping thresholds, $\lambda_{\rm floor}$ for $\mathbf{S}^{(4)}$ and~$\varepsilon$ for $\mathbf{S}^{(3)}$; the negative eigenvalue of the first bin is drawn as an open circle at its absolute value.}
    \label{fig:cov_spectrum}
\end{figure}

The eigenvalue spectrum of the correlation matrix has a characteristic structure, illustrated in Fig.~\ref{fig:cov_spectrum}.
In every redshift bin, \emph{exactly one} eigenvalue lies at or below the error floor of Eq.~\eqref{eq:eps_floor}, with~$\lambda_{\rm min} \in \left[ -6.9\times10^{-3},\, +1.6\times10^{-3} \right]$ across bins, while the next eigenvalue sits at~$\simeq 2.5\times10^{-3}$ and the third at~$\gtrsim 1.4\times10^{-2}$: the lowest eigenvalue is untrusted, the second is marginal, and both are separated from the rest of the spectrum by an order of magnitude.
This is not an artifact of the choice of the fourth kernel: the orthogonalization of Sec.~\ref{sec:shot_kernels} is an invertible linear change of basis of the data vector, which maps a near-null direction of the covariance onto a near-null direction, so the number of untrusted modes is the same for~$\cS_{\rm orth}$ and~$\cS_{\rm shot}$, and the Fisher forecasts obtained with the two coincide.
Moreover, the untrusted eigenvalue cannot be traced back to a single kernel or~$k$-bin: the corresponding eigenvector is a \emph{collective} mode that, in the orthogonalized basis, has~$36$ to~$56\%$ of its weight in the density block and the remainder spread over the other three kernels, with the balance between displacement kernel and shot sector kernel shifting with redshift, and always spread across all~$k$-bins.

\subsection{Consequences of direct inversion, and the projection}
\label{app:covreg_projection}

Writing the correlation matrix as~$r = \sum_i \lambda_i\, \vec{v}_i \vec{v}_i^{\,\top}$, the inverse weights each eigen-direction by~$1/\lambda_i$.
Whenever a computed eigenvalue~$\lambda$ lies below the numerical floor, its true value is unknown at our accuracy and the computed one may even be negative.
A direct (LU) inversion then assigns this mode an essentially arbitrary statistical weight of magnitude~$\left|1/\lambda\right| \gtrsim 10^{3}$, which corrupts the Fisher matrix at order unity.
This corruption shows up in our forecasts as an inconsistent amount of information, in two ways.
A small positive~$\lambda$ contributes spuriously \emph{inflated} information: the skew-spectra forecasts yield more information on the Fisher parameters than the full-bispectrum ones, a clear violation of the information inequality for a compressed statistic.
A negative~$\lambda$ assigns to the corresponding mode a negative weight, so that adding the fourth channel apparently \emph{degraded} the constraints, violating the monotonicity of Fisher information under the addition of data.

We therefore replace the direct inverse by the spectral pseudo-inverse
\begin{equation}
\label{eq:eigclip}
    r^{-1}_{\rm reg} = \sum_{\lambda_i \,\geq\, \lambda_{\rm floor}} \frac{1}{\lambda_i}\, \vec{v}_i \vec{v}_i^{\,\top} \, , \qquad \lambda_{\rm floor} = 5\times10^{-3} \, .
\end{equation}
The floor was placed in the middle of the spectral gap: it removes the below-floor direction together with the marginal one (two modes per redshift bin, out of~$N_c N_k = 16$--$28$), rather than mis-weighting them.
Statistically, this amounts to declaring the corresponding modes as unmeasured -- their noise levels are not known at the accuracy of our covariance -- and conservatively discarding their information.
The projection can only lower the information of the compressed statistic; it does not by itself guarantee the two inequalities, which we therefore verified: with the regularized covariance, the four-kernel constraints are never tighter than the bispectrum ones, and never looser than the three-kernel ones, in every redshift bin and for every parameter, to within the numerical precision of~$\simeq 2\%$ quoted in Sec.~\ref{sec:results_grav} (the largest violation is the~$0.13\%$ of~$\sigma(b_1^{[L^2]})$ in Table~\ref{tab:constraints}).
The procedure is insensitive to the choice of~$\lambda_{\rm floor}$ within the gap: lowering the floor to~$2\times10^{-3}$, which retains the marginal mode, changes the forecasted constraints by less than~$0.7\%$ (while re-introducing sub-percent violations of the compression bound traceable to that mode's uncertain weight, which is why we adopt the mid-gap value as our baseline, rather than the measured numerical error floor).
For the three-kernel data vector, which exhibits no spectral gap, we set the threshold at the error floor~$\varepsilon$ itself; in practice, we verified that in a DESI-like galaxy clustering analysis~\cite{MoradinezhadDizgah:2019xun} the clipped modes ($\lambda \sim 2$--$4 \times 10^{-4}$, 1--2 per bin) change the $z$-summed constraints by only~$0.01$--$0.08\%$, while in our line-intensity mapping analysis no eigenvalue falls below the floor and no mode is removed.
Recovering the information carried by the discarded modes would require covariance entries accurate at the~$10^{-5}$--$10^{-6}$ level, attainable by evaluating the covariance integrals with the FFTLog method of Appendix~\ref{app:fftlog} together with proper~$k$-bin averaging; but, as shown in Sec.~\ref{sec:results_grav}, the regularized four-kernel forecasts already recover the full-bispectrum constraints to a few percent, leaving little to gain.

We emphasize that the near-saturation of the bispectrum information is a result, not an artifact of the regularization: the projection can only discard information, so the true (unregularized)~$\mathbf{S}^{(4)}$ forecast is bracketed between the bispectrum bound and the regularized result, within the~$2$--$3\%$ quoted in the main text.
This residual few-percent gap is thus an upper bound on the information lost in compressing the LIM bispectrum to four skew spectra.

\end{document}